\documentclass[useAMS, usenatbib]{mnras}
\pdfoutput=1
\usepackage{wjh-packages}
\usepackage{booktabs}
\hypersetup{colorlinks=True, linkcolor=blue!50!black, citecolor=black,
  urlcolor=blue!50!black}
\usepackage{etoolbox}
\robustify\bfseries
\robustify\itshape

\usepackage[shortlabels]{enumitem}

\makeatletter
\renewcommand{\fps@figure}{pt}
\renewcommand{\fps@table}{pt}
\makeatother

\DeclareSIUnit\msun{\text{M\ensuremath{_\odot}}}
\DeclareSIUnit\lsun{\text{L\ensuremath{_\odot}}}

\newcounter{ionstage}
\renewcommand{\ion}[2]{\setcounter{ionstage}{#2}%
  \ensuremath{\mathrm{#1\,\scriptstyle\Roman{ionstage}}}}

\newcommand\nii{[\ion{N}{2}]}
\newcommand\oiii{[\ion{O}{3}]}
\newcommand\oii{[\ion{O}{2}]}
\newcommand\Wav[1]{\ensuremath{\lambda #1}}
\newcommand*\chem[1]{\ensuremath{\mathrm{#1}}}

\newcommand\Fion{\ensuremath{F_{\text{ion}}}}
\newcommand\ionpar{\ensuremath{U_{\text{ion}}}}

\newcommand{\sii}{[\ion{S}{2}]}
\newcommand{\heii}{\ion{He}{2}}

\newcommand\NIIlam{[\ion{N}{2}]\,6583\,}

\newcommand\OIIIlam{[\ion{O}{3}]\,5007\,\AA\@}

\newcommand\Ha{\ensuremath{\mathrm{H}\alpha}}
\newcommand\Hb{\ensuremath{\mathrm{H}\beta}}

\newcommand{\teff}{\ensuremath{T_\mathrm{eff}}}

\title{
  The five axes of the Turtle: symmetry and asymmetry in NGC~6210
}

\author[Henney et al.]{
  William J. Henney,\(^1\)\thanks{
    w.henney@irya.unam.mx,
    jal@astro.unam.mx,
    tere@astro.unam.mx,
    richer@astro.unam.mx
  }
  J. A. López,\(^2\)\footnotemark[1]
  Ma.\ T. García-Díaz,\(^2\)\footnotemark[1]
  and M. G. Richer\(^2\)\footnotemark[1]
  \\
  \(^1\)\foreignlanguage{spanish}{
    Instituto de Radioastronomía y
    Astrofísica, Universidad Nacional Autónoma de México, Apartado
    Postal 3-72, 58090 Morelia, Michaoacán, Mexico}
  \\
  \(^2\)\foreignlanguage{spanish}{
    Instituto de Astronomía,
    Universidad Nacional Autónoma de México,
    Ensenada, Baja California, 22800, México}
}
\date{Accepted XXX. Received YYY; in original form ZZZ}

\pubyear{2020}

\begin{document} 
\label{firstpage}
\pagerange{\pageref{firstpage}--\pageref{lastpage}}
\maketitle

\begin{abstract}
  We carry out a comprehensive kinematic and morphological study
  of the asymmetrical planetary nebula:
  NGC~6210, known as the Turtle.
  The nebula's spectacularly chaotic appearance has led to proposals that it was shaped by mass transfer in a triple star system.
  We study the three-dimensional structure and kinematics
  of its shells, lobes, knots, and haloes
  by combining radial velocity mapping from multiple long-slit spectra
  with proper motion measurements from multi-epoch imaging.
  We find that the nebula has five distinct ejection axes.
  The first is the axis of the bipolar, wind-blown inner shell,
  while the second is the axis of the lop-sided, elliptical, fainter, but more massive intermediate shell.
  A further two axes are bipolar flows that form the point symmetric,
  high-ionization outer lobes, all with inclinations close to the plane of the sky.
  The final axis, which is inclined close to the line of sight,
  traces collimated outflows of low-ionization knots.
  We detect major changes in outflow directions during the planetary nebula phase, 
  starting at or before the initial ionization of the nebula 3500~years ago.
  Most notably,
  the majority of redshifted low-ionization knots have kinematic ages greater than 2000~years,
  whereas the majority of blueshifted knots have ages younger than 2000~years. 
  Such a sudden and permanent 180-degree flip in the ejection axis
  at a relatively late stage in the nebular evolution is a challenge to models of planetary nebula formation and shaping.
\end{abstract}


\maketitle

\section{INTRODUCTION}
\label{sec:introduction}
NGC~6210 is a bright and relatively large planetary nebula (PN) in the northern constellation of Hercules, first described by Friedrich Georg Wilhelm von~\citet{Struve:1827a}.\footnote{
  Facsimile available at \url{https://books.google.com.mx/books?id=Rs9JAAAAcAAJ&pg=PA88}.
}
Early photographic images \citep{Curtis:1918a, Duncan:1937a} already revealed the complex shape of the nebula, described by \citeauthor{Duncan:1937a} as having a ``\textit{bizarre form}''.
Other descriptions of the nebula include ``\textit{resembling the popular artist's conception of an atom}'' \citep{Feibelman:1971a},
or ``\textit{a mid-air collision of two intergalactic warships}'' \citep{OMeara:2007a},
or, more succinctly, ``\textit{messy}'' \citep{Soker:2004b}.

\begin{figure}
\centering
\includegraphics[width=\linewidth]{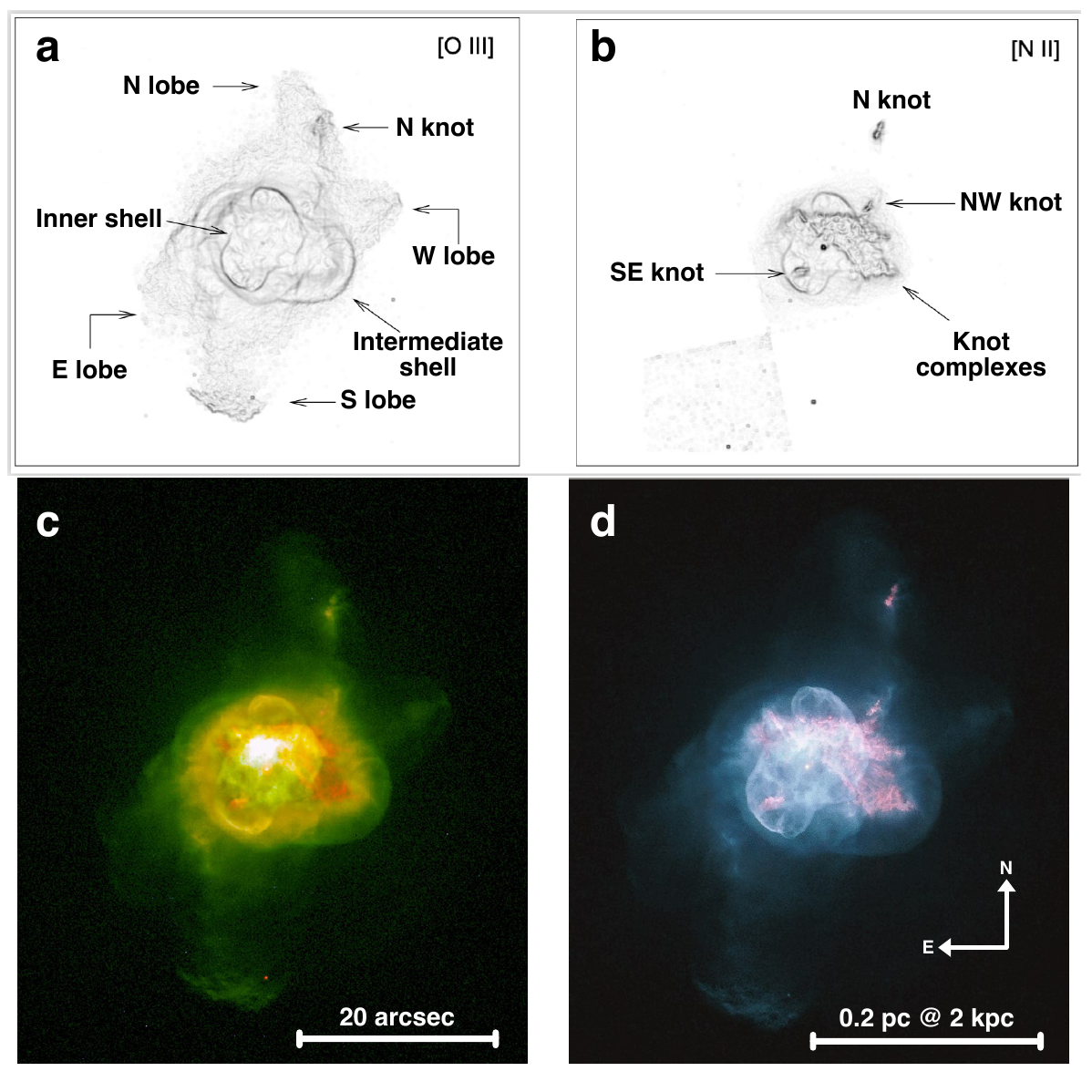}
\caption{
  Images from the \textit{HST} WFPC2 camera of NGC~6210.
  (a)~A logarithmic stretch and an edge-detection algorithm are applied to an  \oiii{} \Wav{5007} image,
  in order to accentuate small-scale structure and remove large-scale radial brightness gradients.
  The most salient nebular features are labeled for reference.
  (b)~Same, but for \nii{} \Wav{6583}.
  (c)~Two-color composite image of \oii{} \Wav{3727} (red) and \oiii{} \Wav{5007} (green).
  Image credit: Robert Rubin and Christopher Ortiz (NASA/ESA Ames Research Center), Patrick Harrington and Nancy Jo Lame (University of Maryland), Reginald Dufour (Rice University), and NASA/ESA.
  (d)~Three-color composite image of \nii{} \Wav{6583} (red), broad \(V\) band (green), and \oiii{} \Wav{5007} (blue). Image credit: ESA/Hubble and NASA.}
\label{fig:hst}
\end{figure}

To the best of our knowledge,
the first use of the name Turtle with regard to NGC~6210
was in a 1998 NASA press release%
\footnote{\url{https://hubblesite.org/contents/news-releases/1998/news-1998-36.html}}
that described images of the nebula obtained with the
\textit{Hubble Space Telescope} (\textit{HST}).
Those and more recent \textit{HST} images%
\footnote{
  \url{https://www.spacetelescope.org/images/opo9836f/}
  and \url{https://www.spacetelescope.org/images/potw1026a/}.}
are shown in Figure~\ref{fig:hst},
with the most important morphological features labelled in the upper panels.
In most emission lines,
such as \Ha{} \Wav{6563} and \oiii{} \Wav{5007},
the brightest part of the nebula is the inner shell, which is elongated approximately NNW--SSE,
with a peanut-like shape.
The larger and fainter intermediate shell is elongated on a roughly perpendicular axis
and is not centered on the progenitor star, but is offset towards the WSW.
Outside these shells are the four lobes of the nebula, which form the ``flippers'' of the Turtle.
In low-ionization lines,
such as \nii{} \Wav{6583},
the shells emit only faintly and the lobes are not seen at all.
Instead, the brightest emission comes from a series of individual knots and knot complexes (seen as pink in Figure~\ref{fig:hst}d).
The N~knot is the farthest from the star and is superimposed on the N~lobe,
with no counterpart to the S.
The NW knot and SE knot form a symmetrical pair, although the SE knot is closer to the star.
The brightest of the more diffuse knot complexes are concentrated to the N and W of the star.

Spectrophotometry of NGC~6210 at UV, optical, and infrared wavelengths
\citep{Rubin:1997a, Kwitter:1998a, Liu:2004a, Pottasch:2009a, Bohigas:2015c}
have provided ionic and elemental abundances,
together with electron densities and temperatures in the bright inner shell.
Distance estimates range from \num{1.5} to \SI{2.1}{kpc} \citep{Hajian:1995a, Frew:2016a},
and we adopt \SI{2}{kpc} in this paper.
There have been many previous studies of the internal kinematics of the nebula
\citep{Osterbrock:1966a, Weedman:1968a, Becker:1984a, Icke:1989a, Rechy-Garcia:2020a}
but the lack of a single dominant symmetry axis has made it difficult to derive a comprehensive spatio-kinematic model.
The same lack of symmetry has led to speculation \citep{Soker:2004b, Bear:2017a}
that the nebula has been shaped by an interacting triple star system.


In this work we present a thorough spectral mapping of all the morphological elements of the PN obtained at high spectral resolution,
which allows us to disentangle the various spatio-kinematic components
and provides an overall view of its structure and evolution.
In \S~\ref{sec:observations}, we describe our spectroscopic observations and corresponding data-reduction steps.
In \S~\ref{sec:proper-motions} we describe the proper motions that we derive from archival \textit{HST} images.
In \S~\ref{sec:kinematic-components} we extract kinematic components from the slit spectra and classify them into different systems under the categories of shells, knots, lobes and halo, which are each analysed.
In \S~\ref{sec:three-dimens-struct} we combine the proper motion and spectroscopic observations to reconstruct the three-dimensional structure of the nebula and derive its mass-loss history.
In \S~\ref{sec:discussion} we discuss our results in the broader context of planetary nebula studies and post-AGB stellar evolution.
Finally, in \S~\ref{sec:conclusions} we summarise our conclusions.
Technical details of the density calibration
and ionization diagnostics are given in appendices.

\section{Longslit Spectroscopy}
\label{sec:observations}
 \begin{figure}
   \centering
   \includegraphics[width=0.38\textwidth]{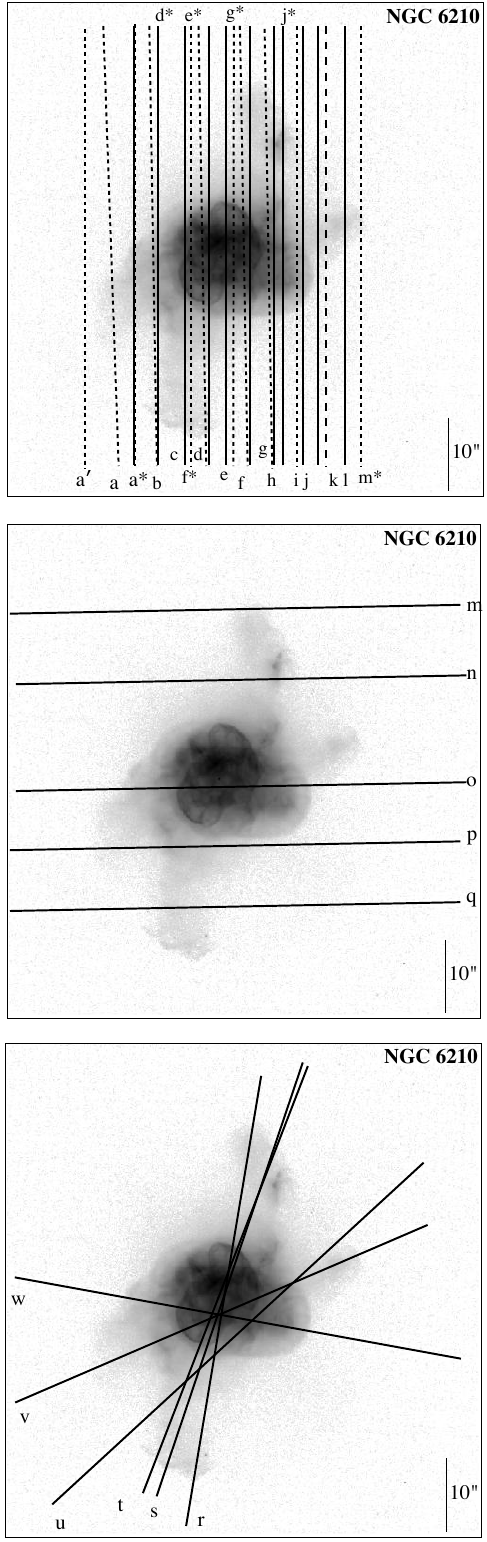}
   \caption{
     Spectrograph slit positions for vertical orientations (upper panel),
     horizontal orientations (middle panel),
     and diagonal orientations (lower panel).
     Background image in each case is an \textit{HST} \oiii{} image of the nebula.
     In the top panel, slit positions from the 2015 season are indicated as solid lines
     and positions from other seasons are shown as dashed lines (see Table~\ref{table:pa5} for details). 
   }
  \label{fig:slit-positions}
\end{figure}

\begin{table}
\centering
\caption{OAN MES-SPM high-resolution spectroscopy of NGC~6210}
\label{table:pa5}
\begin{tabular}{@{}l@{} rrrr l@{}} \toprule
  Epoch&   Exposure & Spectral  & Width & P.A.   & Slit Label\\
       &    (s) & Range  &  (\si{\um})    &    (\(^\circ\))  &  \\
  \addlinespace[1pt]
  {(1)} & {(2)} & {(3)} & {(4)} & {(5)} & {(6)} \\
  \midrule
1998/06/28 & 30 & \Ha  & 150 & 90 & o\\
1998/06/28 & 300 & \Ha  & 150 & 90 & n\\
1998/06/28 & 1200 & \Ha   & 150 & 90 & m,p,q \\
2003/06/05 & 1800 &  \Ha   & 70 &0 &     a,b,d*,e*,f,h,i,k\\ 
2003/10/16 &  1800 &   \Ha    & 70 & $-$21 & t\\
2003/10/16 &  1800 &   \Ha   & 70 & $-$68 & v\\
2003/10/17 &  1800 &   \Ha  & 70 & 77 & w\\
2004/06/13 & 1800 &   \oiii  & 70 & $-$9 & r  \\  
2004/06/14 & 1800 &  \sii &150 & $-$19 & s  \\
2004/06/13 & 1800 &   \oiii  & 70 & $-$19 & s  \\  
2004/06/13 & 1800 &   \oiii  & 70 & $-$56 & u  \\  
  2004/06/14 & 1800 &   \oiii  & 150 & 0 & a\(\prime\),l  \\
2011/05/21 & 1800 & \Ha  & 150 & 0 & g* \\
2011/05/21 & 600 & \oiii & 150 & 0  & g \\
2013/07/06 &  1800 & \Ha   & 150 & 0 & c,i,j \\
2015/08/18 &  1800 &   \Ha, \oiii  & 70 & 0 & c,d,e,f,g\\
2015/08/19 &  1800 &   \Ha, \oiii   & 70 & 0 & b,a*,i\\
2015/08/20 &  1800 &   \Ha, \oiii & 70 & 0 & h,j,k\\
2019/09/18 &  1800 &   \Ha    & 150 & 56 & x\\
  \bottomrule
  \addlinespace
  \multicolumn{6}{@{}p{\columnwidth}@{}}{
  \textsc{Columns:}
  (1)~Date of each observing run.
  (2)~Length of each spectroscopic exposure.
  (3)~Spectral range of each observation.  \Ha{} includes \nii{} \Wav{6548, 83} and \heii{} \Wav{6560}.  \sii{} includes both doublet lines \Wav{6716, 31}.
  (4)~Slit width in microns.
  (5)~Position angle of slit orientation in degrees.
  (6)~Labels assigned to each slit position, see Fig.~\ref{fig:slit-positions}.
  }
\end{tabular}
\end{table}

\begin{figure*}
  \centering
  \includegraphics[width=0.95\textwidth]{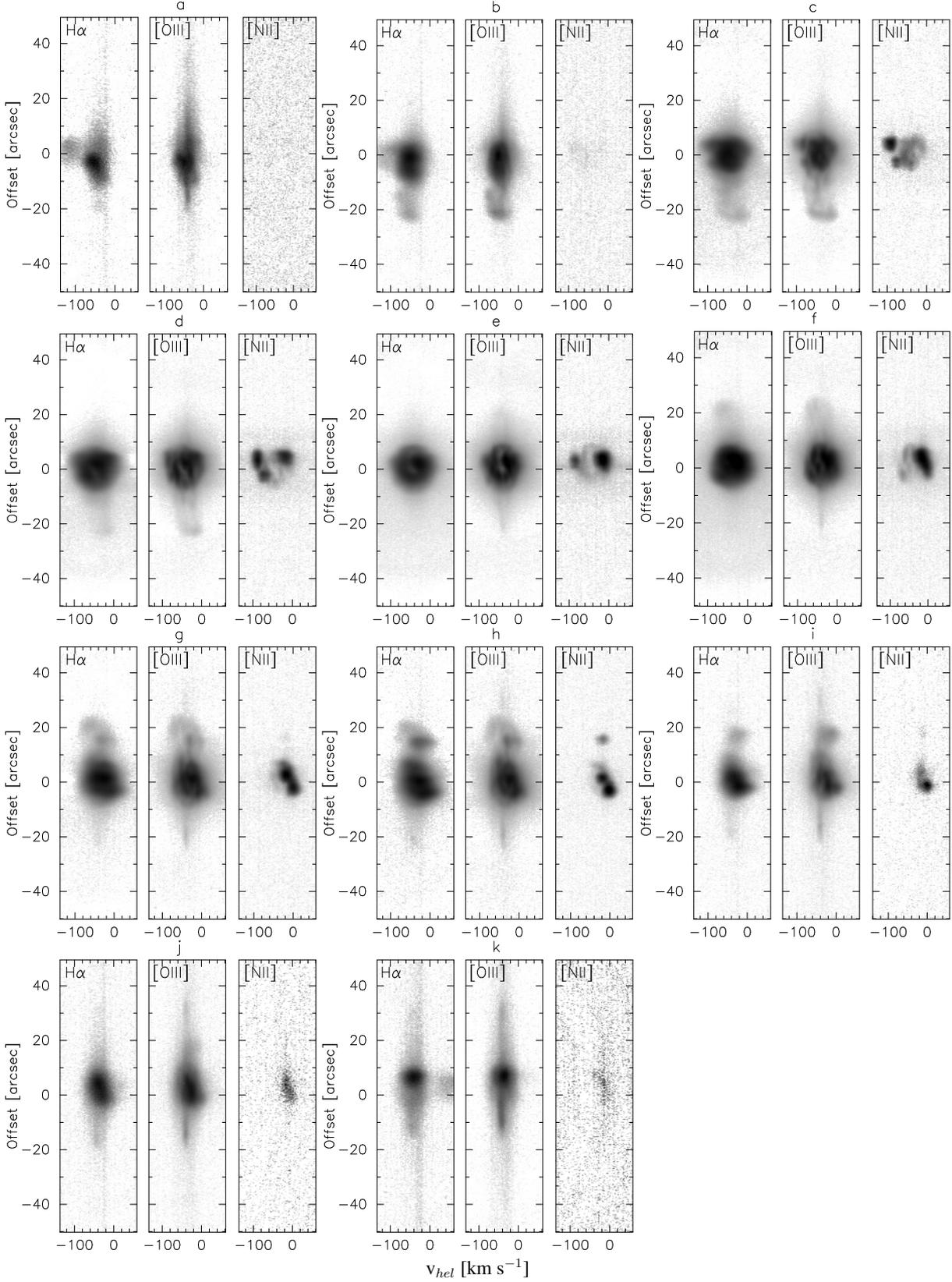}
  \caption{Position--velocity line profiles for the slit positions from the 2015 run are shown in sets where
the H$\alpha$ bi-dimensional line profile is on the left panel of the set, the \oiii{} line profile is on the central panel 
and the \nii{} line profile is on the right panel. The corresponding slit positions are indicated at the top of
each set of line profiles. The velocity scale is heliocentric and the spatial scale is indicated in arcsec 
with respect to the central star.}
  \label{fig:pv-array}
\end{figure*}

\begin{figure}
  \centering
  \includegraphics[width=\linewidth]{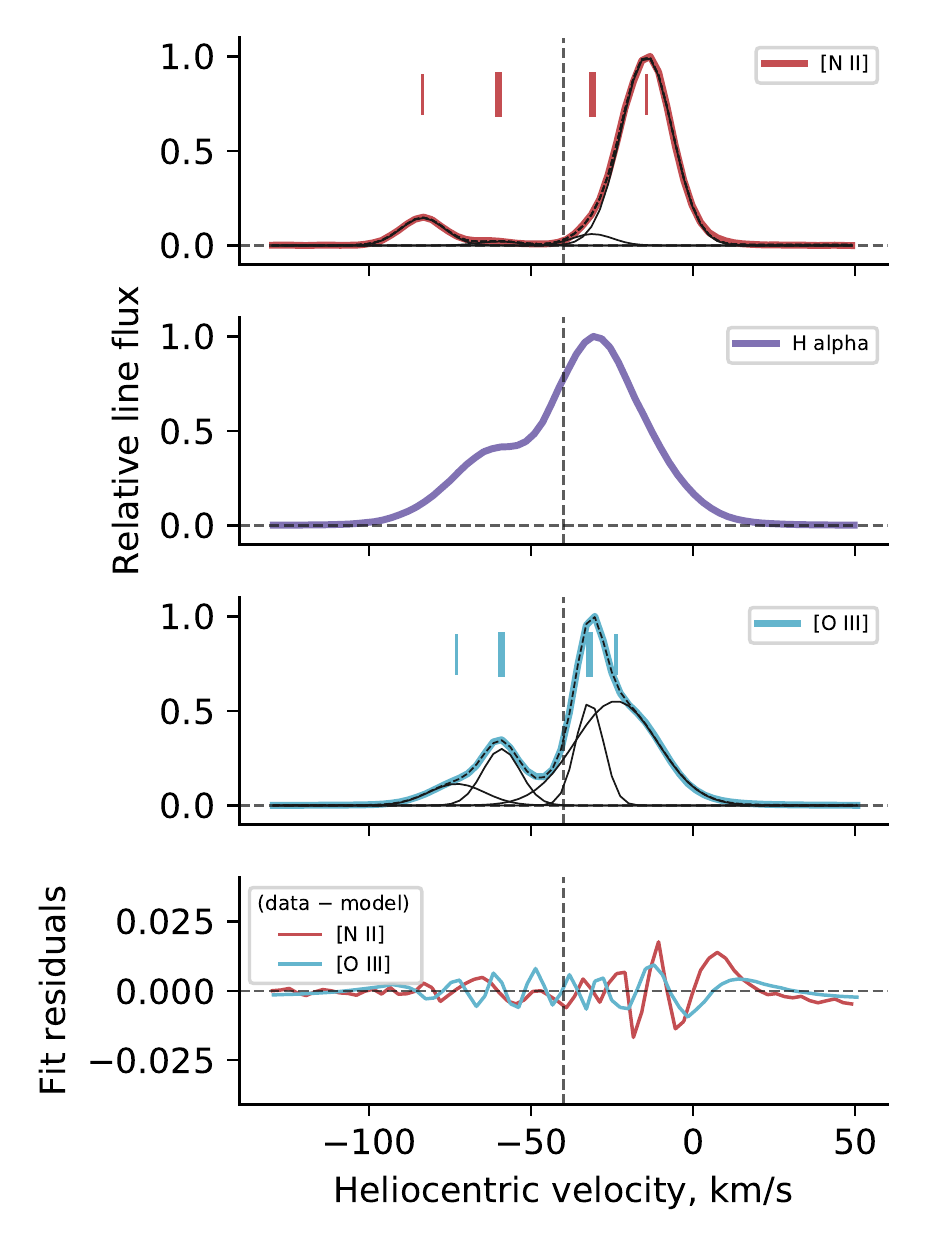}
  \caption{
    Thick colored lines show one-dimensional extracted velocity profiles
    of (top to bottom) \nii{} (red), \Ha{} (purple) and \oiii{} (light blue)
    for a  short section of slit~e
    (length \SI{2}{arcsec},
    centered \SI{3}{arcsec} north of the central star).
    For the \nii{} and \oiii{} lines, four-gaussian fits are shown
    (continuous thin black lines show the individual components,
    while thin dashed line shows the total profile).
    The bottom panel shows the residuals of these fits on an expanded scale.
    Short vertical colored lines mark the centroid velocity of each of the four components,
    with thicker lines corresponding to the high-ionization shell
    and thinner lines low-ionization knot complexes.
    The vertical dashed line shows the systemic velocity.
    No fits are performed to the \Ha{} line
    because the larger thermal broadening means that
    the four components cannot be reliably separated.
  }
  \label{fig:spec-1d}
\end{figure}

\begin{figure}
  \centering
  \includegraphics[width=\linewidth]{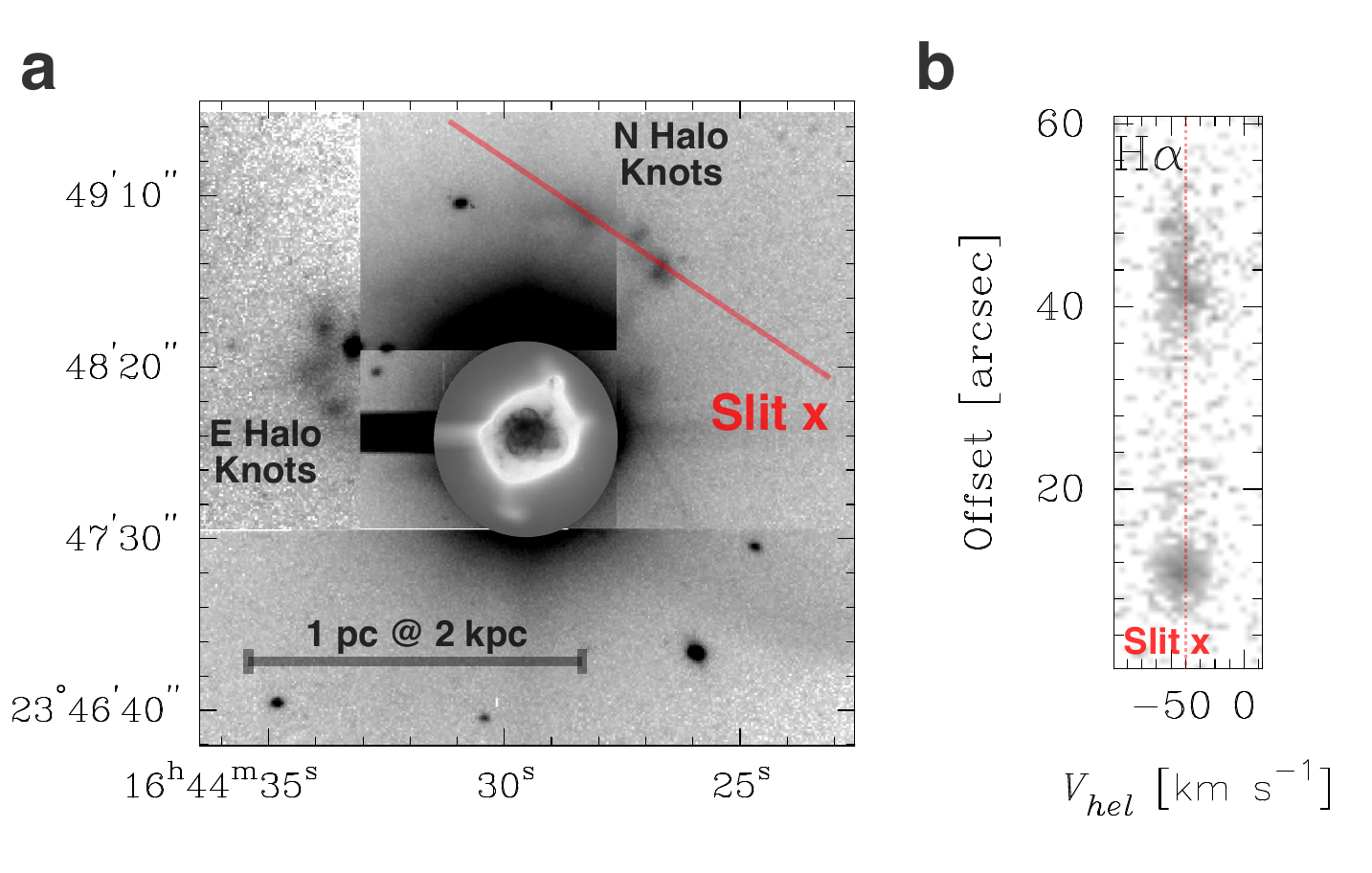}
  \caption{
    Deep images and spectra of the outer halo of NGC~6210, showing knotty structure.
    (a)~Composite mosaic of \oiii{} images,
    each with an exposure time of 1800~s,
    giving a combined field of view of \(3' \times 3'\) centered on the nebula.
    The position of slit~x is shown, for which a deep \Ha{} spectrum was obtained.
    Note that different sections of the mosaic are shown with different brightness normalizations,
    leading to abrupt jumps at the boundaries.
    The images were taken with an occulting bar that blocks out the bright core of the nebula,
    allowing the faint halo to be seen.
    (b)~\Ha{} spectrum from slit~x, showing emission from two of the halo knots.
    The red dotted line shows the nebular nominal systemic velocity.
  }
  \label{fig:halo-knots}
\end{figure}

Longslit echelle spectroscopic observations of the nebula NGC~6210 were performed with the
2.1~m telescope at the Observatorio Astronómico Nacional at San Pedro Mártir (OAN-SPM), Baja California, México.
We used the Manchester Echelle Spectrometer (MES-SPM) \citep{Meaburn:2003a} on the 2.1 m telescope in its $f$/7.5 configuration.
The MES-SPM is a long-slit, echelle spectrometer that has no cross-disperser; it isolates single orders using interference filters.
We used a 90\,\AA\, and 50\,\AA\, bandwidth filter to isolate the 87th and 114th orders containing the \Ha\,$+$\,\NIIlam\, and \OIIIlam\, nebular emission lines, respectively.
Spectroscopic data on NGC 6210 was collected within our regular observing programs over nine observing runs from 1998 to 2019.
Table~\ref{table:pa5} shows the log of observations,
and the location of each slit is shown on an \oiii{} \textit{HST} archive image in Figure~\ref{fig:slit-positions}.
In the top panel the vertical solid lines represent the set of 11 slits obtained in the 2015 run.
The vertical dashed lines represent 12 slit positions from other runs, in some cases they are highly coincident in location with the 2015 set but selected for reasons such as different exposure times or spectral range.
The middle panel shows 5 slit positions oriented E-W and the bottom panel 6 slit positions obtained at position angles that align with specific features, such as collimated outflows or the twisted arms of the nebula.
In order to establish the exact position with respect to the nebula of the slit in each pointing, we use an automatic procedure available in MES-SPM prior to the spectroscopic exposure.

During the 21 year span of the observations, the spectrograph was fitted with a variety of CCD detectors: \mbox{TEK-1} (1998), \mbox{SITE-3} (2003--2004), Marconi E2V-4240 (2011-2019).
The \mbox{TEK-1} and \mbox{SITE-3} detectors consisted of \(1024 \times 1024\) square pixels, 
each \SI{24}{\um} on a side,
used with \(2\times2\) binning to give a plate scale of \SI{0.624}{arcsec} per double-binned pixel
and a projected slit length of \SI{5.32}{arcmin}.
The spectral resolution was \SI{11.5}{km.s^{-1}} for the \SI{150}{\um} (\SI{1.9}{arcsec}) slit
and \SI{4.6}{km.s^{-1}} for the \SI{70}{\um} (\SI{0.95}{arcsec}) slit.
The Marconi detector has \(2048 \times 2048\) square pixels, each \SI{13.5}{\um} on a side.
Again, \(2\times2\) binning was used to give a plate scale of \SI{0.352}{arcsec} per double-binned pixel
and a projected slit length of \SI{5.47}{arcmin},
with a spectral resolution of \SI{11.9}{km.s^{-1}} (\SI{150}{\um} slit)
or \SI{5.9}{km.s^{-1}} (\SI{70}{\um} slit).


All data was reduced using standard IRAF\footnote{IRAF was originally
  distributed by the National Optical Astronomy Observatory,
  \url{http://ast.noao.edu/data/software},
  and is currently available from \url{https://iraf-community.github.io}.}
routines.
Bias and cosmic-ray removal was followed by rectification and first-order wavelength 
calibration of the two-dimensional spectra
based on the comparison spectrum
of a Th/Ar lamp, yielding an accuracy in velocity units of \SI{\pm 1}{km.s^{-1}}.
Figure~\ref{fig:pv-array} shows example position-velocity arrays of (left-to-right)
\Ha{} \Wav{6563}, \oiii{} \Wav{5007}, and \nii{} \Wav{6583} for the 11 individual positions from the 2015 run.
The velocity scale is heliocentric 
and the scale along the slit is specified as arcsecond offsets from the central star.
Figure~\ref{fig:spec-1d} shows example one-dimensional spectra,
extracted from a short section of slit~e.
An additional slit (x, shown in Figure~\ref{fig:halo-knots}) was placed on the northern outskirts of the nebula 
in the outer halo of NGC~6210 where filaments had been clearly detected in some of our deep images.\footnote{For this spectrum only, \(3 \times 3\) pixel on-chip binning was used.}
Most of the spectra presented here are publicly available from the \mbox{San Pedro Mártir} Kinematic Catalogue of Galactic Planetary Nebulae\footnote{
  \url{http://kincatpn.astrosen.unam.mx}
}
\citep{Lopez:2012a}.

% ...


\begin{figure}
  \centering
  \includegraphics[width=\linewidth]{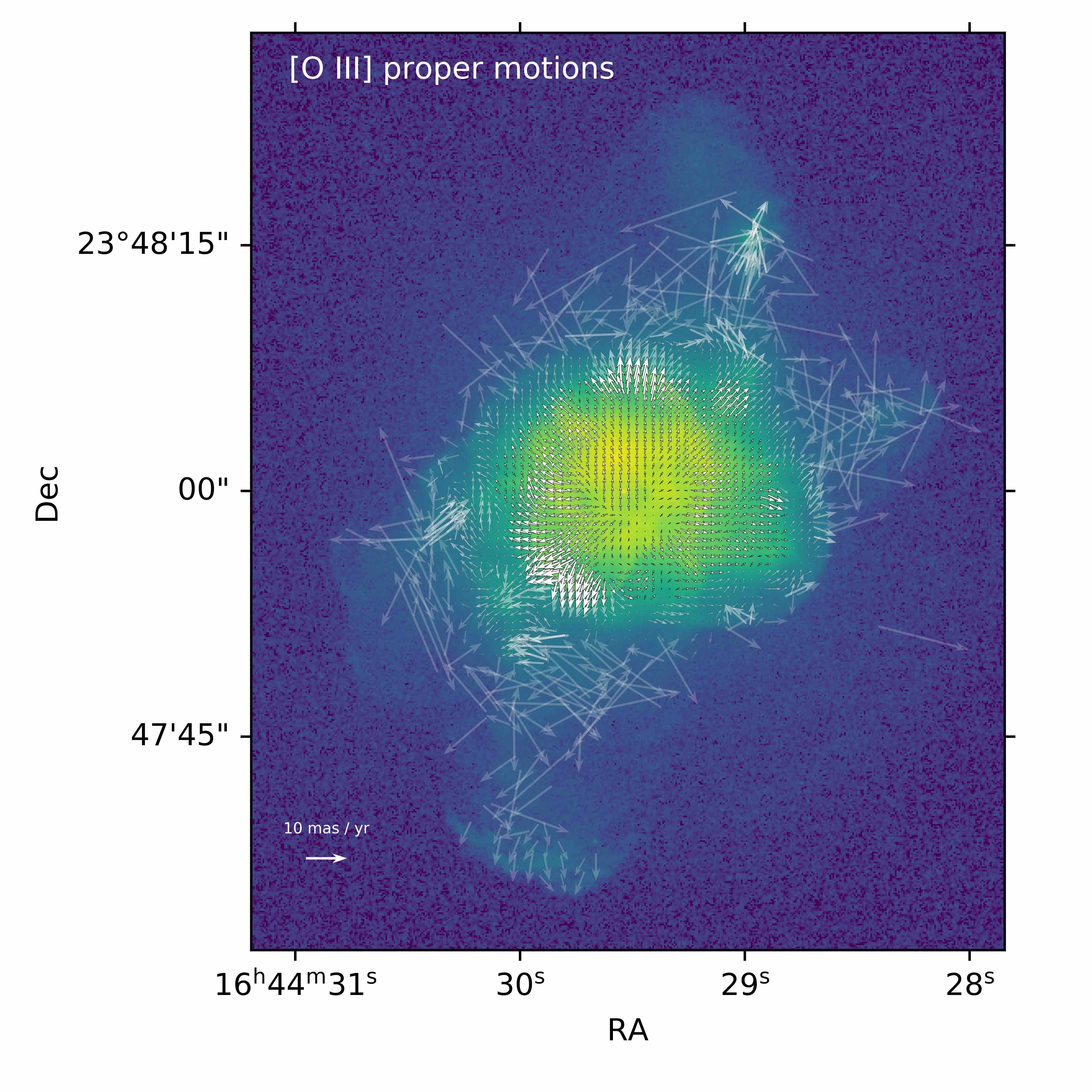}
  \caption{Proper motions derived from two HST \oiii{} images (F502N
    filter) separated by 10.45 years, using the FLCT algorithm with a
    Gaussian window width of 10~pixels. The key at bottom left shows a
    proper motion of \SI{10}{mas.yr^{-1}}, corresponding to
    \SI{95}{km.s^{-1}} for an assumed distance of \SI{2}{kpc}.}
  \label{fig:proper-motions-oiii}
\end{figure}

\begin{figure}
  \centering
  \includegraphics[width=\linewidth]{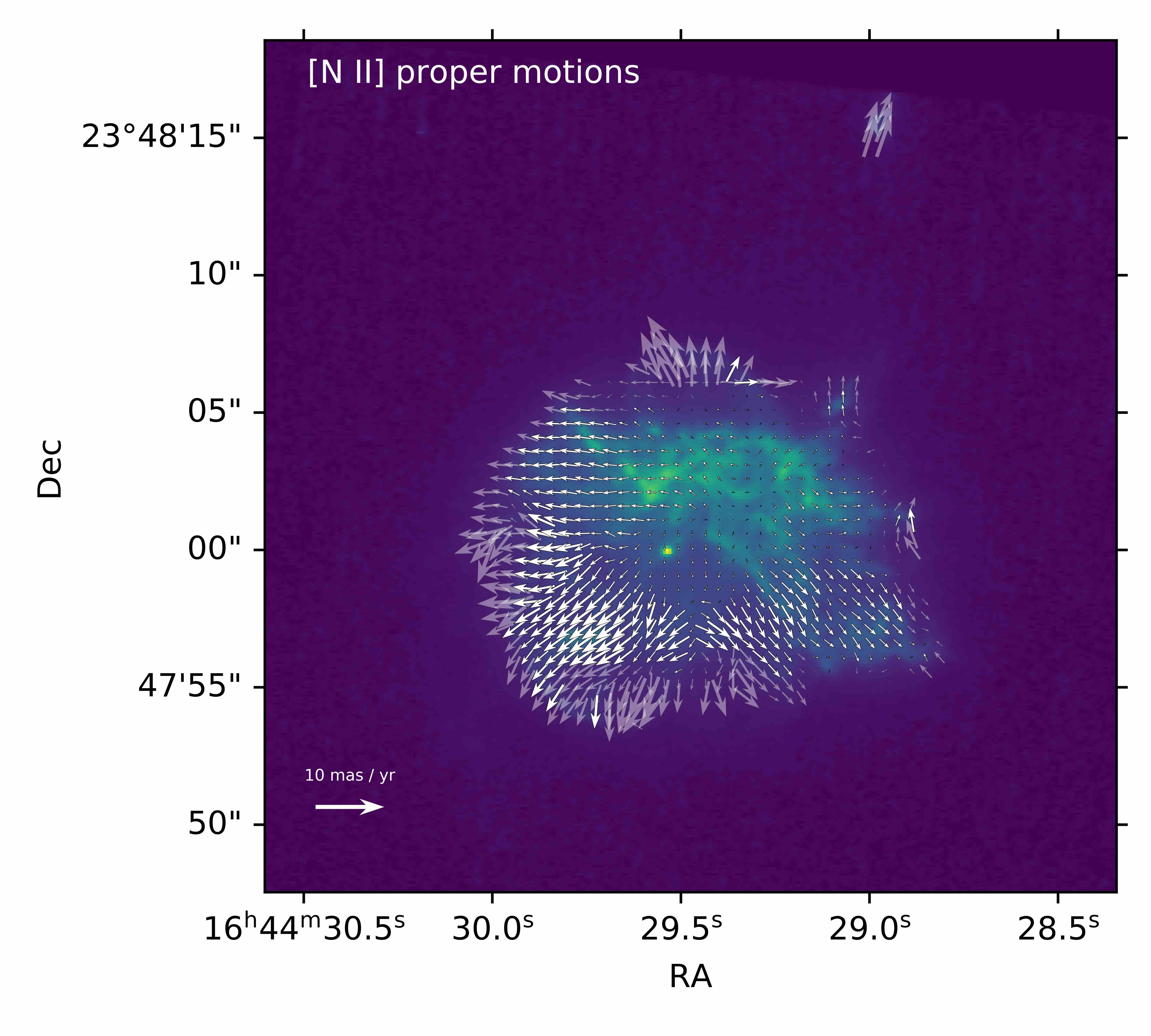}
  \caption{As Fig.~\ref{fig:proper-motions-oiii} but for two HST
    \nii{} images (F658N filter). Note that the field of view is
    cropped slightly smaller than for \oiii{}.}
  \label{fig:proper-motions-nii}
\end{figure}

\begin{figure*}
  \centering
  \includegraphics[width=\linewidth]{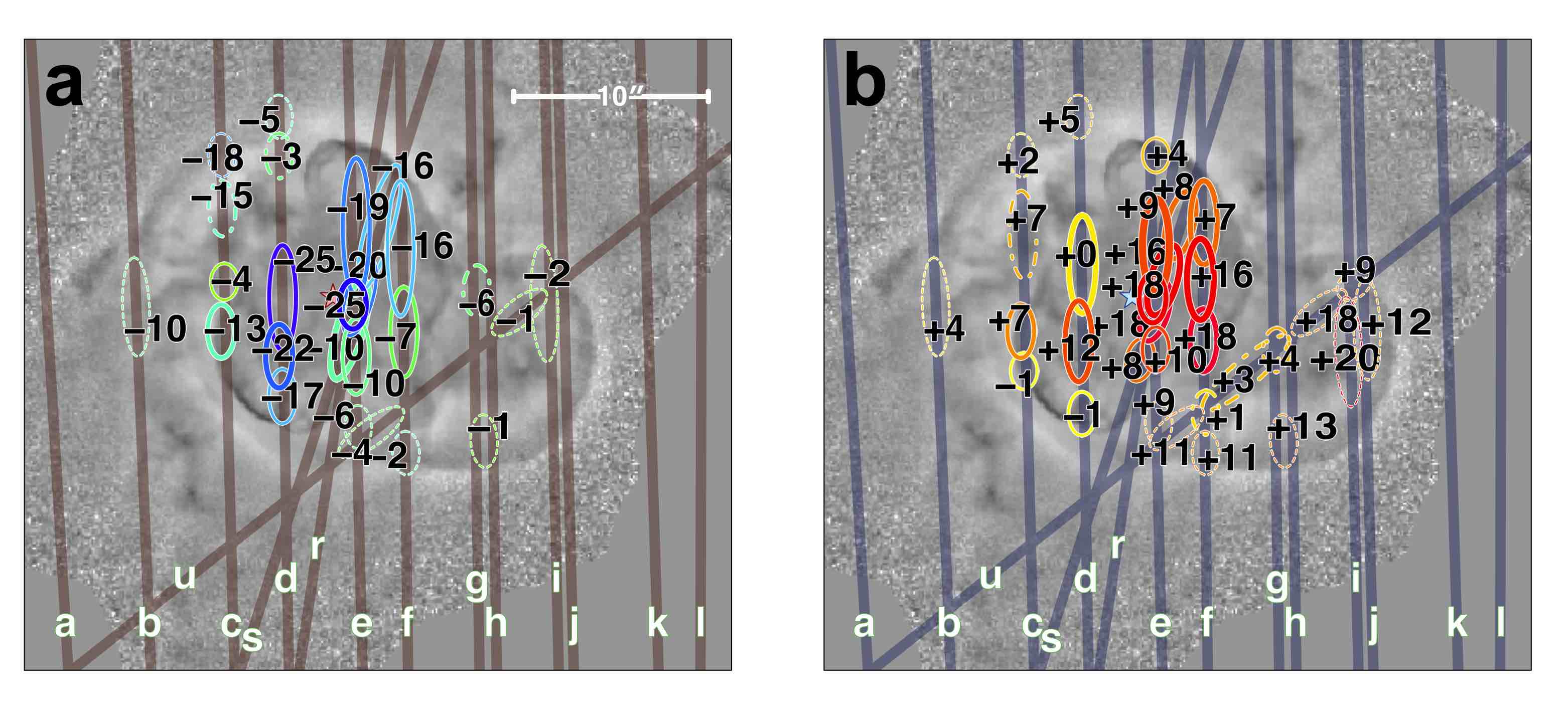}
  \caption{
    Velocity features in the high-ionization shells,
    which have been identified in the \oiii{} slits.
    (a)~Blue-shifted features. 
    (b)~Red-shifted features.
    Each are labelled with their line-of-sight velocity
    with respect to the nominal systemic velocity of \SI{-40}{km.s^{-1}}.
    Solid lines show features in the inner shell,
    dashed lines show features in the intermediate shell,
    and dot-dashed lines show miscellaneous features between the two shells.
    The line width is a qualitative indicator of the brightness of each feature.
    The background grayscale shows a high-pass filtered version of the HST \oiii{} image.
    The star symbol in this and subsequent figures indicates the position of the central star.
  }
  \label{fig:shell-velocity-components}
\end{figure*}

\begin{figure*}
  \centering
  \includegraphics[width=\linewidth]{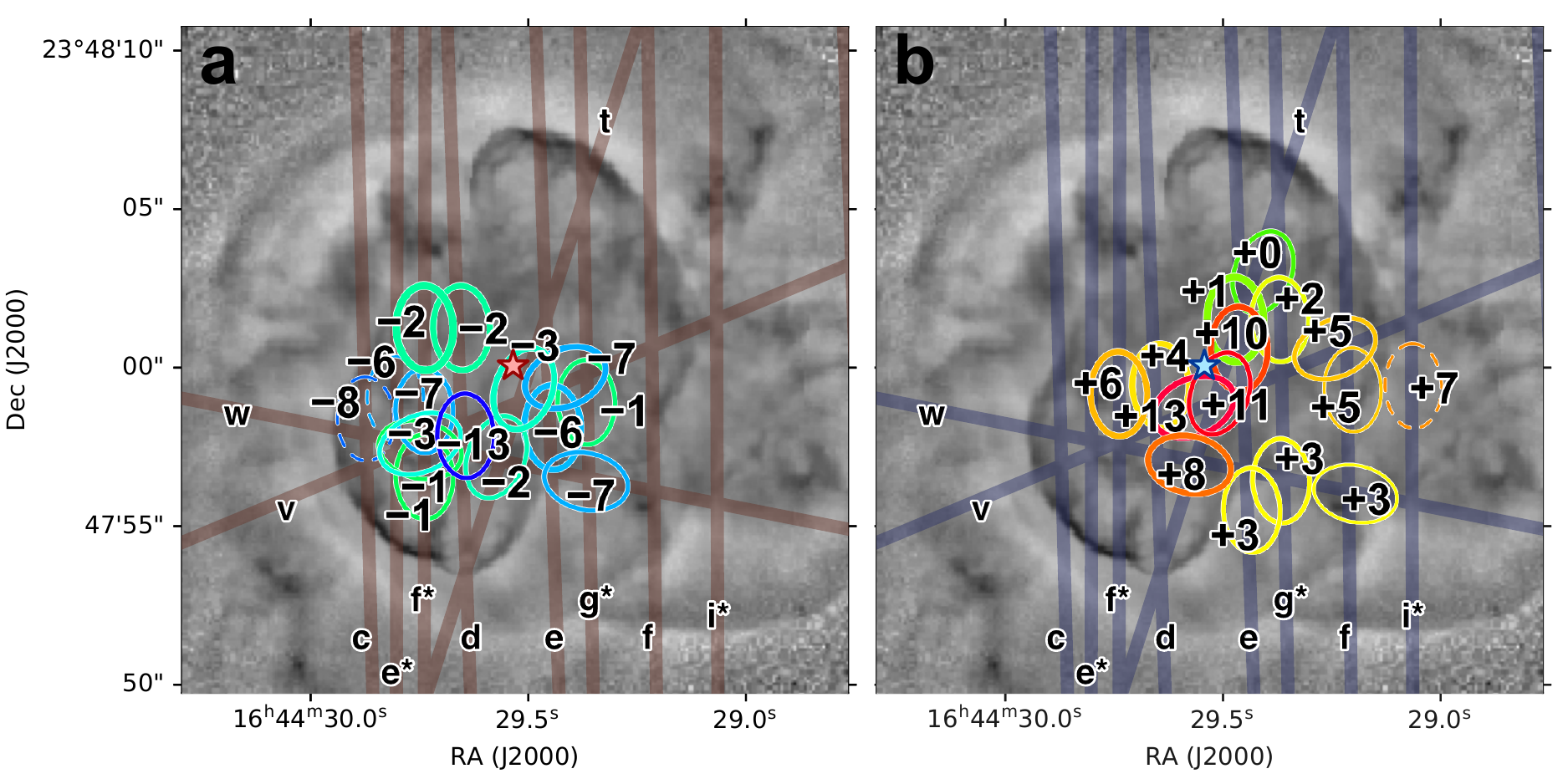}
  \caption{
    Velocity features in the highest-ionization gas,
    which have been identified in the \heii{} slits.
    (a)~Blue-shifted features. (b)~Red-shifted features.
    Each are labelled with their line-of-sight velocity
    with respect to the nominal systemic velocity of \SI{-40}{km.s^{-1}}.
    The line width is a qualitative indicator of the brightness of each feature,
    with dashed lines showing the very faintest features.
    The background grayscale image is the same as in Fig.~\ref{fig:shell-velocity-components}.
  }
  \label{fig:heii-shell-components}
\end{figure*}

\begin{figure*}
  \centering
  \includegraphics[width=\linewidth]{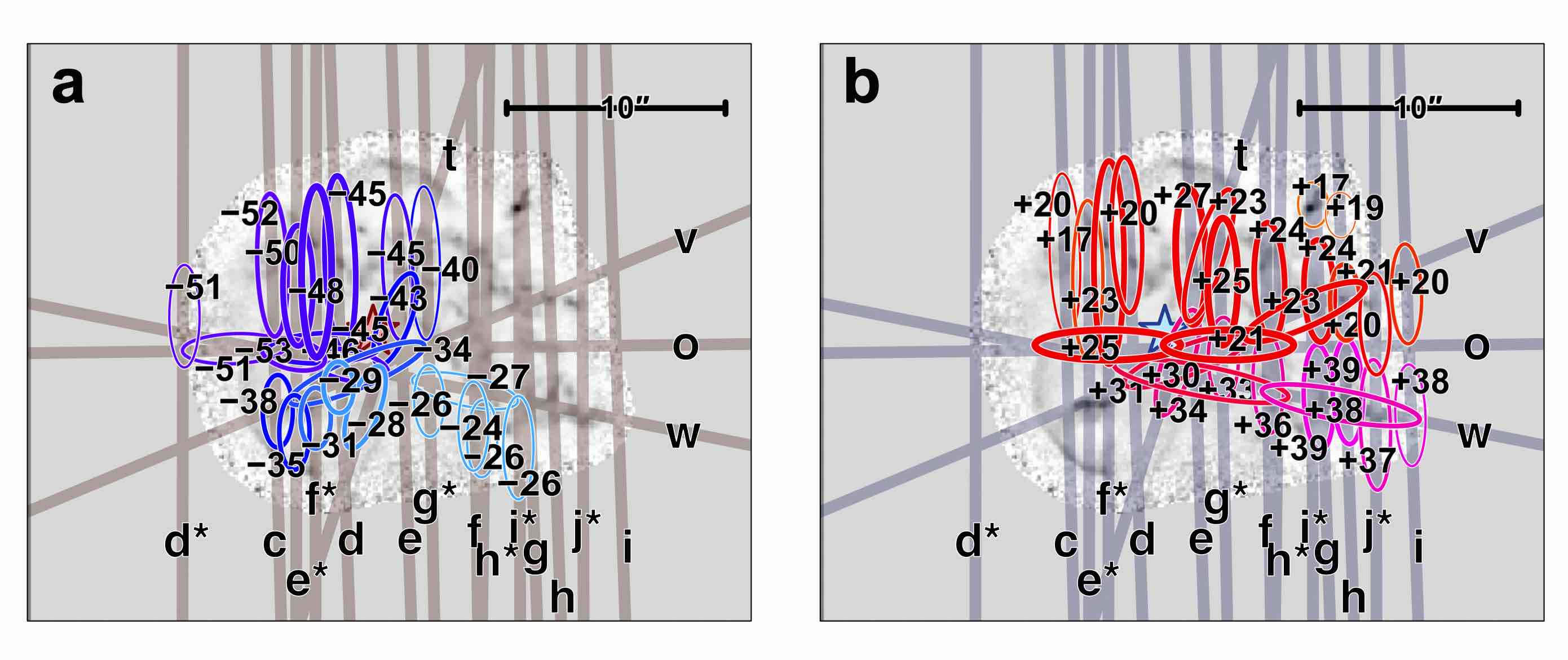}
  \caption{
    Velocity features in the low-ionization knot complexes,
    which have been identified in the \nii{} slits.
    (a)~Blue-shifted features.
    (b)~Red-shifted features.
    Each is labelled with their line-of-sight velocity
    with respect to the nominal systemic velocity of \SI{-40}{km.s^{-1}}.
    The line width is a qualitative indicator of the brightness of each feature.
  }
  \label{fig:knot-complex-map}

  \bigskip
  \includegraphics[width=\linewidth]
  {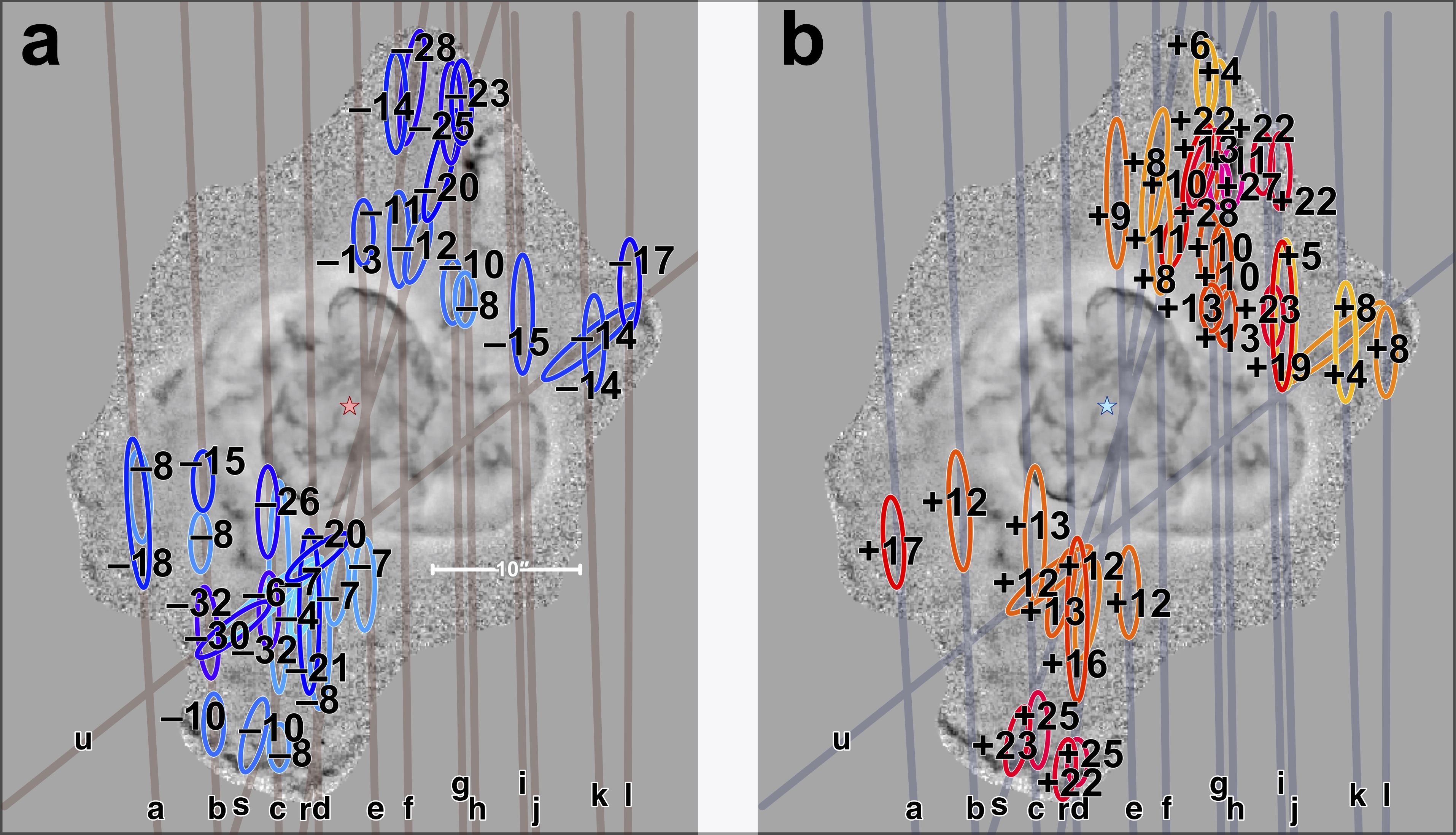}
  \caption{
    Velocity components in the outer lobes,
    which have been identified in the \oiii{} slits.
    (a)~Blue-shifted components identified in individual slits.
    (b)~Red-shifted components identified in individual slits.
  }
  \label{fig:outer-lobe-components}
\end{figure*}

\begin{figure}
  \centering
  \includegraphics[width=\linewidth]{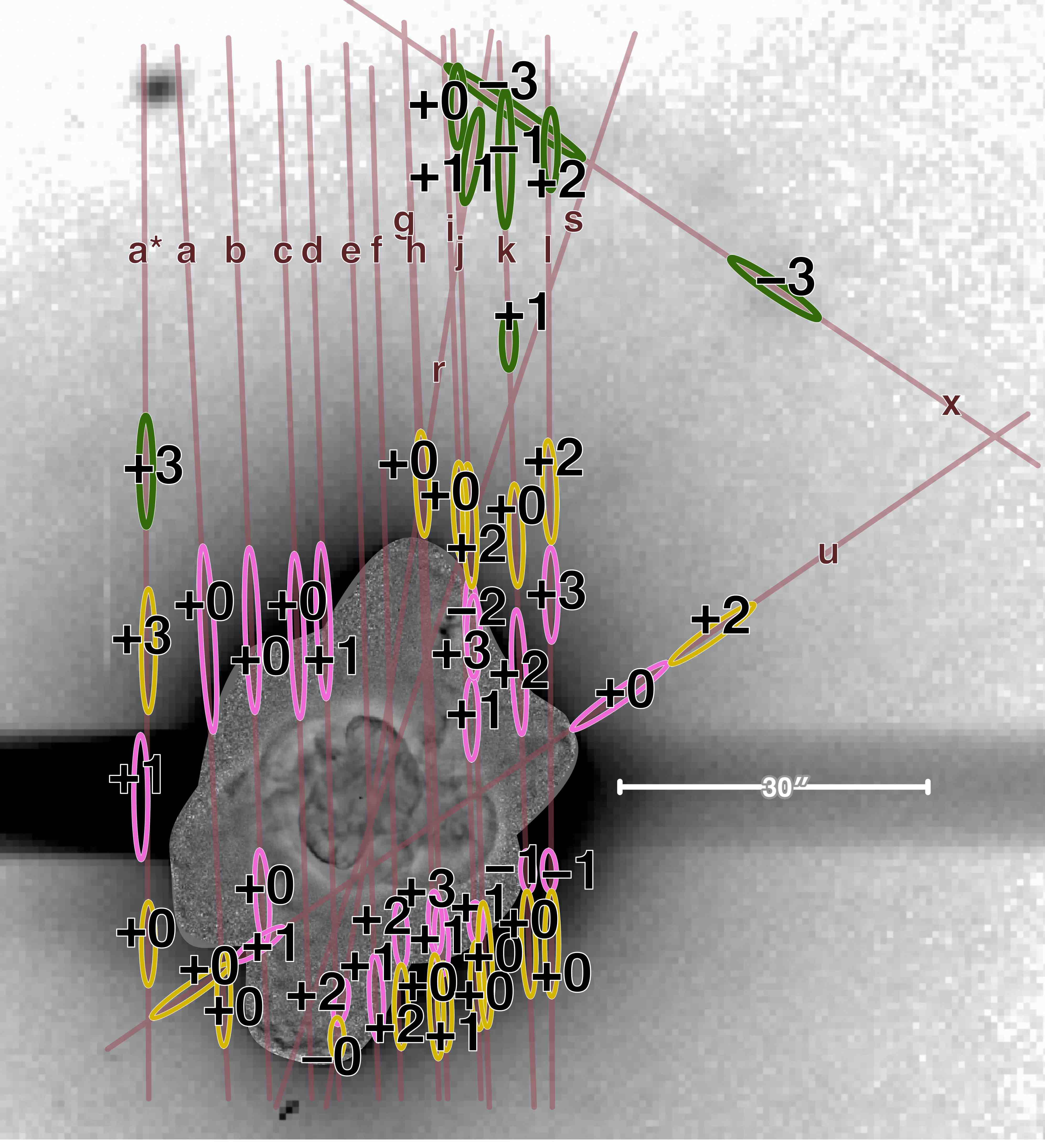}
  \caption{
    Velocity components in the halo, all measured from the \oiii{} spectra with slit positions as indicated.
    Components are divided into three classes:
    inner halo (pink ellipses), outer halo (yellow ellipses),
    and halo knots (green ellipses).
    The inner portion of the background image is
    the same high-pass-filtered \textit{HST} image
    shown in Figs.~\ref{fig:shell-velocity-components} and~\ref{fig:outer-lobe-components}.
    The outer portion of the background image is a deep \oiii{} exposure of the halo
    obtained with MES-SPM in direct imaging mode with an occulting bar (see Fig.~\ref{fig:halo-knots}).
  }
  \label{fig:halo-components}
\end{figure}

\begin{figure}
  \centering
  \includegraphics[width=\linewidth]{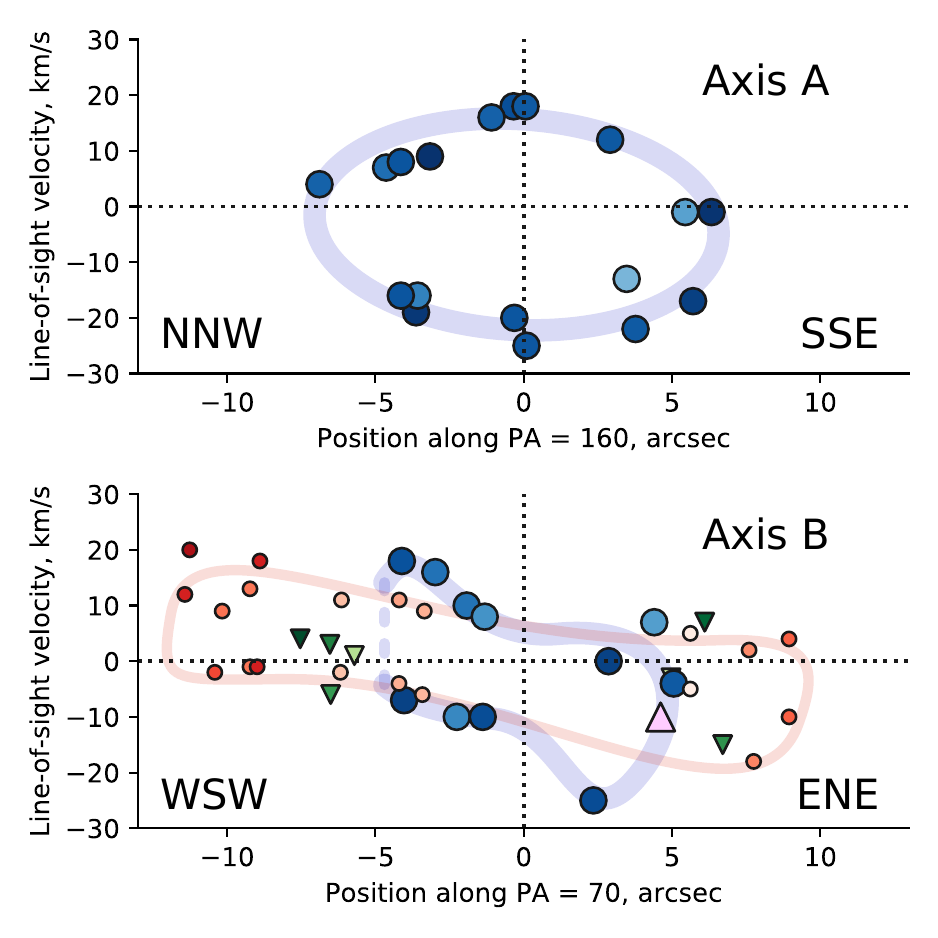}
  \caption{
    Radial velocity versus position
    for the shell features shown in Fig.~\ref{fig:shell-velocity-components}.
    Results are shown along two axes:
    Axis~A (upper panel) is the apparent projected major axis of the inner shell,
    while Axis~B (lower panel) is perpendicular to this.
    Large blue circles show the inner shell,
    small red circles show the intermediate shell,
    and green triangles show miscellaneous features between the two shells.
    Darker colors indicate features that are closer to each respective axis.
    Colored lines are merely to guide the eye,
    and show possible interpretations of the shell kinematics along the two axes.
  }
  \label{fig:shell-velocity-axes}
\end{figure}

\section{Proper motions}
\label{sec:proper-motions}

We employ archival emission line images from the Wide-Field Planetary Camera~2 \citetext{WFPC2, \citealp{Holtzman:1995a}} of the \textit{Hubble Space Telescope} (\textit{HST}) to derive plane-of-sky motions in the nebula.
The images, which we downloaded from the Barbara~A. Mikulski Archive for Space Telescopes,%
\footnote{MAST, \url{https://archive.stsci.edu/}}
come from observations made between 1997 (Cycle 6) and 2008 (Cycle~16) via proposal IDs 6792, 7501, 8773, and 11122.
Original images had the source centered on the
WFPC2 camera's WFC3 chip (first epoch)
or PC chip (second epoch). 
Images were aligned to the celestial coordinate frame using Astrodrizzle%
\footnote{\url{https://drizzlepac.readthedocs.io}},
which also corrected for the optical distortions of the camera
and remapped all the images to a common uniform square pixel grid
at the WFC resolution of \SI{0.1}{arcsec.pix^{-1}}. 
Proper motions were then estimated from pairs of images
separated by 10 to 12 years,
using the Fourier Local Correlation Tracking (FLCT) method \citep{Welsch:2004a, Fisher:2008a}\footnote{
  We used version 1.07 of FLCT, obtained from \url{http://cgem.ssl.berkeley.edu/cgi-bin/cgem/FLCT/home},
  together with version 1.04 of the Python wrapper pyflct,
  obtained from \url{https://github.com/PyDL/pyflct}.}
with a kernel width of 10~pixels.
Results are shown in Figures~\ref{fig:proper-motions-oiii} and~\ref{fig:proper-motions-nii} for \oiii{} (filter F502N) and \nii{} (filter F658N), respectively.
The resultant per-pixel motions between the two epochs are found to be of order \SI{0.5}{pix} (\(\approx \SI{5}{mas.yr^{-1}}\))
and these raw results were then corrected by applying a global shift to force the motion of the central star to be zero.
The systematic error from the global alignment of the two epochs
and residual uncertainties in the distortion correction
is estimated to be \SI{1.5}{mas.yr^{-1}},
which is expected to dominate the proper motion uncertainties in the brighter parts of the nebula.
In fainter and more featureless regions of the nebula, the proper motions are increasingly affected by random noise, 
which we indicate by partially transparent arrows in the figures, 
such as can be seen in parts of the lobes in Figure~\ref{fig:proper-motions-oiii}.
To convert the angular motions into transverse velocities, we assume a distance of \SI{2}{kpc},
so that \SI{10}{mas.yr^{-1}} is equivalent to \SI{95}{km.s^{-1}}.

In both lines, we find motions that are predominantly radial from the central star.
Although some non-radial vectors are seen in the figure,
these tend to be in the faintest parts of the nebula
and are likely to be dominated by random noise.
From the \oiii{} images (Fig.~\ref{fig:proper-motions-oiii}),
the fastest plane-of-sky motions are of order \SI{60}{km.s^{-1}},
and are chiefly along the NNW--SSE direction,
including the projected major axis of the inner peanut shell,
the NW~knot, the N~jet, and the end-cap of the S~lobe.
Motions along the perpendicular ENE--WSW direction are typically slower,
of order \SI{30}{km.s^{-1}}.
Note that proper motions are unavailable for the end cap of the N lobe since the second epoch HST image does not cover this region.
The \nii{} images (Fig.~\ref{fig:proper-motions-nii}) show a similar expansion pattern for the features that are visible in both lines.
Remarkably low plane-of-sky velocities of \(\le \SI{15}{km.s^{-1}}\) are seen for the \nii{}-bright knot complexes immediately north and west of the central star. 

Previously, \citet{Schonberner:2018a} have also studied the proper motions
in the Turtle Nebula, based on the same HST observations as those used here.
They employ a magnification method, which implicitly assumes a homologous expansion,
finding a magnification factor over the 10 to 11-year baseline
of \(M = 1.0085 \pm 0.0005\) for the inner high-ionization shell
(which they call the cavity rim)
and \(M = 1.0055 \pm 0.0005\) for the low-ionization knot complexes (see their Table~2).
After conversion to velocity units (assuming a distance of \SI{2}{kpc}),
these are consistent within the stated errors with our own average results for each of these two nebular components.

\section{Kinematic components from slit spectra}
\label{sec:kinematic-components}

In order to investigate the kinematics of the nebula in detail,
we have measured the velocities of distinct emission components in each slit spectrum
and organized them into broad systems based on their location, morphology and degree of ionization.
The velocities were determined by fitting gaussians to one-dimensional line profiles,
as illustrated in Figure~\ref{fig:spec-1d}.
The higher ionization features are measured in \oiii{},
while the lower ionization features are measured in \nii{}.

Maps of the individual slit positions,
marked with the position and velocity of each component
are presented in Figures~\ref{fig:shell-velocity-components} to \ref{fig:halo-components}.
All line-of-sight velocities are given with respect to a nominal systemic heliocentric velocity of \SI{-40}{km.s^{-1}},
which we choose as a typical velocity of the planetary nebula halo. 
The figures are organized according to the different nebular features shown in Figure~\ref{fig:hst}:
shells (Figure~\ref{fig:shell-velocity-components} for \oiii{} and Figure~\ref{fig:heii-shell-components} for \heii{}),
knot complexes (Figure~\ref{fig:knot-complex-map}),
outer lobes (Figure~\ref{fig:outer-lobe-components}),
and halo (Figure~\ref{fig:halo-components}).
Each of these features is discussed in more detail below.

\subsection{High-ionization shells}
\label{sec:high-ioniz-shells}

These systems represent the majority of the \oiii{} emission in the core of the nebula
and show a nested elliptical shell morphology.
Figure~\ref{fig:shell-velocity-components} shows the \oiii{} emission components associated with these shells,
as derived from the longslit spectra,
separated into negative and positive velocities
with respect to the nominal systemic velocity
(panels~a and~b, respectively).
The inner shell, with a radius of \(5''\) to \(7''\), is the brightest
and is indicated by thick-lined colored ellipses. 
The edge of the more extended intermediate shell, with a radius of \(8''\) to \(12''\), is 10 to 100 times fainter than the inner shell
and is indicated by thinner dashed ellipses.
Additional miscellaneous emission features located in between these shells are indicated by dot-dashed ellipses.
Further features that seem to be associated with the low-ionization knots discussed below in \S~\ref{sec:knot-complexes} are omitted from the figure.

The inner shell has an irregular peanut-like shape
with an apparent elongation along \(\text{PA} \approx \ang{160}\).
The intermediate shell is elongated roughly perpendicular to this, along \(\text{PA} \approx \ang{70}\).
In Figure~\ref{fig:shell-velocity-axes} we plot the velocity of each shell component
against position along each of these axes,
which we denote axis~A and axis~B.  
Each component was assigned to only one axis (A or B), according to its location,
but this assignment is unavoidably subjective for components near the center,
where the two axes cross.

Along axis~A a closed velocity ellipse can be seen for the inner shell,
with a maximum splitting of \(\pm \SI{22}{km.s^{-1}}\) close to the central star
and velocities close to zero at either end.
The pattern is not entirely symmetric,
showing a slight gradient  of \(\pm \SI{3}{km.s^{-1}}\)
over \(\pm 7''\) along the length of the axis,
in which the more negative velocities are at the SSE end.
The centroid of the ellipse is also shifted by \(\SI{-3.5}{km.s^{-1}}\)
with respect to the systemic velocity.

Along axis~B,
which is the apparent minor axis of the inner shell (\(\pm 5''\)),
the ellipse is distorted and the velocity differential is much more pronounced:
\(\pm \SI{11}{km.s^{-1}}\),
with the more negative velocities at the ENE end
(large blue circle symbols in lower panel of Fig.~\ref{fig:shell-velocity-axes}).
Velocity splitting of \(\pm \SI{9}{km.s^{-1}}\) is seen near both ends,
but it is not clear from the \oiii{} spectra if the ends are closed or open,
since none of the \oiii{} slits are aligned with this axis.
However, one of the \nii{} slits (slit~w) is indeed oriented close to axis~B and,
although the shell emits only weakly in \nii{},
the distorted ellipse is clearly closed at the ENE end
(large pink triangle symbol in Fig.~\ref{fig:shell-velocity-axes}).
The situation is not so clear at the WSW end
since any \nii{} emission from the shell is swamped by brighter emission from the knot complexes.
However, the evidence from \textit{HST} imaging suggests that the shell may be closed in this direction also (indicated by thin dashed blue line in the lower panel of Fig.~\ref{fig:shell-velocity-axes}).

The intermediate shell along axis~B repeats a similar kinematic pattern to the inner shell,
but at larger radii.
It is represented by dashed ellipses in Figure~\ref{fig:shell-velocity-components} and small red circle symbols in Figure~\ref{fig:shell-velocity-axes}.
The gradient (\(\pm \SI{9}{km.s^{-1}}\) over \(\pm 11''\))
and splitting (\(\pm \SI{8}{km.s^{-1}}\))
are both marginally smaller than for the inner shell.
Note that, unlike the inner shell, the intermediate shell is markedly lop-sided,
extending \(12''\) to the WSW, but only \(10''\) to the ENE.

The miscellaneous high-ionization components lie outside the inner shell,
and are represented by dot-dashed ellipses in Figure~\ref{fig:shell-velocity-components}
and small green triangle symbols in Figure~\ref{fig:shell-velocity-axes}.
They do not show any marked kinematic pattern,
but are broadly compatible with the velocities of nearby portions of the intermediate shell.

\begin{figure}
  \centering
  \includegraphics[width=\linewidth]{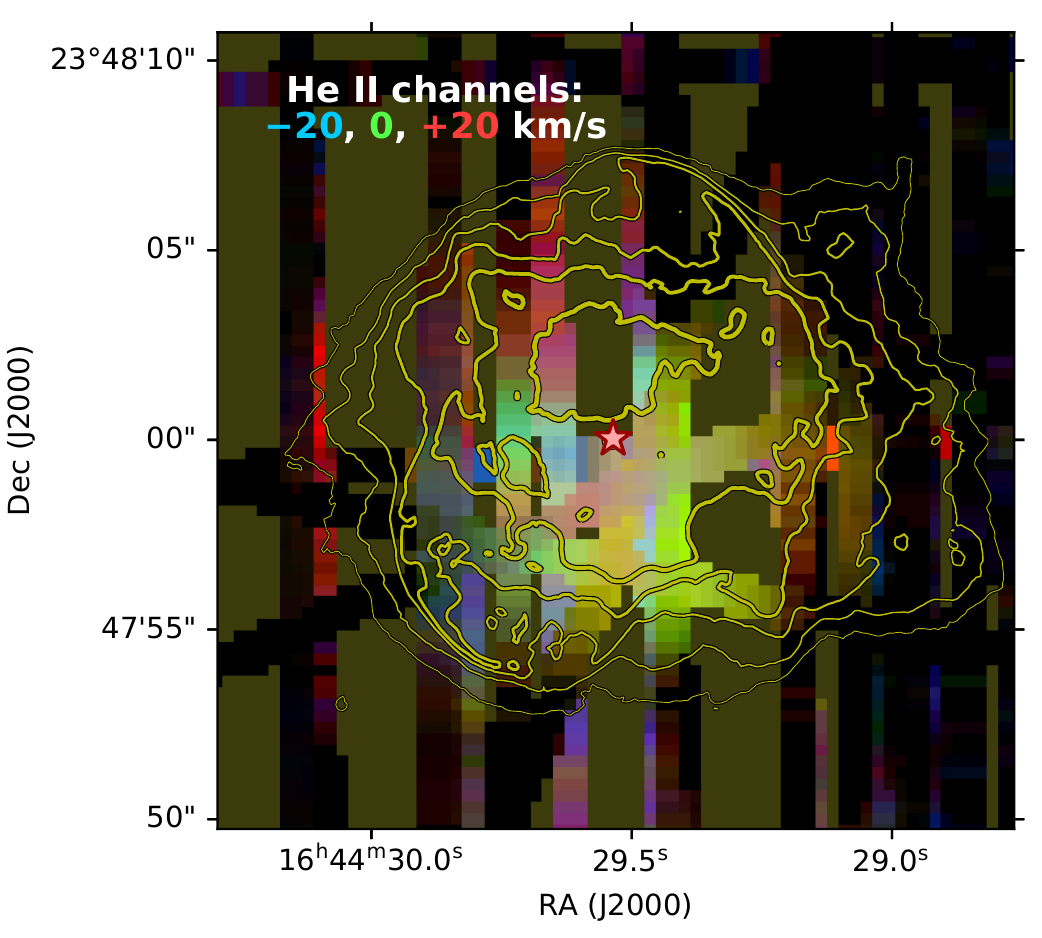}
  \caption{
    Reconstructed velocity channel maps from the \heii{} slit spectra,
    showing the highest ionization gas in the nebula.
    The color image is constructed from 3 channels, each of width \SI{20}{km.s^{-1}},
    as indicated in the figure.  Contours show the \oiii{} HST image.
    No spatial interpolation is performed, so regions between the slits are lacking data,
    indicated by olive green color.
  }
  \label{fig:heii-shell-annotated}
\end{figure}

\begin{figure}
  \centering
  \includegraphics[width=\linewidth]{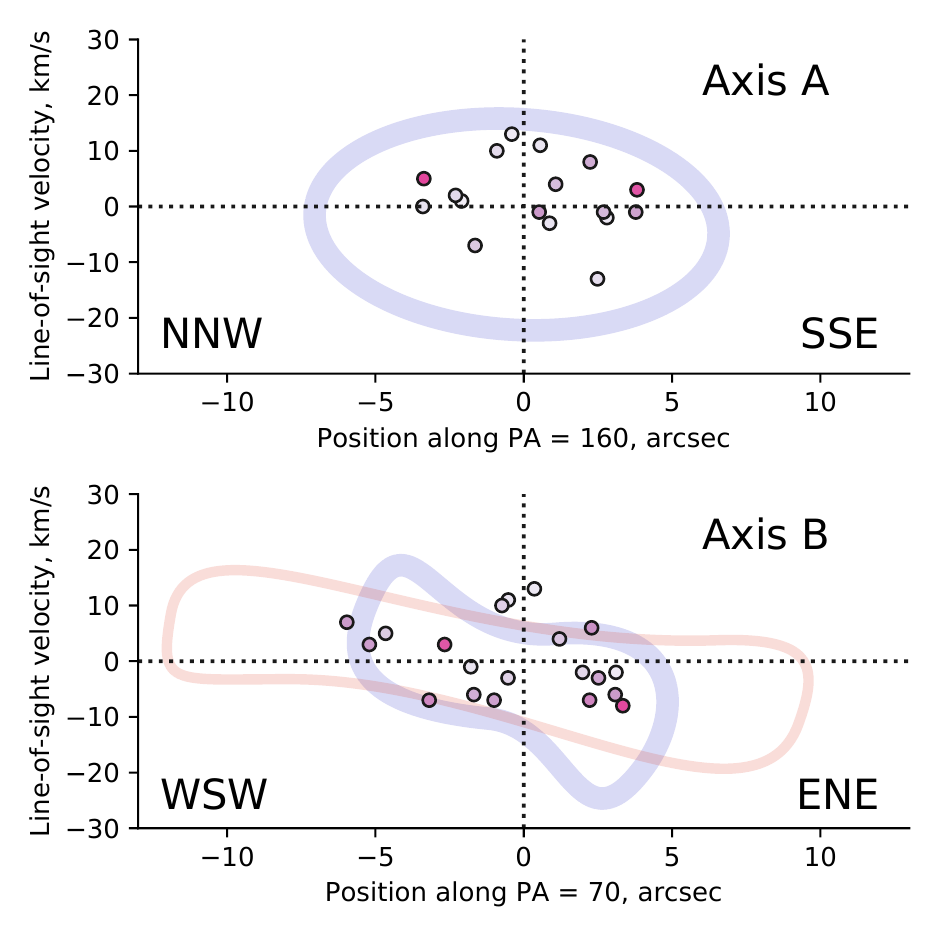}
  \caption{
    Radial velocity versus position
    for the \heii{} shell features shown in Fig.~\ref{fig:heii-shell-components}.
    Results are shown projected along the same two axes
    as in Fig.~\ref{fig:shell-velocity-axes}:
    Axis~A (upper panel) and Axis~B (lower panel).
    Colored lines show the \oiii{} shells from Fig.~\ref{fig:shell-velocity-axes}.  
  }
  \label{fig:heii-shell-velocity-axes}
\end{figure}

In order to trace the more highly photo-ionized gas in the nebula,
we also analyze the weak \heii{} \SI{6560.10} line,
which is seen in the blue wing of the \Ha{} line,
displaced in velocity units by \SI{-123}{km.s^{-1}}.
Figure~\ref{fig:heii-shell-annotated} shows a three-color combination of channel maps in this line, reconstructed from our slit spectra.
It can be seen that the emission is more centrally concentrated than other emission lines
(contours show the HST \oiii{} image for comparison)
and is mainly confined to the inner shell.
Note that the red channel is partially contaminated by \Ha{} emission in some slits.
It is also notable that the \heii{} emission is lop-sided in the opposite sense to the other emission lines,
with a brightness peak that is displaced \(\approx 2.5''\) to the south of the central star.
All three of the lines \Ha{}, \nii{}, and \oiii{} have a brightness peak that is displaced \(\approx 2''\) to the north of the central star.

Further details of the \heii{} kinematics are illustrated in Figure~\ref{fig:heii-shell-components},
which shows velocity features identified in the slit spectra.
For the slits that pass closest to the central star,
a full velocity ellipse is visible in the spectrum,
in which case four velocity components are measured,
two corresponding to the maximum spatial extension
and two to the maximum velocity splitting.
For other slits, only a partial ellipse is seen and fewer velocity components are measured.
As expected for an expanding shell structure, the largest absolute velocities (blue and red)
are seen close to the center,
with smaller absolute values around the periphery.
On the other hand, there is also a clear tendency for bluer velocities towards the east
and redder velocities towards the west, see also Figure~\ref{fig:heii-shell-annotated}.

Figure~\ref{fig:heii-shell-velocity-axes} shows the \heii{} velocity components projected on to the same two axes,
A and B,
which were found above from analysis of the \oiii{} shell kinematics.
The velocity and spatial scales are the same as in Figure~\ref{fig:shell-velocity-axes}
and the approximate loci of the \oiii{} shells are indicated by solid lines,
which allows the \oiii{} and \heii{} kinematics to be easily compared.
The maximum \heii{} velocity splitting of \(\pm \SI{13}{km.s^{-1}}\) is a little more than half as large as is seen in \oiii.
Along axis~A, the \heii{} emission is also more spatially compact than in \oiii{}.
Along axis~B, on the other hand, the \heii{} emission has a similar spatial extent to the inner \oiii{} shell,
even extending slightly beyond the inner shell on the WSW side
(for instance, the \heii{} component detected in slit~i\(^*\), as shown in the right panel of Fig.~\ref{fig:heii-shell-components}).
A gradient of \(\pm \SI{7.5}{km.s^{-1}}\) over \(\pm 5''\) is seen along axis~B,
which has the same sense as the slightly larger gradient seen in \oiii{} along the same axis.

\subsection{Low-ionization knot complexes}
\label{sec:knot-complexes}

\begin{figure*}
  \centering
  \includegraphics[width=\linewidth]{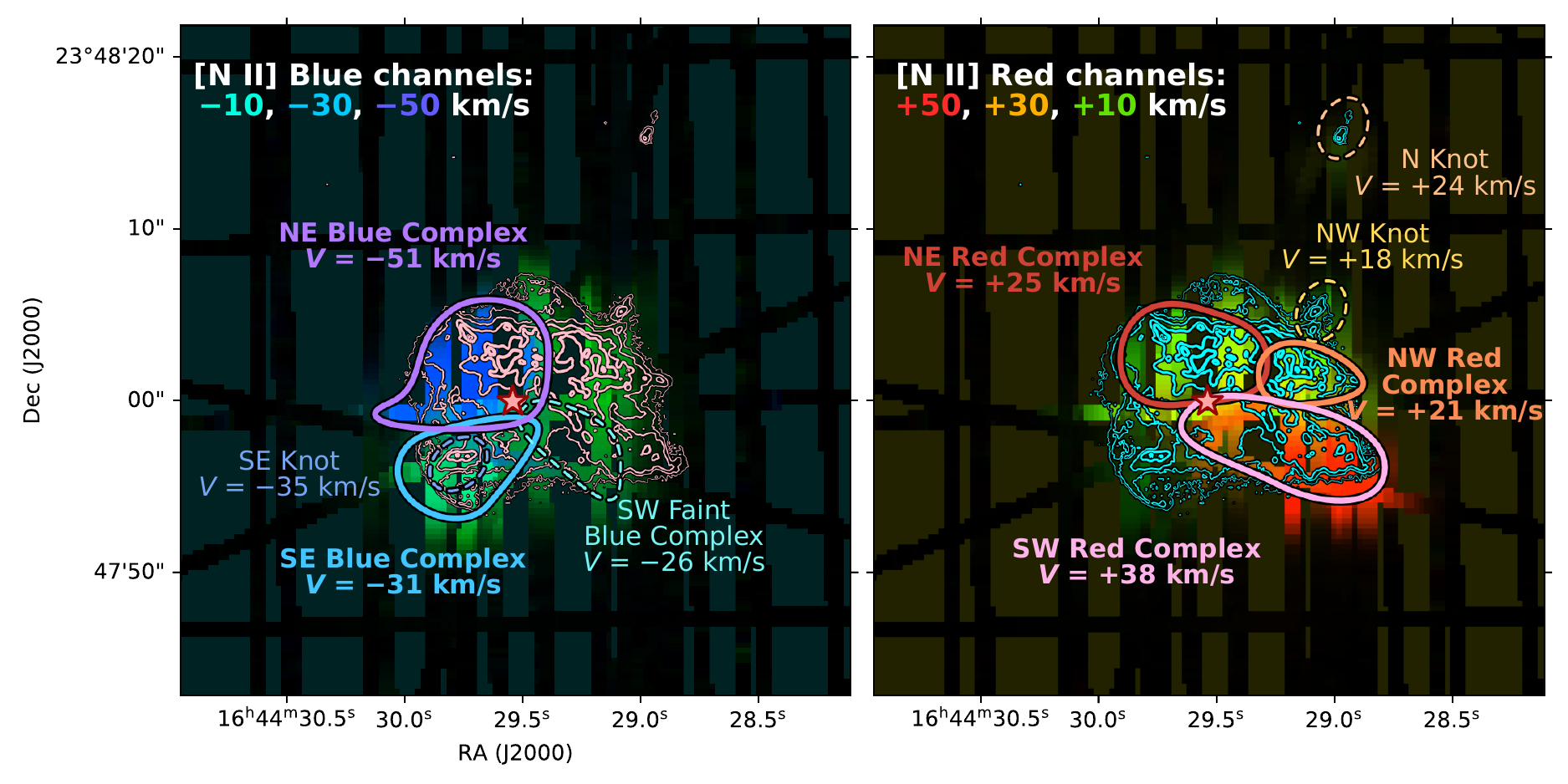}
  \caption{
    Reconstructed velocity channel maps from the \nii{} slit spectra,
    showing the blue-shifted (left panel) and red-shifted (right panel) knot complexes.
    Note that channel maps have not been spatially interpolated,
    so that the individual slit positions can be seen.
    Each color image is constructed from 3 channels, each of width \SI{20}{km.s^{-1}},
    as indicated on the figure.
    All velocities are with respect to the nominal heliocentric systemic velocity of \SI{-40}{km.s^{-1}}.
    The low-ionization emission components from Fig.~\ref{fig:knot-complex-map}
    have been classified into 6 knot complexes,
    which are shown as colored outlines.
    Regions that lack data are shown in dark turquoise (left panel) and olive green (right panel).
  }
  \label{fig:knot-complexes}
\end{figure*}

\begin{figure}
  \centering
  \includegraphics[width=\linewidth]{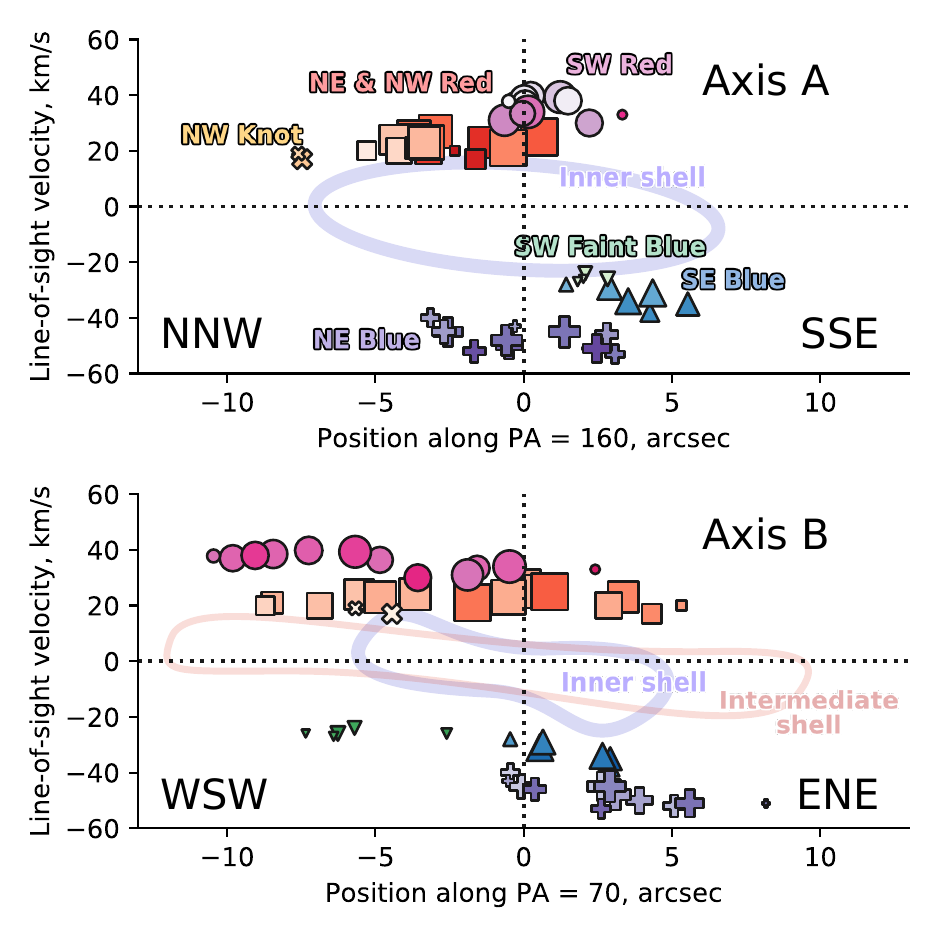}
  \caption{
    Radial velocity versus position
    for the low-ionization features shown in Fig.~\ref{fig:knot-complex-map}.
    Results are shown projected along the same two axes,
    Axis~A (upper panel) and Axis~B (lower panel),
    as in Fig.~\ref{fig:shell-velocity-axes},
    but this time each feature is shown projected along both axes.
    The features are divided into different knot complexes,
    as shown in Fig.~\ref{fig:knot-complexes},
    which are indicated by symbol type and color.
    Symbol size is proportional to feature brightness (log scale)
    and symbol shade indicates position along the other axis (darker is more positive).
    Continuous lines show the same high ionization shells
    as in Fig.~\ref{fig:shell-velocity-axes}.
  }
  \label{fig:knot-complex-velocity-axes}
\end{figure}

The \nii{} emission from the nebula is dominated by small-scale knots,
as can be appreciated on the \textit{HST} images (Figure~\ref{fig:hst}).
These knots have typical sizes \(< 1''\) and so are not spatially resolved
in our ground-based spectroscopy.
A small number of isolated knots are individually detected in our spectra,
but in general we detect only the combined emission of extended knot complexes.
Figure~\ref{fig:knot-complex-map} shows all the \nii{} emission components
identified from our spectra, but excluding those associated with the high-ionization
shells discussed above.
As previously, the components are divided into negative (left panel)
and positive (right panel) velocities.  
In Figure~\ref{fig:knot-complexes}
we classify the emission components into six knot complexes,
plus three individual knots, whose plane-of-sky distribution
and typical line-of-sight velocity are shown
superimposed on the \textit{HST} \nii{} image (contours)
and isovelocity channel maps reconstructed from our slit spectra (color images).

Figure~\ref{fig:knot-complex-velocity-axes} shows the velocity of each \nii{} component
as a function of position along the two axes that characterize the high-ionization shells:
axis~A at \(\text{PA} = \ang{160}\)
and axis~B at \(\text{PA} = \ang{70}\) (see \S~\ref{sec:high-ioniz-shells}).
The kinematics of the shells themselves are shown by faint continuous lines for comparison.
Note that the extent of the velocity axis in this figure
is twice as large as in Figure~\ref{fig:shell-velocity-axes}. 


The upper panel of Figure~\ref{fig:knot-complex-velocity-axes} indicates
that there is no clear velocity gradient along axis~A,
but the low-ionization knots have significantly faster radial velocities
than the high-ionization shells, by roughly a factor of two.
In addition, there is an asymmetry between the blue and red components:
the blue-shifted knot complexes tend to have a larger radial velocity magnitude,
but to have fainter \nii{} emission, as compared with the red-shifted knot complexes.

From the lower panel of Figure~\ref{fig:knot-complex-velocity-axes},
it is apparent that there is a significant gradient of \(\approx \SI{20}{km.s^{-1}}\) over \(\pm 8''\)
along axis~B, which is seen in both blue-shifted and red-shifted components.
The sense of this gradient is the same as that seen in the high-ionization shells
along this axis.
It is mainly due to velocity differences
\emph{between} complexes rather than within them,
although both the SW Red and NE Blue complexes
show significant internal gradients of \(\approx \SI{10}{km.s^{-1}}\) over \(10''\) in the same direction.
There is an approximate kinematic and spatial symmetry
between pairs of opposite knot complexes:
SW Red with NE Blue, NW Red with SE Blue,
and NE Red with SW Faint Blue.

The knot complexes are also visible in the \oiii{} spectra,
although they are fainter than the shells in this line.
There is also a slight difference in velocity,
with \oiii{} showing expansion velocities that are \num{2} to \SI{5}{km.s^{-1}} smaller than \nii{},
as can be seen in Figure~\ref{fig:spec-1d}.
If the expansion velocities within the knot complexes are assumed to increase with radius,
then this is evidence for ionization stratification along the line of sight
with the \oiii{} emission arising slightly closer to the star than \nii{}.
This is also consistent with the HST images (e.g., Figure~\ref{fig:hst}d),
which show that the \oiii{} emission associated with the complexes seems to arise from the inner surface of the\nii{} knots.

\subsection{Outer lobes}
\label{sec:outer-lobes}

\begin{figure*}
  \centering
  \includegraphics[width=\linewidth]
  {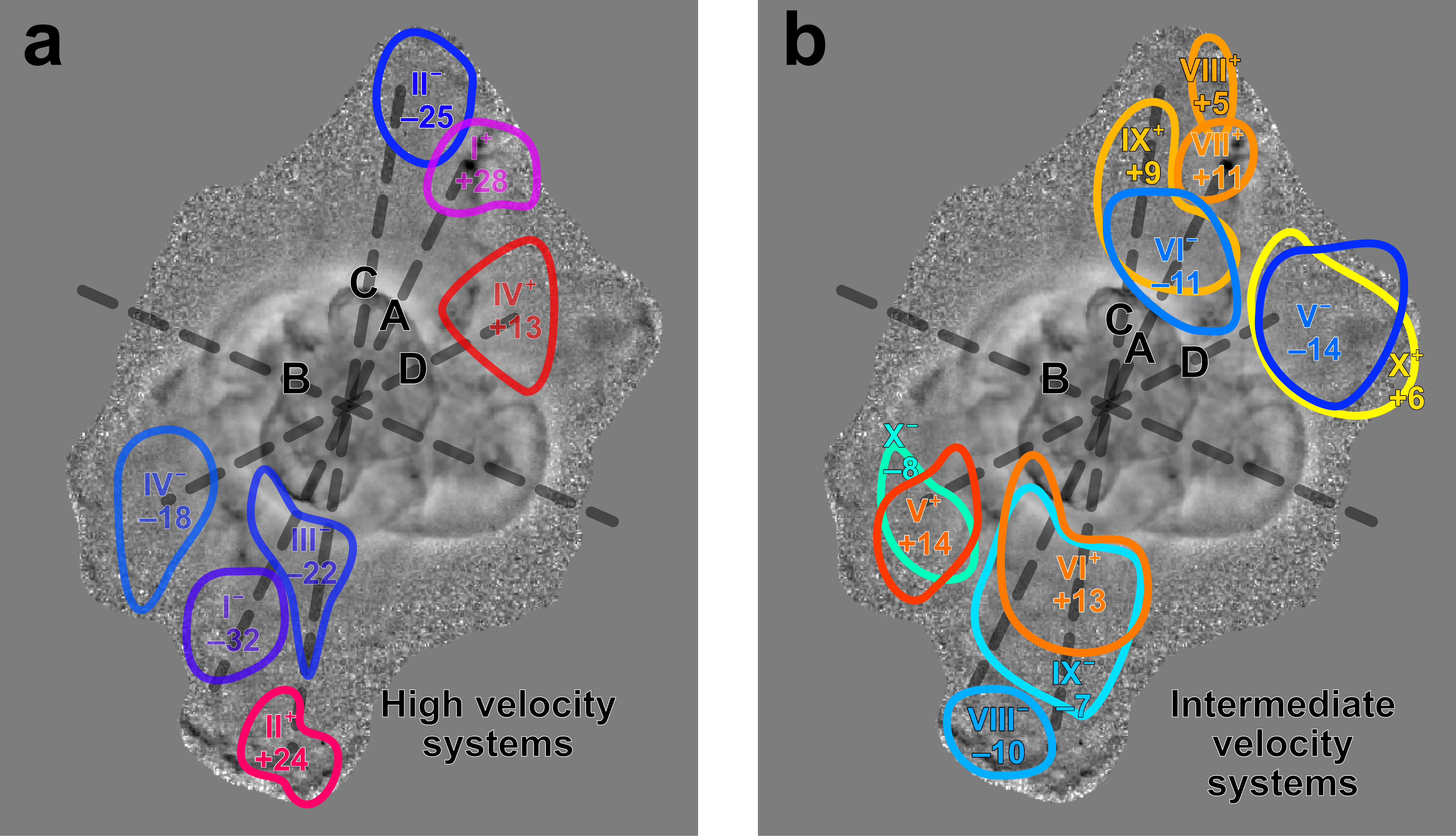}
  \caption{
    Velocity systems in the outer lobes.
    (a)~Classification of components into seven high-velocity systems (colored shapes)
    along three different axes (heavy dashed lines).
    (b)~Same but for eleven intermediate velocity systems.
  }
  \label{fig:outer-lobe-systems}
\end{figure*}

\newcounter{Syscounter}
\newcommand\Sys[1]{%
  \setcounter{Syscounter}{#1}%
  \ensuremath{\mathrm{\Roman{Syscounter}}}%
}
\newcommand\SysP[1]{\ensuremath{\Sys{#1}^+}}
\newcommand\SysM[1]{\ensuremath{\Sys{#1}^-}}

The emission from the outer lobes is much fainter than the inner shell,
and is most easily detected in \oiii{}.
Figure~\ref{fig:outer-lobe-components} shows all the outer lobe velocity components
that we have been able to measure from the \oiii{} slits:
blue-shifted components in panel~a and red-shifted components in panel~b.
We have not included all of the \oiii{} components that lie outside the inner shell,
since some of these are closely associated with the low ionization knot complexes discussed above in \S~\ref{sec:knot-complexes}.
Such \oiii{} components are omitted from the figure 
with the exception of the N knot and the NW knot (see Fig.~\ref{fig:knot-complexes}).

As with the low-ionization knot complexes,
there are many apparent pairings between blue and red outer lobe components on opposite sides of the nebula.
This is illustrated in Figure~\ref{fig:outer-lobe-systems},
where we classify the components into four high-velocity systems,
\Sys{1} to \Sys{4}
with \(|V| = \text{\SIrange{17}{25}{km.s^{-1}}}\) (panel~a)
and six intermediate-velocity systems,
\Sys{5} to \Sys{10}
with \(|V| = \text{\SIrange{5}{14}{km.s^{-1}}}\) (panel~b).
Nearly all of these systems have a blue and a red component
which are approximately symmetrically arranged about the central star.
For instance, the blue-shifted \SysM{2} component in the north
is paired with the red-shifted \SysP{2} component in the south.
Note, however, that \SysM{3} and \SysP{7} have no opposite counterparts.

The high-velocity systems seem to define three separate flow axes (Figure~\ref{fig:outer-lobe-systems}a).
Most notably, system~\Sys{1}, with \(\text{PA} = \ang{155 \pm 5}\),
is closely aligned with axis~A of the high-ionization shells at \(\text{PA} \approx \ang{160}\)
(see \S~\ref{sec:high-ioniz-shells}),
and also has the same sense of inclination (receding to the north).
Although system~\Sys{2}, with  \(\text{PA} = \ang{171 \pm 5}\),
is also marginally consistent in projection with axis~A,
the sense of inclination is opposite (receding to the south),
implying that the axis is distinct,
and we name it axis~C.
System~\Sys{4}, with  \(\text{PA} = \ang{120 \pm 15}\), defines yet another axis,
which we name axis~D.
Note that none of the outer lobe components are aligned with axis~B of the high-ionization shells.

The intermediate velocity systems (Fig.~\ref{fig:outer-lobe-systems}b)
are less well collimated than the high-velocity components,
but seem to be aligned with the same three axes: A, C, and D.
This is most clearly seen in the East and West Lobes,
where systems~\Sys{5} and \Sys{10} overlap the higher velocity system \Sys{4} along axis~D.
They may represent slower moving entrained material or bow shock wings from the same flow.
The same is true of the North and South Lobes,
where systems~\Sys{6}, \Sys{7}, \Sys{8}, and \Sys{9}
overlap the higher velocity systems \Sys{1} and \Sys{2}.
However, the fact that axis~A and axis~C have similar projected position angles
makes it hard to assign each intermediate velocity system to one axis or the other.

\subsection{Haloes}
\label{sec:haloes}

Figure~\ref{fig:halo-components} shows the velocity components
from the \oiii{} slits that are close to the systemic velocity,
and which are mainly associated with the nebula halo.
These components can be divided into three different types,
according to the width and location of the component.
The \textit{inner halo} (pink ellipses in the figure) is characterised by broad lines,
while the \textit{outer halo} (yellow ellipses in the figure) is characterised by narrower lines.
After correcting for instrumental broadening,
the full-width half-maximum line widths are \(W = \SI{28 \pm 2}{km.s^{-1}}\) for the inner halo,
but only \(W = \SI{14 \pm 1}{km.s^{-1}}\) for the outer halo.

The outer halo shows a smooth drop in brightness with distance
between about \(20''\) and \(30''\) from the central star,
beyond which it generally becomes undetectable in our spectra,
although it can still be seen in the very deep image shown as the background to Figure~\ref{fig:halo-components}.
Some bright emission patches are detected at larger radii
(\(40''\) to \(70''\)),
which we denote \textit{halo knots} (dark green ellipses in the figure).
These have a corrected velocity width intermediate between the inner and smooth outer halo,
\(W = \SI{19 \pm 1}{km.s^{-1}}\) but the signal-to-noise is low.

There is no clear pattern to the small velocity deviations between individual halo components,
and these are probably largely due to residual wavelength calibration uncertainties.
The mean and standard deviation for both the inner and outer halo are identical at \(V = \SI{+0.7 \pm 0.3}{km.s^{-1}}\),
which gives a more refined estimate for the heliocentric systemic velocity as
\(V_\odot = \SI{-39.3 \pm 0.3}{km.s^{-1}} \),
instead of the nominal \(\SI{-40}{km.s^{-1}}\) that we have been assuming.

Comparison of Figures~\ref{fig:outer-lobe-components} and~\ref{fig:halo-components}
shows that some inner-halo components overlap spatially with the high-velocity and intermediate-velocity systems of the outer lobes.
This is particularly true on the western side of the north and south lobes.
In these regions, there is a selection effect whereby
the halo components are more easily identifiable in regions where the outer lobe emission is weaker. 

\section{Three dimensional nebular structure}
\label{sec:three-dimens-struct}

\subsection{Velocities and positions of nebular features}
\label{sec:veloc-posit-three}

\newcommand\AxP[1]{\ensuremath{\mathrm{#1}^+}}
\newcommand\AxM[1]{\ensuremath{\mathrm{#1}^-}}
\begin{table*}
  \caption{Positions and velocities of nebular features in three dimensions}
  \label{tab:3d}
  \centering
  \sisetup{
    table-format = >2.0(1),
    table-auto-round,
  }
  \begin{tabular}{
    l 
    S[table-format = 3.0] 
    l 
    S 
    S 
    S 
    S 
    S 
    S[table-format = >1.2(2)] 
    S[table-format = >1.1(2)] 
    }
    \toprule
    & {PA} &        & {\(V_\text{los}\)} &  {\(V_\text{pos}\)} &  {\(V_\text{tot}\)} & {\(i\)} & {\(R_\text{proj}\)} & {\(R\)} & {\(t_\text{kin}\)}\\
    {Feature} & {\si{\degree}} & {Axis} & {\si{km.s^{-1}}}   &  {\si{km.s^{-1}}}  & {\si{km.s^{-1}}} & {\si{\degree}} & {\si{\arcsecond}} & {pc} & {\SI{1000}{yr}}\\
    \addlinespace[1pt]
    {(1)} & {(2)} & {(3)} & {(4)} & {(5)} & {(6)} & {(7)} & {(8)} & {(9)} & {(10)} \\
    \midrule
    Inner shell A NNW & 340 & \AxP{A} & +8 \pm 3 & 48 \pm 7 & 49 \pm 7 & 11 \pm 4 & 7 \pm 1 & 0.07 \pm 0.01 & 1.4 \pm 0.2\\
    Outer lobe \SysP{1}/\SysP{7} & 340 & \AxP{A} & +28 \pm 2 & 82 \pm 29 & 89 \pm 27 & 22 \pm 7 & 16 \pm 1 & 0.17 \pm 0.01 & 1.9 \pm 0.7\\
    North knot & 340 & \AxP{A} & +24 \pm 1 & 60 \pm 4 & 67 \pm 4 & 26 \pm 2 & 17 \pm 1 & 0.19 \pm 0.01 & 2.7 \pm 0.2\\
    Inner shell A SSE & 160 & \AxM{A} & -17 \pm 3 & 51 \pm 7 & 55 \pm 7 & -22 \pm 4 & 7 \pm 1 & 0.08 \pm 0.01 & 1.2 \pm 0.2\\
    Outer lobe \SysM{1} & 160 & \AxM{A} & -32 \pm 1 & 71 \pm 29 & 81 \pm 26 & -28 \pm 10 & 15 \pm 2 & 0.17 \pm 0.03 & 2.0 \pm 0.8\\
    Outer lobe \SysM{8} & 160 & \AxM{A} & -10 \pm 1 & 67 \pm 9 & 68 \pm 9 & -10 \pm 2 & 23 \pm 1 & 0.23 \pm 0.01 & 3.2 \pm 0.4\\
    Outer lobe \SysP{2} & 170 & \AxP{C} & +24 \pm 2 & 58 \pm 12 & 65 \pm 11 & 26 \pm 5 & 24 \pm 1 & 0.27 \pm 0.02 & 3.9 \pm 0.9\\
    Outer lobe \SysM{2} & 355 & \AxM{C} & -25 \pm 3 &   & > 25.0 & < 0.0 & 21 \pm 3 & > 0.21 & \\
    \addlinespace
    Inner shell B ENE & 65 & \AxM{B} & -25 \pm 3 & 31 \pm 5 & 43 \pm 4 & -44 \pm 6 & 4 \pm 1 & 0.06 \pm 0.01 & 1.2 \pm 0.4\\
    NE Blue complex & 60 & \AxM{E}(iii) & -51 \pm 1 & 24 \pm 3 & 66 \pm 2 & -69 \pm 2 & 5 \pm 1 & 0.14 \pm 0.01 & 2.0 \pm 0.2\\
    Intermediate shell ENE & 50 & \AxM{B} & -18 \pm 3 & 27 \pm 15 & 35 \pm 12 & -39 \pm 16 & 9 \pm 2 & .12 \pm 0.04 & 3.2 \pm 2.0\\
    Inner shell B WSW & 240 & \AxP{B} & +18 \pm 3 & 29 \pm 4 & 36 \pm 4 & 37 \pm 6 & 5 \pm 1 & 0.06 \pm 0.01 & 1.5 \pm 0.4\\
    Intermediate shell WSW & 260 & \AxP{B} & +20 \pm 3 & 29 \pm 26 & 38 \pm 20 & 40 \pm 26 & 12 \pm 1 & 0.16 \pm 0.06 & 3.8 \pm 3.5\\
    SW Red complex & 250 & \AxP{E}(iii) & +38 \pm 1 & 17 \pm 4 & 49 \pm 2 & 70 \pm 4 & 7 \pm 2 & 0.20 \pm 0.04 & 3.9 \pm 0.9\\
    NE Red complex & 30 & \AxP{E}(ii) & +25 \pm 2 & <8 & 31 \pm 1 & > 75 & 4 \pm 1 & > 0.15  & >5\\
    SW Faint Blue & 225 & \AxM{E}(ii) & -26 \pm 1 & 17 \pm 10 & 36 \pm 5 & -61 \pm 14 & 7 \pm 1 & 0.14 \pm 0.02 & 3.9 \pm 2.2\\
    \addlinespace
    Outer lobe \SysP{10} & 285 & \AxP{D} & +6 \pm 2 & 48 \pm 43 & 49 \pm 43 & 9 \pm 8 & 19 \pm 2 & 0.19 \pm 0.02 & 3.7 \pm 3.3\\
    Outer lobe \SysP{4} & 315 & \AxP{D} & +13 \pm 1 & 33 \pm 7 & 37 \pm 6 & 25 \pm 5 & 10 \pm 1 & 0.11 \pm 0.01 & 2.9 \pm 0.6\\
    NW knot  & 310 & \AxP{E}(i) & +18 \pm 1 & < 14 & 23 \pm 2 & >57 & 8 \pm 1 & > 0.15  & > 5\\
    NW Red complex & 300 & \AxP{E}(i) & +21 \pm 2 & 7 \pm 4 & 26 \pm 3 & 74 \pm 9 & 6 \pm 2 & 0.22 \pm 0.12 & > 4\\
    Outer lobe \SysM{10} & 120 & \AxM{D} & -8 \pm 4 & 24 \pm 14 & 26 \pm 13 & -22 \pm 15 & 13 \pm 3 & 0.14 \pm 0.04 & 5.2 \pm 3.3\\
    SE Blue complex & 135 &  \AxM{E}(i) & -31 \pm 3 & 32 \pm 11 & 49 \pm 8 & -49 \pm 10 & 3 \pm 2 & 0.05 \pm 0.02 & 0.9 \pm 0.3\\
    SE knot & 135 & \AxM{E}(i) & -35 \pm 3 & 42 \pm 3 & 59 \pm 3 & -45 \pm 3 & 5 \pm 1 & 0.07 \pm 0.02 & 1.1 \pm 0.1\\
    \bottomrule
    \multicolumn{10}{p{15cm}}{
    \textsc{Columns:}
    (1)~Name of nebular feature.
    (2)~Position angle of feature with respect to central star.
    (3)~Outflow axis that feature is best aligned with.
    (4)~Line-of-sight velocity, derived from longslit spectra.
    (5)~Plane-of-sky velocity, derived from proper motions assuming \(D = \SI{2}{kpc}\).
    (6)~Total pattern velocity: \(V_{\text{tot}} = [ (1.2 V_{\text{los}})^2 + V_{\text{pos}}^2 ]^{1/2}\) (see text for explanation of the factor 1.2).
    (7)~Inclination of velocity vector to line of sight: \(\tan i = 1.2 V_{\text{los}} / V_{\text{pos}}\).
    (8)~Projected radius of feature from central star.
    (9)~True radius, assuming that velocity vector is strictly radial: \(R = R_{\text{pos}} / \cos i\).
    (10)~Kinematic timescale: \(t_{\text{kin}} = R_{\text{pos}} / V_{\text{pos}}\).
    }
  \end{tabular}
\end{table*}

For those nebular features where both proper motion and radial velocity measurements exist,
it is possible to estimate the full three-dimensional velocity vector.
Under the assumption that all motions are strictly radial,
one can also find the three-dimensional position.
These are given in Table~\ref{tab:3d} for a wide variety of nebular features
in the inner shell, intermediate shell, low-ionization knots, and outer lobes.\footnote{
  Note that the uncertainties given in the table are derived from observational errors only,
  and they do not include the systematic uncertainty in the distance.
}
When combining radial velocities and proper motion measurements,
it is necessary to account for the fact that radial velocities measure the material speed,
whereas proper motions measure pattern speeds of shocks or ionization fronts \citep{Mellema:2004a}. 
The exact correction factor (material speed divided by pattern speed)
is model-dependent (e.g., Appendix~A of \citealp{ODell:2009c}),
but is predicted to be generally larger than unity
\citetext{see Fig.~8 of \citealp{Jacob:2013a}}
unless the nebula is in the recombination phase.
For simplicity, we adopt a constant correction factor of 1.2 for all entries in Table~\ref{tab:3d},
which is consistent with the predicted value for expansion velocities greater than \SI{30}{km.s^{-1}}
\citetext{see Fig.~3 of \citealp{Schonberner:2019a}}.

\begin{figure*}
  \includegraphics[width=\linewidth]
  {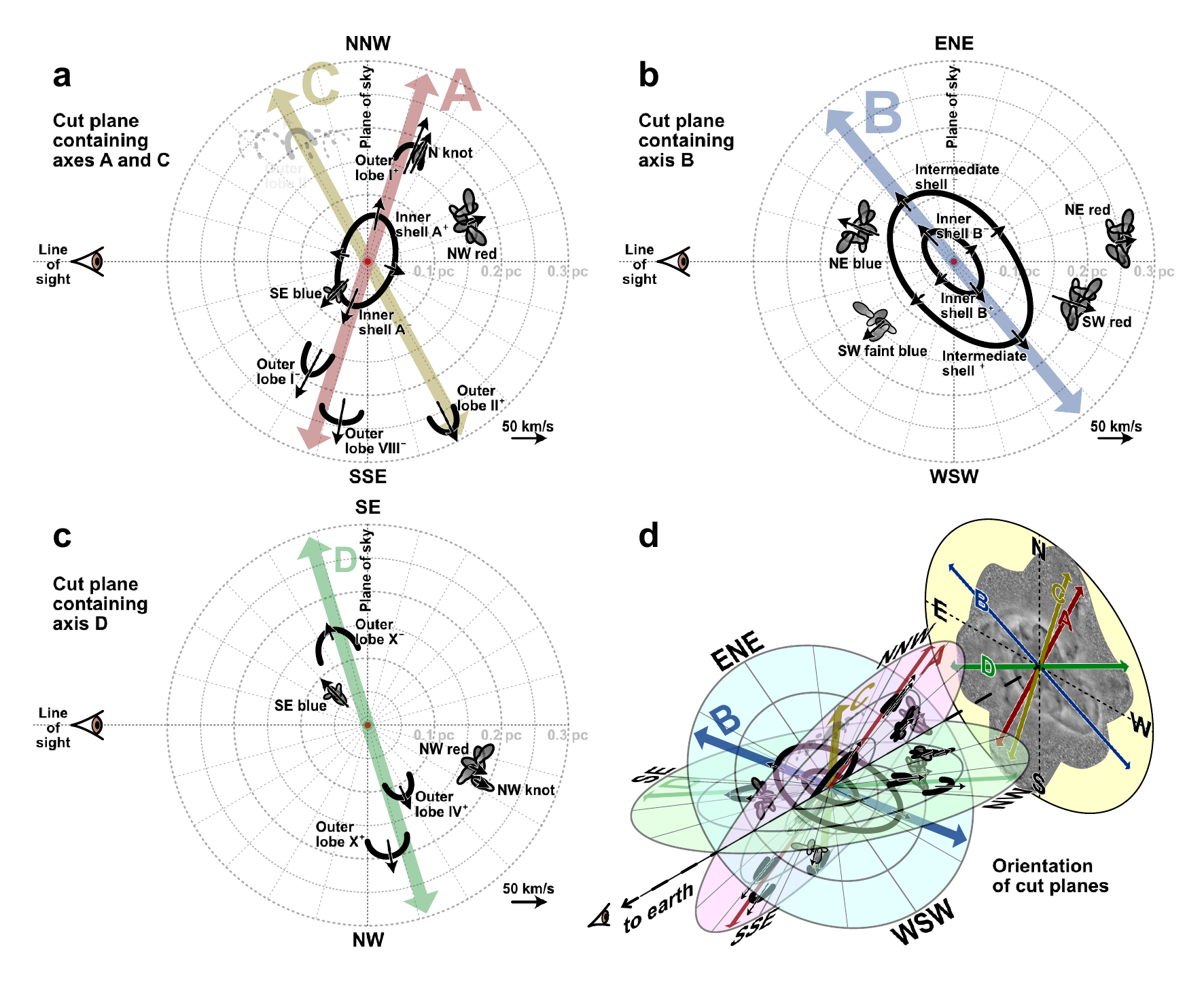}
  \caption{
    Three dimensional reconstruction of the nebular structure and kinematics,
    assuming all outflows are strictly radial.
    Panels (a), (b), and (c) show cut planes containing the line of sight
    and the symmetry axes of the shells and lobes.
    Some small artistic license is taken, since axes A and C are not strictly coplanar with the line of sight.
    Knots and knot complexes do not share these axes, but are also shown projected onto the closest planes.
    Arrows show the magnitude and direction of the derived three-dimensional velocity of each feature.
    Faint gray dashed lines for Outer lobe system \SysM{2} show the range of possibilities for this feature given that no proper motions are available for it.
  }
  \label{fig:cut-axis-3d}
\end{figure*}

An approximate three-dimensional visualization of the nebular structure is given in Figure~\ref{fig:cut-axis-3d}.
The high-ionization shells have two distinct symmetry axes, A and B, which are close to orthogonal.
Axis~A is shared with the outer lobes but the lobes also present outflows along two additional axes, C and~D.
The low-ionization knot complexes and individual knots do not share any of these four axes,
with the notable exception of the N~knot, which lies along axis~A.

\begin{figure}
  \centering
  \includegraphics[width=\linewidth]{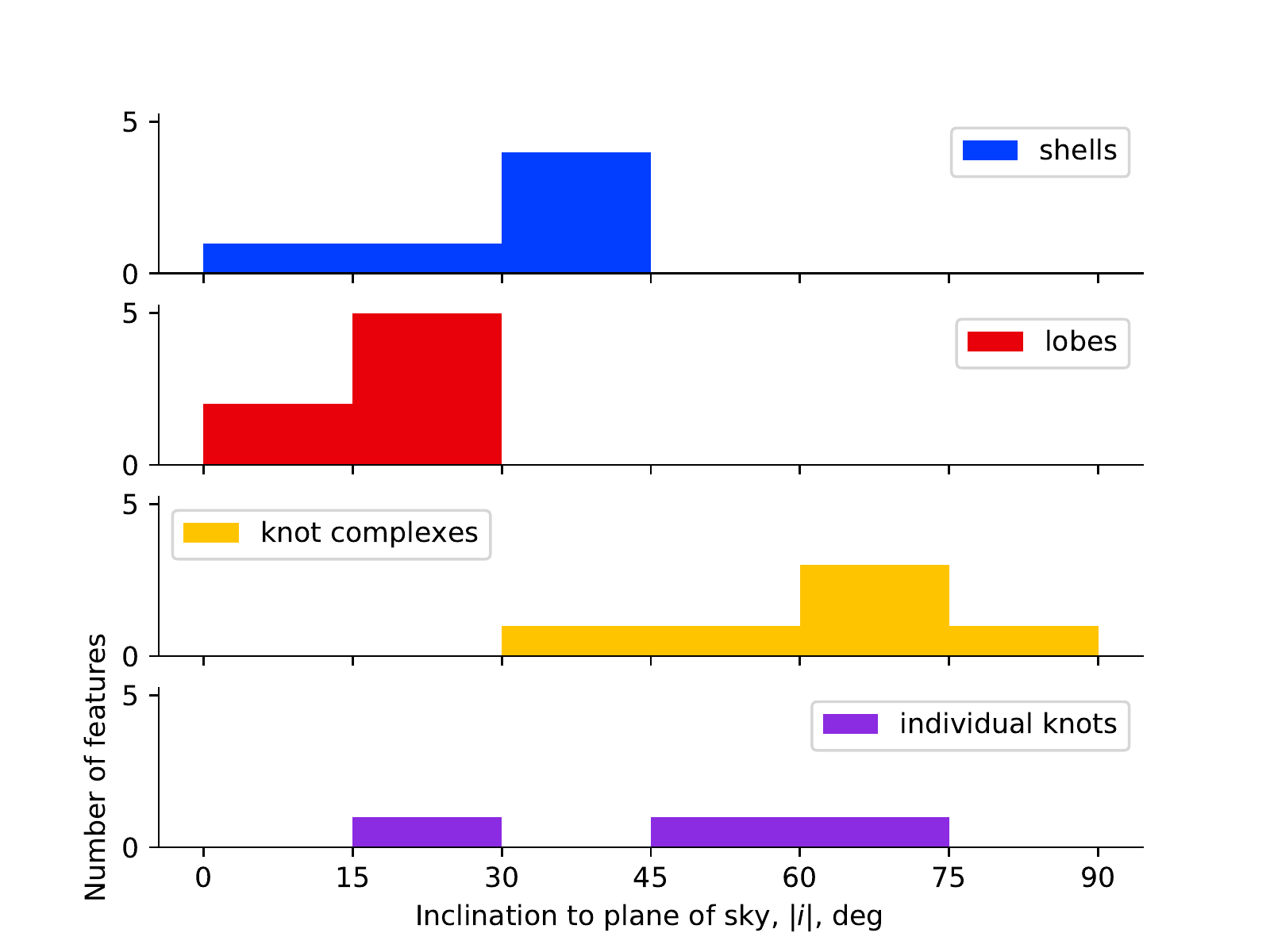}
  \caption{Distribution of outflow inclinations with respect to the plane of the sky for the different features.}
  \label{fig:inclinations}
\end{figure}

Despite the apparent complexity of the multi-axis structure, there are some clear systematic patterns,
one of which is illustrated in Figure~\ref{fig:inclinations}.
This shows histograms of the outflow inclination angle from the plane of the sky (absolute value)
for all of the individual features of Table~\ref{tab:3d}, classified by nebular component.
The inclination distribution for the outer lobes is very different from that of the knot complexes.
The lobes are concentrated at low inclinations, \(|i| = \ang{20 \pm 3}\),
whereas the knot complexes are concentrated at high inclinations, \(|i| = \ang{63 \pm 6}\).
The shells show intermediate inclinations, \(|i| = \ang{32 \pm 3}\),
while the individual knots are diverse
(the SE and NW~knot have high inclinations like the knot complexes,
while the N~knot has a low inclination like the lobes).

The concentration of high inclinations for the knot complexes imply the presence of a fifth axis in the nebula, which we denote axis~E.
Since this axis is close to the line of sight,
its signature is not obvious on images of the nebula but can only be discerned from spectroscopy.
If we take the average vector outflow direction of the knot complexes,
then axis~E has an inclination \(i = \ang{69}\)
and a position angle \(\mathrm{PA} = \ang{331}\).
The individual knot complexes have outflows that deviate by only \ang{16} to \ang{29} from this average direction,
confirming that it is indeed a well-defined axis,
albeit with a substructure of three point symmetric pairs of outflows,
as discussed in \S~\ref{sec:knot-complexes},
which we denote E(i), E(ii), and E(iii) in Table~\ref{tab:3d}.

\begin{figure}
  \includegraphics[width=\linewidth]
  {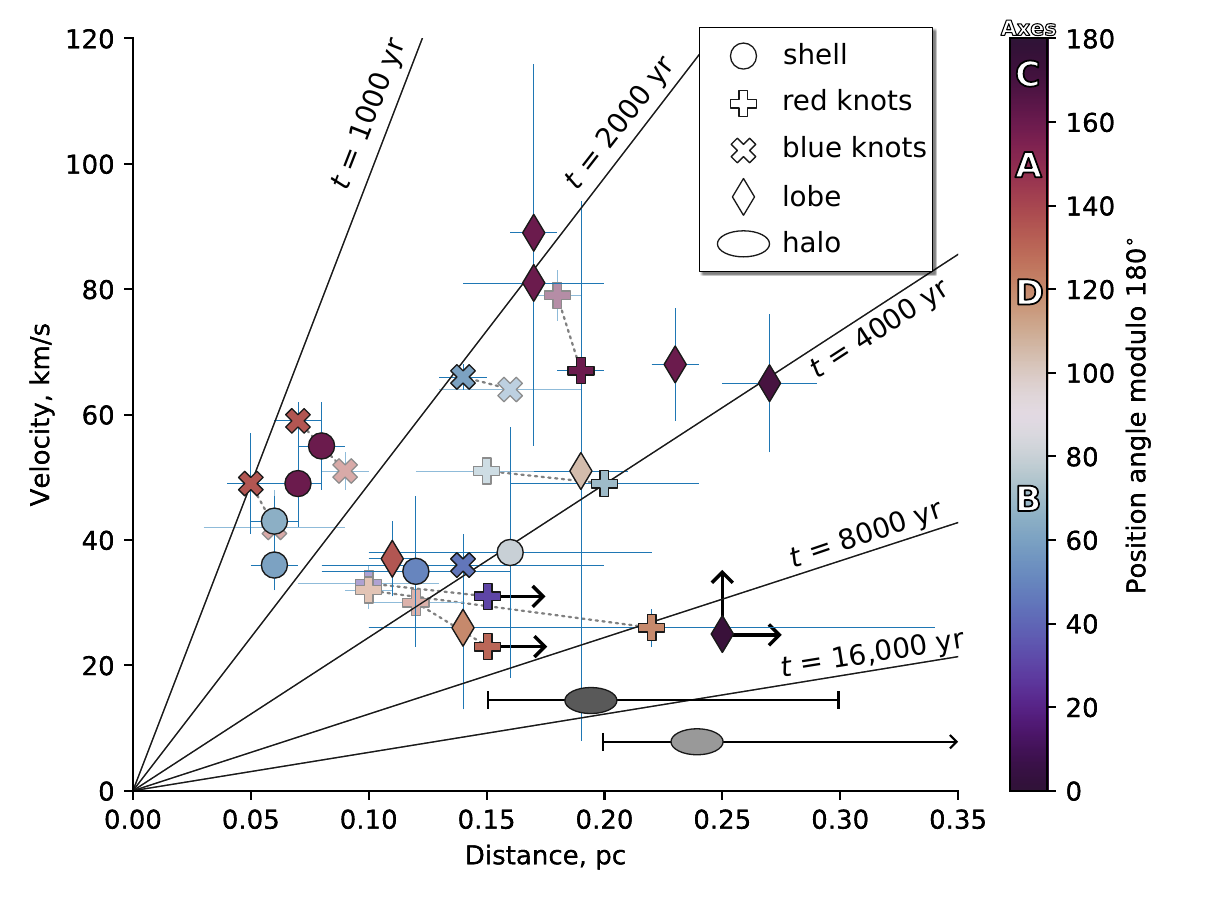}
  \caption{
    True velocity, \(V_{\text{tot}}\), versus true radius, \(R\), for all the features listed in Table~\ref{tab:3d}.
    Diagonal lines show dynamical times between \num{1000} and \SI{16000}{years}, as marked.
    Symbol shape indicates the type of feature:
    circle symbols are high-ionization shells (\S~\ref{sec:high-ioniz-shells}),
    cross symbols are low-ionization knots or knot complexes (\S~\ref{sec:knot-complexes}, plus symbol for redshifted and times symbol for blueshifted),
    diamond symbols are high-velocity and intermediate-velocity systems in the outer lobes (\S~\ref{sec:outer-lobes}).
    Error bars show the random errors on the proper motion measurements,
    with black arrows indicating lower limits.
    Fainter symbols show re-analysis of the knot systems under a
    worst-case scenario for systematic errors in the proper motions,
    which has opposite effects on the red and blue knots.
    The color of each symbol corresponds the position angle of the feature, as shown by the key at right,
    which also indicates the values for each of the four axes: A, B, C, and D.
    The position angle is calculated modulo \ang{180} so that the positive and negative arm of each axis have the same color.
    Also shown (gray ellipses) are the expansion velocity and range of radii of the inner (dark) and outer (light) halo features (\S~\ref{sec:haloes}).
  }
  \label{fig:ages}
\end{figure}

\begin{figure}
  \includegraphics[width=\linewidth]
  {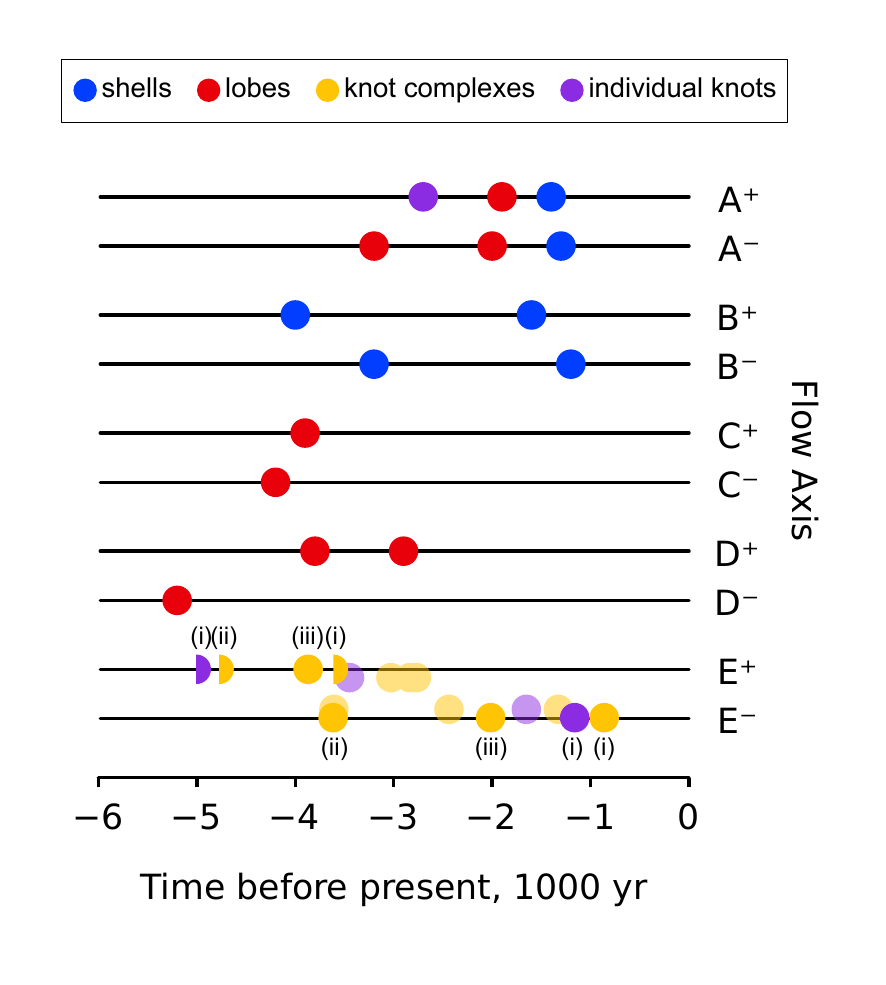}
  \caption{
    Ejection history of each flow axis,
    assuming no acceleration or deceleration of the features.
    Lower-case roman numerals indicate sub-axes of axis~E.
    Uncertainties are typically of order \(\pm 30\%\),
    see Table~\ref{tab:3d} for details,
    but are larger for \AxM{D} and \AxM{E}(ii).
    For Axis~E, semi-circular symbols indicate lower limits,
    while fainter symbols show the results of a re-analysis
    assuming a worst-case systematic uncertainty in the proper motions.
  }
  \label{fig:axis-history}
\end{figure}

Figure~\ref{fig:ages} plots the deprojected velocity versus position for each feature,
classified by type (symbol shape) and position angle (symbol color).
In the case of the halo, no line splitting is observed,
so we estimate the expansion velocity as \(V = 0.5 W\),
where \(W\) is the intrinsic FWHM of the line (section~\ref{sec:haloes}).
Diagonal lines show constant values of kinematic time (distance divided by velocity)
with values between \num{1000} and \SI{16000} years.
There is no clear trend for the velocity to increase as a function of distance,
but rather a wide variety of kinematic times are found among the different features.
Figure~\ref{fig:axis-history} shows a timeline of the history of ejections along each axis,
assuming that the kinematic time reflects the true age of each feature.
Axes~A to~D tend to show matched pairs of ejections with similar ages along the positive and negative arms.
There is evidence that the ejections along axes~A and B are more recent than those along C and~D.\@
Axis~E is very different, with all the ejections along the positive arm occurring first,
followed by all the ejections along the negative arm.

This dichotomy between the two arms of axis~E
is such a surprising result that it is important to establish how robust it is in the face of the uncertainties in our analysis.
The kinematic ages depend on the plane-of-sky positions and proper motions (columns 8 and 5 of Table~\ref{tab:3d}),
and the uncertainties in the proper motions are of two types: random and systematic.
The random uncertainties for individual features are quite large,
greater than 50\% for the \AxP{E} axis and \(\approx 30\%\) for the \AxM{E} axis,
but these can be reduced by aggregating multiple independent measurements.
For instance, by averaging all of the features
associated with each arm of the axis,
we find \(t_{\text{kin}}(\AxP{E}) = \SI{6.7 \pm 1.6}{kyr}\) 
versus \(t_{\text{kin}}(\AxM{E}) = \SI{1.6 \pm 0.2}{kyr}\),
which differ by more than \(3\sigma\).
However, this does not account for the systematic uncertainty
in the image alignment between observational epochs,
which we estimate to be equivalent to
\SI{1.5}{arcsec.yr^{-1}} or \SI{\pm 14}{km.s^{-1}}
(see section~\ref{sec:proper-motions}).
We therefore consider a scenario in which a shift of this magnitude
is applied between the two epochs of \textit{HST} observations.
The shift is applied along \(\text{PA} = \ang{315}\),
which is the average projected axis of the knot outflow,
and in the sense that would tend to reduce the difference between the two sides of the outflow.
The proper motion measurements are repeated under this worst-case scenario,
yielding \(t_{\text{kin}}(\AxP{E}) = \SI{3.7 \pm 0.5}{kyr}\) 
versus \(t_{\text{kin}}(\AxM{E}) = \SI{2.3 \pm 0.3}{kyr}\).
The difference between the timescales of the positive and negative axes is now smaller, \(\SI{1.4 \pm 0.6}{kyr}\),
but it is still significant at the \(2\sigma\) level.
We therefore conclude that the redshifted knots truly were ejected
earlier than the blueshifted knots on average,
although,
after taking into account all sources of uncertainty,
we cannot exclude some temporal overlap between individual features. 
Note that any breakdown of our auxiliary assumptions
(strictly radial flow, no acceleration or deceleration of knots)
would not change this conclusion unless it acted differentially on the two sides of the outflow.

\subsection{Density structure of the nebula}
\label{sec:density-structure}

\begin{table*}
  \caption{Physical parameters of nebular components}
  \label{tab:summary}
  \begin{tabular}{
    l 
    r 
    r 
    S 
    r 
    S 
    S 
    S 
    r 
    }
    \toprule
    {}          & {\Ha{} flux} & {Ionized Density} & {Ionized Mass}    & {Ionization Parameter} & {\(t_{\text{start}}\)} & {\(t_{\text{end}}\)} & {\(\dot{M}\)} & {\(V\)}\\
    {Component} & {\% of total}& {\si{H.cm^{-3}}} & {\(M_\odot\)} &          & {\SI{1000}{yr}}      & {\SI{1000}{yr}}    & {\si{\msun.yr^{-1}}} & {\si{km.s^{-1}}}\\
    \addlinespace
    Inner shell & 70\% & 4000--5000 & 0.078 & 0.0036--0.0070 & -4 & -1.5 & 3.1e-5 & 40--55\\
    Intermediate shell & 13\% & 700--1000 & 0.135 & 0.0080--0.0140 & -10 & -4 & 2.3e-5& 35--38\\
    Red knots & 11\% & 2000--5000 & 0.038 & 0.0004--0.0020 & -5 & -3 & 9.5e-6& 25-40\\
    Blue knots & 4\% & 1500--2000 & 0.025 & 0.0040--0.0170 & -3 & -1 & 1.2e-5& 40--65\\
    Outer Lobes & 1\% & 160--320 & 0.024 & 0.0086--0.0224 & -5 & -2 & 8e-6 & 40--90\\
    Inner Halo & 0.5\% & 80--160 & 0.030 & 0.0200--0.0340 & -20 & -10 & 3e-6 & 15: \\
    Outer Halo & 0.2\% & 15--40 & 0.042 & 0.0540--0.0760 & -40 & -20 & 2.1e-6 & 7: \\
    \addlinespace
    NW knot & 0.01\% & 6700 & 1e-5 & 0.0006 & & & & 24\\
    SE knot & 0.08\% & 8200 & 6e-5 & 0.0034 & & & & 54\\
    N knot &  0.02\% & 3000 & 3e-5 & 0.0012 & & & & 67\\
    \bottomrule
  \end{tabular}
\end{table*}

To further investigate the structure of NGC~6210,
we have used HST WFPC2 narrow-band images to derive the surface brightness in the lines \Ha{} \Wav{6563}, \Hb{} \Wav{4861}, \nii{} \Wav{6583}, and \oiii{} \Wav{5007}.
Flux calibration is carried out following the procedure outlined in \citet{Rubin:2002a}, which accounts for contamination of each emission line filter by non-target lines and atomic continuum emission \citetext{see also \citealp{Ueta:2019a}}.
Foreground dust extinction is corrected for using the observed Balmer decrement, \(\Ha/\Hb\),
assuming an intrinsic value of 2.85 and the reddening law of \citet{Cardelli:1989a},
yielding an average extinction of \(C(\Hb) = 0.13\) for the nebula.
The total extinction-corrected \Ha{} flux for the nebula is then found by summing over the entire HST image,
yielding \(F(\Ha) = \SI{3.176e-10}{erg.s^{-1}.cm^{-2}}\),
which is very close to that found by previous studies \citep{Liu:2004a}.

For each of the nebular features listed in Table~\ref{tab:3d},
we have measured the \Ha{} surface brightness \(S(\Ha)\) and flux \(F(\Ha) = S(\Ha)\,d\Omega\),
where \(d\Omega\) is the solid angle of each feature.
Results aggregated by type of feature are presented in Table~\ref{tab:summary}.
Column~2 gives the fraction of the total \Ha{} flux due to each component,
which is dominated by the high-ionization inner shell.

\begin{figure}
  \includegraphics[width=0.8\linewidth]
  {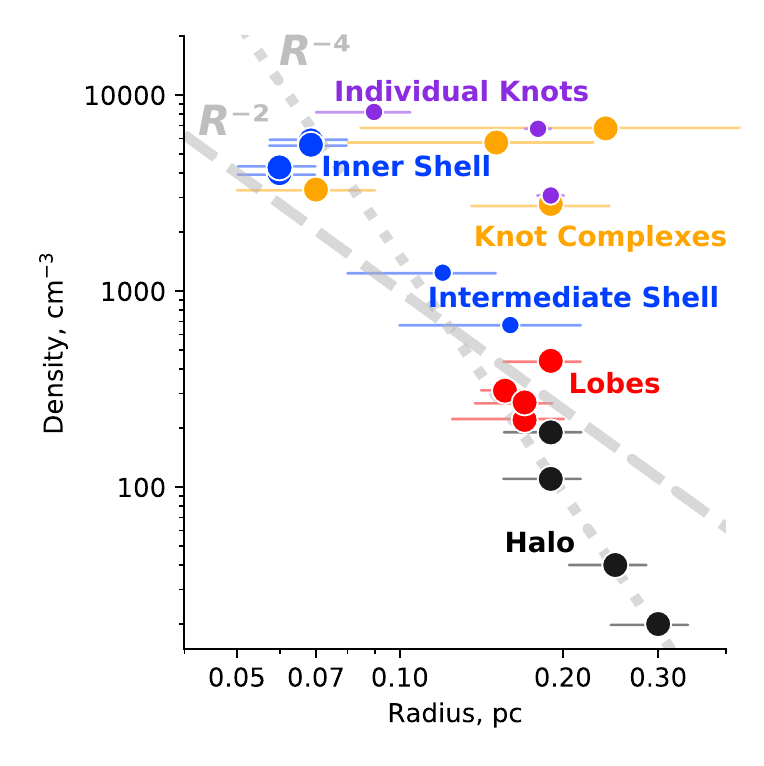}
  \caption{
    Ionized density of different nebular features as a function of true distance from the central star.
    Densities are derived from the \Ha{} surface brightness, as explained in the text,
    and distances are deprojected using the kinematic information of Table~\ref{tab:3d}.
    Gray lines show power-law density distributions of \(n \propto R^{-2}\) (dashed)
    and \(n \propto R^{-4}\) (dotted).
  }
  \label{fig:density-plots}
\end{figure}

Assuming a homogeneous fully-ionized emission region of hydrogen density \(n\) and line-of-sight depth \(dz\), the \Ha{} surface brightness due to recombinations is predicted to be 
\begin{equation}
  \label{eq:sha-density}
  S(\Ha) = \frac{1.0932\, \alpha_{\text{eff}}\, E(\Ha) \, n^2 \, dz} {4\pi}
\end{equation}
where \(\alpha_{\text{eff}} = \SI{8.6e-14}{cm^{3}.s^{-1}}\) is an effective recombination coefficient \citep{Osterbrock:2006a}%
\footnote{We assume \(T = \SI{9500}{K}\) and that the Lyman lines are optically thick, while the Lyman continuum is optically thin, as appropriate for a matter-bounded nebula.}
and \(E(\Ha) = h c / \lambda = \SI{3.027e-12}{erg}\).
The factor \num{1.0932} accounts for the free electron contribution from Helium, assumed to be singly ionized.
Therefore, if \(dz\) can be determined for each feature, then the ionized density follows as \(n \propto [S(\Ha) / dz]^{1/2}\).
We estimate \(dz\) differently for knot-like and shell-like features.
For the first, we take \(dz = (dr\, ds)^{1/2}\), where \(dr\) is the projected width of the knot in the radial direction and \(ds\) is the width in the transverse direction.
For the second we take the maximum chord length of the shell, calculated as \(dz = 2 [R_c^2 - (R_c - dr)^2]^{1/2}\), where \(R_c\) is the shell radius of curvature and \(dr\) is the shell thickness
(limiting cases are \(dz \approx 2 R_c\) for thick shells and \(dz \approx 2 (2 R_c\, dr)^{1/2}\) for thin shells).

Resultant densities are shown in column~3 of Table~\ref{tab:summary} and are plotted against true (deprojected) radius in Figure~\ref{fig:density-plots}.
In Appendix~\ref{sec:density-calibration}, we compare the \Ha{}-derived densities with measurements of the \sii{} doublet ratio, finding good agreement.
It can be seen that a very wide range of densities is present in the nebula, from \(\approx \SI{4000}{cm^{-3}}\) in the inner shell and knot complexes down to \(\approx \SI{30}{cm^{-3}}\) in the outer halo.
For most components, there is a consistent steep decline of density with radius,
which is well approximated by \(n \propto R^{-4}\) (gray dotted line in the figure).
The exception to this rule is the outlying low-ionization knots,
located at about \SI{0.2}{pc} from the star, which have a density 20--50 times higher than the other components at a comparable radius.

\subsection{Reconstructed mass-loss history}
\label{sec:reconstr-mass-loss}

Once the density is known, then the column density of ionized hydrogen follows as proportional to \(S(\Ha)/n\) and 
the ionized mass of each component can be calculated as
\begin{equation}
  \label{eq:mass}
  M = \frac{4\pi D^2\, \bar{m}\, F(\Ha) }{\alpha_{\text{eff}}\, E(\Ha)\, n} \ , 
\end{equation}
where \(D\) is the distance from Earth (assumed \SI{2}{kpc}) and \(\bar{m} \approx \SI{2.17e-24}{g}\) is the mean mass per hydrogen nucleon.
The total ionized mass of the nebula \emph{including the halo} is found to be \SI{0.372}{\msun} and results for individual components are shown in column~4 of Table~\ref{tab:summary}.
It is interesting to note that the high-density components (inner shell plus knot complexes)
only represent about one-third of the total ionized mass,
despite dominating the emission line flux.
The intermediate shell is the most massive single component (\SI{0.135}{\msun}),
also roughly one-third of the total,
with the remaining third corresponding to the outer lobes and halos.

\begin{figure}
  \includegraphics[width=\linewidth]
  {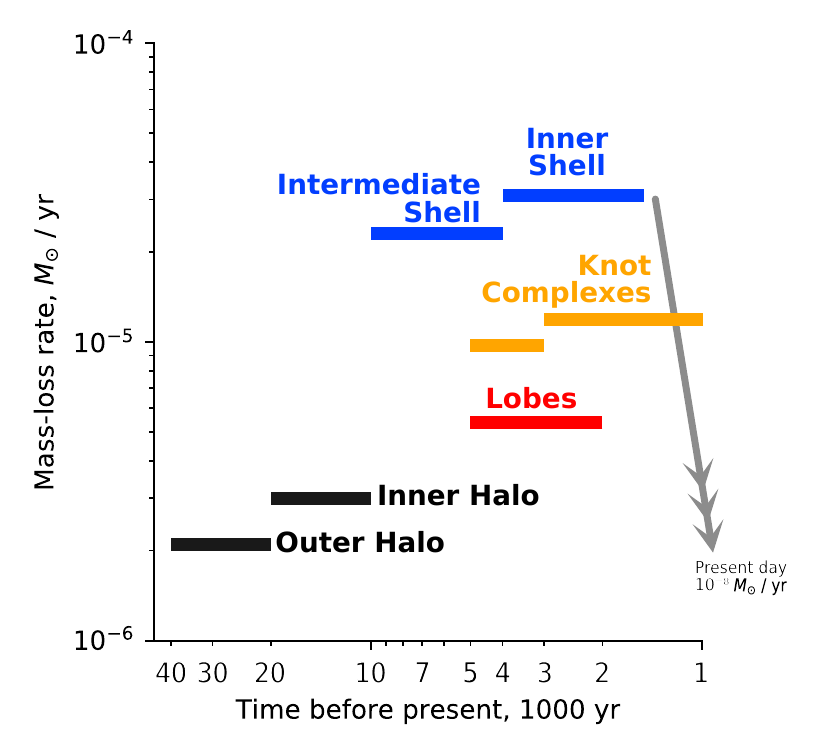}
  \caption{
    Reconstruction of the mass-loss history of NGC~6210.
    Horizontal bars show the time period and mass-loss rate that led to the formation of each nebular component, as labeled.
    Only the average mass-loss rate is shown,
    although for the cases of the knot complexes and lobes the ejection was highly intermittent and non-isotropic during the indicated period.
    The period of high mass-loss ended approximately \SI{1500}{yr} ago.
    Since then, the mass loss rate has declined to a much lower value, as indicated by the gray arrow.
  }
  \label{fig:mass-loss-history}
\end{figure}

The masses can be combined with the kinematic ages (column~10 of Table~\ref{tab:3d}) to estimate the mass loss history of the nebula.
This is given in columns~6 to~8 of Table~\ref{tab:summary} and illustrated in Figure~\ref{fig:mass-loss-history}.
For the start and end times of each component, we take the full range of kinematic times of discrete features in the case of the lobes and knot complexes.
In the case of shells, the start time is taken as the kinematic age of material just outside the shell (for instance, the innermost halo material in the case of the intermediate shell).
In all cases, it is implicitly assumed that no significant acceleration or deceleration has taken place since the material was ejected from the star.
If instead the material was initially ejected at a lower velocity and later accelerated to its currently observed velocity, then the ages would be underestimated.
In general, the mass loss is seen to increase monotonically over the last \SI{40000}{yr} from about \SI{2e-6}{\msun.yr^{-1}} to about \SI{3e-5}{\msun.yr^{-1}} in the last few thousand years.
Note that the values shown in Figure~\ref{fig:mass-loss-history} are average values over the respective periods,
so the sudden jump that is apparent \SI{10000}{yr} ago may have been a more gradual transition.
Also, this method can only estimate mass loss before \SI{1500}{yr} ago, which is the kinematic timescale of the youngest components,
which are the blue knot complexes.
Since that time, the mass-loss rate is expected to have declined significantly
due to the central star transitioning to higher effective temperature and smaller radius,
eventually reaching the present-day value of \(\dot{M} = \SI{9.12e-9}{\msun.yr^{-1}}\) at a wind velocity of \(V = \SI{2150}{km.s^{-1}}\) \citep{Herald:2011a}.
This is indicated on the figure by the gray downward pointing arrow
in Figure~\ref{fig:mass-loss-history}.

\setlist{leftmargin=*}

\section{Discussion}
\label{sec:discussion}

Despite the complex structure of NGC~6210 revealed by our observations,
the broad division into different components (shells, knots, lobes, halo)
is similar to that seen in many other nebulae
\citep{Balick:1987b}
and needs to be analyzed in the light of theories of planetary nebula formation and evolution.
The basic framework for interpreting planetary nebulae has long been the
Generalised Interacting Stellar Winds (GISW) paradigm \citep{Kwok:1978a, Kahn:1985a}:
a slow, dense wind is emitted while the central star is in the cool
Asymptotic Giant Branch phase, which is subsequently ionized after the central star has lost the majority of its outer envelope,
revealing the hot stellar core.
At the same time, a fast, tenuous stellar wind from the core interacts with the envelope to form a dense, shocked shell.
This scenario has been elaborated in great detail in subsequent studies that include additional physics and couple the nebular evolution to the evolution of the central star \citep[for example,][]{Frank:1994b, Garcia-Segura:1997a, Villaver:2002a, Perinotto:2004a, Garcia-Segura:2006a, Steffen:2013a, Toala:2014a}.
Despite the success of the GISW paradigm,
there is much that it cannot explain \citep{Soker:1997a}
and it has become increasingly clear that binary interactions play a vital role in the shaping of planetary nebulae
\citetext{\citealp{Boffin:2019a} and references therein}.
This includes not only stellar companions, but also planets.
For sufficiently close orbits,
a common-envelope phase and Roche lobe overflow can lead to
the ejection of a bipolar or elliptical nebula \citep{Garcia-Segura:2018a}.
For wider orbits, the situation is less clear but various processes can lead to non-spherical mass ejection in this case also \citep{Kim:2012c, Chen:2017b, Chen:2020a}.

A significant fraction of planetary nebulae show deviations even from axisymmetry \citep{Soker:2001a}.
In the cases where these deviations are moderate,
then they may be explained by binary interactions in which the stars are on eccentric orbits,
or the presence of hot or cold spots in the AGB star envelope \citep{Soker:2002b}.
Other cases, where the nebular asymmetries occur only at the outer boundary,
are probably due to interaction with the interstellar medium \citep{Ali:2012a}.
However, a significant minority (\(\sim 10\%\)) of nebulae show such large deviations from bipolar symmetry \citep{Bear:2017a}
that it has been proposed that interacting triple systems are responsible for their shaping \citep{Soker:2004b, Glanz:2020a}.
NGC~6210 falls into this category and has been proposed to be the outcome of either a ``tight binary in a wide orbit'' or a ``tight binary merger'' scenario \citep{Soker:2016b}.
However, given the multitude of evolutionary pathways available in hierarchical triple systems \citep{Toonen:2016a, Toonen:2020a},
other scenarios are also possible.
Sub-arc second imaging observations of young planetary nebulae \citep{Sahai:1998a, Sahai:2011a, Hsia:2014a}
and preplanetary nebulae \citep{Sahai:2007a}
show that multipolarities and asymmetries can appear very early in the evolution of the nebula,
even before the central star is hot enough to emit ionizing photons.

\subsection{Comparison with post-AGB stellar evolution tracks}
\label{sec:comparison-with-post}
\begin{figure}
  \centering
  \includegraphics[width=\linewidth]{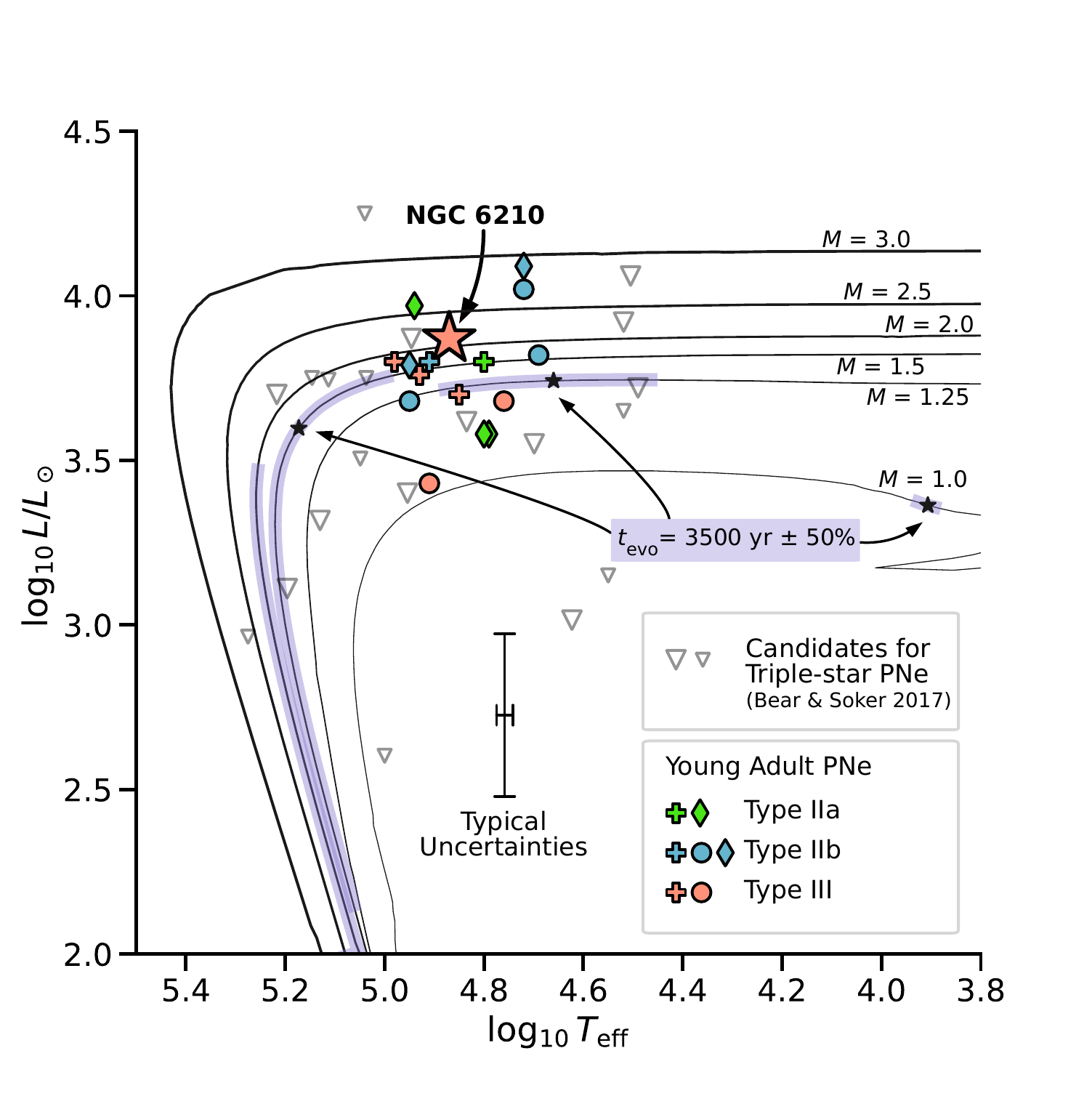}
  \caption{
    HR diagram of planetary nebula central stars,
    showing the observed luminosity and effective temperature of the Turtle
    (large star symbol),
    compared with post-AGB evolutionary tracks from
    \citet{Miller-Bertolami:2016a}
    (solid lines, labelled with initial stellar mass in solar masses).
    Small star symbols show an evolutionary time along each track equal
    to the kinematic age of the intermediate shell of NGC~6210,
    with purple shading indicating 50\% variation about this value.
    Open triangle symbols show sources from \citet{Bear:2017a}
    that are suspected to harbor triple star systems on the basis of the nebular morphology.
    Filled symbols shows the sample of ``young-adult'' nebulae
    that have been selected to be physically similar to NGC~6210
    and which are illustrated in Figure~\ref{fig:cousins}.
    Symbol shape indicates the morphological complexity, from simple (circle) to multipolar (cross).
  }
  \label{fig:hr-diagram}
\end{figure}

Figure~\ref{fig:hr-diagram} shows the Turtle central star on
the Hertzsprung--Russell diagram (large star symbol).
Post-AGB stellar evolutionary tracks for approximately solar metallicity \citep{Miller-Bertolami:2016a} are indicated by solid lines for initial masses between \SI{1}{\msun} and \SI{3}{\msun}
(final masses of \SI{0.532}{\msun} to \SI{0.706}{\msun}),
with evolution proceeding from right to left, followed by top to bottom.

Comparison of the Turtle's central star
(\(\teff = \SI{75 \pm 13}{kK}\), \(L \approx \SI{7400 \pm 3500}{\lsun}\))
with the evolutionary tracks would seem to imply an initial mass of \SI{2}{\msun},
but such an inference is unwarranted since the uncertainty in the stellar luminosity is of the same order as the vertical offset between the different mass tracks.
The evolutionary timescale in this phase is a very steep function of mass
(\(t \sim M^{-5}\); \citealp[Fig.~4]{Miller-Bertolami:2019a}), so comparison with the kinematic expansion timescale of the nebula allows a much more precise mass estimate.
Small black star symbols on the evolutionary tracks indicate evolutionary times of \SI{3500}{yr}, which is the approximate expansion timescale of the intermediate shell.
This shell represents the ``superwind'' material,
initially ejected as a slow neutral outflow at the end of the AGB phase,
and subsequently ionized and accelerated by thermal pressure gradients
once \(\teff > \SI{20000}{K}\) (see \S~\ref{sec:anatomy} for details).
As such, it is the nebular component whose kinematic timescale should most closely match the evolutionary timescale \citetext{see discussion in \S~2.3.4 of \citealp{Schonberner:2005a}}.
Light purple shading shows a variation of \(\pm 50\%\) about this timescale to account for possible shell acceleration/deceleration and observational uncertainties.
It is immediately clear that only the \num{1.25} and \SI{1.5}{\msun} models are at all consistent with the Turtle's central star.
The \SI{2}{\msun} model evolves so fast that within a few thousand years it would be a hot, low luminosity star on the vertical white-dwarf cooling track,
whereas the \SI{1}{\msun} model evolves so slowly that it would still be a B-type star with a low-excitation, ionization-bounded nebula. 
The \SI{1.25}{\msun} model provides the best fit,
which implies a lower luminosity of \SI{5000}{\lsun}
and a slightly closer distance of \SI{1.7}{kpc},
but this is still within the estimated uncertainties and would not significantly effect the analysis in the rest of the paper.

This estimate of the central star mass is consistent with nebular abundance studies
\citep{Liu:2004a, Pottasch:2009a, Delgado-Inglada:2009a, Bohigas:2015c}, which show no evidence for chemical enrichment: \(\chem{O/H} \approx \num{4e-4}\), \(\chem{N/O} \approx 0.15\), \(\chem{C/O} \approx 0.25\).
Spectral signatures of crystalline silicates are seen at mid-infrared wavelengths
\citep{Delgado-Inglada:2014a}, which is further evidence for O-rich chemistry.
Predictions for C and N enrichment as a function of mass
vary significantly with initial metallicity
and between different AGB evolution codes \citep[Figs.~8 and 9 of][]{Henry:2018a},
but all models agree that \(M < \SI{1.5}{\msun}\) is required to maintain \chem{N/O} and \chem{C/O} at or below solar values, as seen in the Turtle.
\citet{Pottasch:2009a} deduced \(M = \SI{0.9}{\msun}\) because the models of \citet{Karakas:2007a} result in noticeable N enrichment for \(M \ge \SI{1}{\msun}\),
contrary to what is observed in the Turtle,
but they noted that such a low mass is inconsistent with the kinematic age of the nebula,
as we discuss above.
However, other stellar evolution simulations \citep{Ventura:2005a, Di-Criscienzo:2016a}
do allow \(\chem{N/O} \approx 0.15\) for initial masses as high as \SI{1.25}{\msun},
which removes this inconsistency.

A mass of \SI{1.25}{\msun} is also consistent with the fact that the Turtle is at a high Galactic latitude, \(b = \ang{37.7}\),
implying a large height \(|z| \sim \SI{1}{kpc} \) above the Galactic plane.
Combined with the large peculiar velocity, \(|\Delta V| = \SI{56}{km.s^{-1}}\),
this leads to a classification as Type~III \citep{Quireza:2007a} according to the scheme of \citet{Peimbert:1978a},
which places the star in the ``thick disk'' stellar population
with ages \SI{> 5}{Gyr}
and masses \(\loa \SI{1.2}{\msun}\) \citep{Maciel:1994b, Stanghellini:2018a}.

Finally, an initial stellar mass of \SI{1.25}{\msun} corresponds to a final mass of \SI{0.566}{\msun} in the \citet{Miller-Bertolami:2016a} models,
meaning that about \SI{0.7}{\msun} should have been lost to mass-loss,
most of which occurs during the AGB phase.
This is consistent with the nebular mass determinations that we make in \S~\ref{sec:reconstr-mass-loss},
where we find a total ionized mass of \SI{0.4}{\msun}.
Our measurement is an underestimate,
since the ionized halo becomes too faint to observe beyond a radius of \SI{1}{pc},
meaning that we are only detecting the last \SI{e5}{years} of mass loss.
We show in Appendix~\ref{sec:ioniz-struct-nebula} that the Turtle is optically thin to H-ionizing radiation and there is no detectable \chem{H_2} emission \citep{Kastner:1996a},
so there is unlikely to be any significant hidden reservoir of neutral gas in the nebula.
Note that our nebular mass determination is much higher than previous estimates for the Turtle,
for instance \citet{Bohigas:2015c} find only \SI{0.046}{\msun}.
This is mainly because those authors only included the emission from the bright,
high-density inner nebula in their estimate,
whereas, as we show in Table~\ref{tab:summary}, the outer faint low-density components
(intermediate shell, lobes, and halo) dominate the total mass.

\subsection{Detailed anatomy of the Turtle}
\label{sec:anatomy}

\subsubsection{Shells}
\label{sec:shells}

The innermost nebular components of NGC~6210 are the nested high ionization shells:
the inner shell, with a radius of about \SI{0.05}{pc}
and the intermediate shell, with a radius of about \SI{0.1}{pc}.
The intermediate shell is the more massive (Table~\ref{tab:summary})
but the inner shell is roughly 5 times denser and dominates the emission line flux from the nebula.
In terms of the GISW model, the intermediate shell represents the detached stellar envelope
that was ejected during a short period of intense mass loss at the end of the AGB phase,
while the inner shell represents material at the inner edge of the detached envelope
that has been compressed by a shock driven by the fast stellar wind from the central stellar remnant.\footnote{%
  Various different names for these components appear in the literature.
  \citet{Villaver:2002a} use ``intermediate shell'' in the same sense that we do
  and use ``main shell'' for the inner shell when it dominates the brightness.
  On the other hand, \citet{Balick:2002a} prefer ``mantle'' and ``rim'',
  while Schönberner and collaborators consistently use ``shell'' for the outer structure
  and ``rim'' for the inner structure
  \citep[e.g.,][]{Perinotto:2004a, Schonberner:2014a, Schonberner:2018a}.
}
The expansion velocity of the inner shell (\num{30} to \SI{50}{km.s^{-1}})
is higher than that of the intermediate shell (\num{20} to \SI{30}{km.s^{-1}}),
which is the opposite to the trend that is typically seen in multiple-shell planetary nebulae \citep{Corradi:2007a}.
In a sample of 20 high-luminosity nebulae, \citet{Schonberner:2014a} find that the intermediate shell expansion velocity lies in the range \num{20} to \SI{40}{km.s^{-1}},
while the inner shell velocity is always slower at \num{10} to \SI{30}{km.s^{-1}}.
In general, the expansion velocities are found to increase with stellar effective temperature
during the high-luminosity phase of central star evolution
\citep{Richer:2008a, Richer:2010a, Pereyra:2016a}.
On the other hand, a faster inner shell such as the Turtle's
is not totally unprecedented,
with a particularly extreme example being NGC~2392,
where the inner shell has an observed expansion velocity of up to \SI{120}{km.s^{-1}} \citep{Garcia-Diaz:2012a}.
In both nebulae the inner shell is non-spherical,
with fastest expansion along the line of sight in NGC~2392,
but in the plane of the sky in the case of the Turtle.
In addition, the Turtle's inner shell is not cylindrically symmetric but rather shows two distinct axes, A and B (\S~\ref{sec:high-ioniz-shells}).
Axis~A corresponds to the projected major axis of the inner shell,
which lies close to the plane of the sky,
while axis~B is inclined at roughly \ang{40} to the plane of the sky
and corresponds to the largest net velocity differential across the shell.
Axis~B also coincides with the major axis of the intermediate shell,
which seems to have a simpler, more elliptical morphology,
albeit slightly lop-sided.

\subsubsection{Lobes and knots}
\label{sec:lobes-knots}

Outside of the nested shell structure lie two highly anisotropic outflow systems with very different characteristics:
the lobes and the knot complexes.
The lobes, located at radii of \num{0.15} to \SI{0.25}{pc},
are high-ionization and show a bow-shock morphology in \oiii{},
with maximum outflow velocities of \SI{90}{km.s^{-1}}.
They are arranged on three bipolar axes: A, C, and D (Figure~\ref{fig:outer-lobe-systems}),
all of which are close to the plane of the sky (Figure~\ref{fig:cut-axis-3d}).
The kinematic ages of the flows along axes C and D are similar to or slightly larger than that of the intermediate shell (Figures~\ref{fig:ages} and \ref{fig:axis-history}),
suggesting that these lobes were ejected contemporaneously with the ``birth'' of the planetary nebula \SI{3500}{yr} ago.
The flow along axis~A is younger (about \SI{2000}{yr}), implying that it was ejected more recently, after the central star had already begun to ionize the nebula.
This is the same axis as the major axis of the inner shell (Figure~\ref{fig:shell-velocity-axes}),
which has a similar kinematic timescale of \SI{1500}{yr}.
Another difference between the axes is that the red-shifted arm of axis~A uniquely shows a low-ionization \nii{}-bright feature, the N~knot,
which shares the kinematics of the more diffuse \oiii{} emission.

The knot complexes are much lower ionization than the lobes and are most prominent in \nii{}.
The redshifted knot complexes are also located at radii of \num{0.15} to \SI{0.25}{pc},
but, unlike the lobes, their outflow axis (axis~E) is close to the line of sight (Figure~\ref{fig:inclinations}),
so that in images of the nebula they appear superimposed on the inner and intermediate shells.
The blueshifted knot complexes lie significantly closer to the star,
at radii of \num{0.07} to \SI{0.15}{pc},
with the closest of them (SE~Blue complex) lying physically inside the intermediate shell.
The outflow velocities of the knot complexes range from \num{30} to \SI{70}{km.s^{-1}},
which are slower than the lobes, but faster than the shells.
As we found with the lobes,
there is evidence for a change in the outflow direction about \SI{3000}{yr} ago,
but in the case of the knot complexes it is a much more dramatic shift of \ang{180},
from predominantly red-shifted ejections before that time to predominantly blue-shifted ejections afterward (see Figure~\ref{fig:axis-history}).
The density of the knot complexes ranges from \num{2000} to \SI{5000}{cm^{-3}},
which is comparable to the density of the inner shell
and is more than an order of magnitude higher than the lobes.
Since the lobes and knots are at similar distances from the star,
they receive the same ionizing flux,
so it is this difference in density that accounts for the low ionization of the knots
(see Appendix~\ref{sec:ioniz-struct-nebula}).

Despite their differences,
the knot complexes and the lobes span a similar range of ages
(Figure~\ref{fig:axis-history}),
and have apparently similar average mass loss rates of \(\sim \SI{1e-5}{\msun.yr^{-1}}\) 
(Figure~\ref{fig:mass-loss-history}).
It is therefore worth asking if they are truly distinct,
or simply two different manifestations of the same collimated outflow phenomenon.
The disparity in their ejection patterns, however,
would argue that the two types of outflow are indeed fundamentally different in nature.
The lobe components tend to occur in point symmetric blue/red pairs along each axis,
with similar kinematic ages.
In contrast,
although the knot complexes also show a partial point symmetry in each of the 3 sub-axes,
E(i), E(ii), and E(iii),
it is very lop-sided, with the blue side invariably being younger.

In addition, the mass loss rate that we calculate does not necessarily reflect the mass loss rate of the collimated outflow,
since it may instead be measuring the mass of the nebula that has been swept up by the outflow.
This might apply to the lobes, whose densities are a few times higher than the density of the inner halo (Figure~\ref{fig:density-plots}),
which is a similar density contrast to that between the inner shell and the intermediate shell.
The mass of the inner shell does not come from the fast stellar wind,
which is very low density,
but rather from ionized slow AGB-wind material that has been shock-compressed by the expansion of this relatively isotropic fast wind.
In a similar way, a fast light jet or collimated wind with a much lower mass loss rate than \SI{1e-5}{\msun.yr^{-1}} may be responsible for driving the lobe flows.

In the case of the knot complexes,
such a scenario is less attractive
because the density contrast between the knots and the inner halo is as high as 100.
A radiative shock with Mach number \(\mathcal{M}\) can produce a maximum compression factor of \(\mathcal{M}^2\),
so that \(\mathcal{M} \sim 10\) is required in order for the knots to be the result of shocks in the AGB outflow.
The relative velocity between the knots and the inner halo yields only \(\mathcal{M} \approx 3\)
in ionized gas,
rendering such a mechanism unlikely.
It is therefore more reasonable to suppose that the knots were directly ejected as dense bullets from the central star system,
in which case the measured mass loss rate is a true reflection of the mass loss in the collimated outflow.

\citet{Guerrero:2020a} studied the kinematics of jets in a sample of 58 planetary nebulae.
They found that the observed distribution of radial velocities can be reproduced by a bimodal distribution of three-dimensional space velocities: 
70\% of nebulae have low-velocity jets with \(V = \SI{66 \pm 26}{km.s^{-1}}\),
while 30\% have high-velocity jets with \(V = \SI{180 \pm 60}{km.s^{-1}}\).
All of the collimated outflows that we find in the Turtle,
with \(V = \num{30}\) to \SI{90}{km.s^{-1}}, are typical of the low-velocity jet population.
When compared with the general taxonomy of collimated outflows seen in planetary nebulae \citep{Lopez:1997a},
the outer lobes are similar to the class of Bipolar Rotating Episodic Jets
\citetext{BRETs, \citealp{Lopez:1993a}}
while the knot complexes are similar to the class of
Fast Low-Ionization Emission Regions \citetext{FLIERs, \citealp{Balick:1994a}}
with the exception that we find no evidence for enhanced N abundance in the knots,
as is often found in FLIERs.

\subsubsection{Halo}
\label{sec:halo}

The outermost component of NGC~6210 is the fully ionized attached halo
(see Fig.~\ref{fig:halo-knots} and Fig.~\ref{fig:halo-components}),
which can be divided into three parts:
the inner halo, outer halo, and halo knots.
The inner halo and outer halo form a single smooth, circularly symmetric structure,
extending from roughly \num{0.1} to at least \SI{0.7}{pc} from the central star.
The distinction between the two is purely kinematic,
with the outer halo showing much narrower lines than the inner halo,
which we interpret as a lower outflow velocity (Table~\ref{tab:summary}).
The halo knots are chains of bright condensations,
visible in the outer halo beyond \SI{0.5}{pc} to the NNW and ENE
(Figures~\ref{fig:halo-knots} and \ref{fig:halo-components}),
each with diameter \(\approx \SI{0.05}{pc}\) and low radial velocities
\(< \SI{5}{km.s^{-1}}\).

In terms of the GISW model, the halo represents the relatively undisturbed remnant of the AGB wind.
The wind is originally neutral,
but it is flash-ionized when the intermediate shell becomes optically thin to EUV radiation.
Since the propagation of the ionization front through the halo is R-type \citep{Kahn:1954a},
the gas initially experiences no mechanical effect and maintains its original density and velocity,
but since the outflow velocity is now subsonic or transonic, the density gradient becomes a pressure gradient that will begin to accelerate the ionized halo.
The acceleration is stronger at smaller radii, which may explain the difference in velocity between the inner and outer halo.
The time required to produce a velocity change \(\Delta v\) is \(t = H \Delta v / c_{\text{s}}^2\),
where \(H\) is the pressure scale height and \(c_{\text{s}} \approx \SI{12}{km.s^{-1}}\) is the isothermal sound speed.
Assuming a power-law density profile, \(\rho \sim R^{-\alpha}\), yields \(H = R / \alpha\), where \(\alpha \approx 4\) is implied by our observations (Figure~\ref{fig:density-plots}).
The transition between the inner and outer halo is observed to occur around \(R \approx \SI{0.22}{pc}\) (Figure~\ref{fig:halo-components}) and the velocity change is \(\Delta v \approx \SI{8}{km.s^{-1}}\), which together imply a halo ionization age of \(t \approx \SI{2500}{yr}\), albeit with considerable uncertainty.

An alternative way of deriving the halo ionization age is to use the observed size of the halo knots,
assuming that they had a very small size when the halo was neutral and have expanded at the sound speed since the halo was ionized.
The result is \(t \approx \SI{2000}{yr}\), which is very similar to the previous estimate. 
This is consistent with the idea that the intermediate shell first became optically thin when the nebula was roughly half its current age and the central star effective temperature was considerably cooler than it is now (\num{40} to \SI{50}{kK}).

\subsubsection{Asymmetries}
\label{sec:asymmetries}
One of the most obvious qualities of the Turtle Nebula is its chaotic and unstructured appearance,
far from the neat symmetries of many other planetary nebulae.
In part this is due to the shear number of symmetry axes in the nebula
(we have found at least five of them),
but there are also asymmetries \emph{along} many of the axes.
In this section we collect together the principle ways in which the nebula is lop-sided:
\begin{enumerate}[1.]
\item The inner shell emission in \oiii{} and \Ha{} shows brightness asymmetries in both position and velocity.
  The redshifted half of the shell is roughly twice as bright as the blueshifted half,
  while both halves show spatial asymmetries along axis~A, but in opposite senses:
  redshifted emission is brighter to the north,
  while blueshifted emission is brighter to the south.
  Because the inner shell dominates the total flux, this leads to a north--south asymmetry in the nebula as a whole,
  with the peak brightness region lying \SI{2}{arcsec} north of the central star. 
\item In the \heii{} line, the inner shell has an opposite asymmetry: it is brighter to the south than to the north, as measured in our slit spectra.
  Such an asymmetry is hard to explain if the \chem{He^{2+}} zone of the nebula is ionization-bounded and the central star emits isotropically.
  Note however that there is a gap in our slit coverage in precisely the region of maximum \Ha{} brightness, so it is possible that this discrepancy is only apparent.
\item The intermediate shell shows an egg-shaped morphology that is lop-sided along its major axis (axis~B),
  extending roughly 30\% further from the star to the WSW than to the ENE.\@ 
\item The outer lobes are probably the \emph{least} asymmetric component of the entire nebula,
  consisting as they do of balanced point-symmetric red-blue pairs of emission components along the axes A, C, and D (Figure~\ref{fig:outer-lobe-systems}).
  However, there is one significant asymmetry: \oiii{} system \SysP{7},
  which corresponds to the N~knot in \nii{} has no corresponding blue-shifted feature to the S.
\item With the low-ionization knot complexes,
  the principle asymmetry is between the redshifted and blueshifted knots.
  The redshifted knots are brighter by a factor of 2.5,
  slower-moving by a factor of 1.5,
  and further from the star by a factor of 2.
\item In the halo, the only significant asymmetry is in the halo knots,
  which are found only to the N and E of the star. 
\end{enumerate}

What is the origin of the asymmetries?
Mechanisms proposed to date \citep{Soker:2004b, Soker:2016b}
involve accretion disk and jet formation in a triple star system.
This may be via accretion of the AGB wind onto a close binary on a wide orbit around the AGB,
or, if the orbit is closer, via a common envelope or grazing envelope phase,
which may or may not result in the merger of the close binary pair.
The descendant system would be binary if a merger or ejection occurred,
or still a triple if not,
but there is no direct evidence that the Turtle's central star is binary or triple.
The fact that the basic parameters of the nebula and its central star can be consistently interpreted in terms of single-star evolution (see \S~\ref{sec:comparison-with-post})
suggests that a triple interaction was not required for the initial ejection of the envelope that forms the intermediate shell,
although it may be responsible for its subsequent shaping.
The possible role of magnetic fields in shaping the nebula and its bipolar outflows
\citep{Garcia-Segura:2000a}
is also unclear.

\subsection{A plausible history of outflows in the Turtle}
\label{sec:poss-hist-outfl}

The complex history of collimated and uncollimated mass-loss episodes
that we have deduced from studying the shells, lobes and knot complexes
should provide important constraints for any future modeling of this nebula:
\begin{enumerate}[1.]
\item Around the end of the central star's AGB phase,
  there were several episodes of unipolar, collimated dense bullet ejections
  along axis \AxP{E},
  which resulted in what we now observe as the redshifted knot complexes. 
\item Either at the same time, or possibly a few thousand years later,
  collimated fast bipolar jets were launched along axes~C and D,
  driving shocks into the slow AGB wind to form what we now observe as the outer lobes.
\item At roughly the same time, approximately \SI{3500}{yr} before the present,
  the central star began to ionize the nebula,
  driving a D-type ionization/shock front into the slow AGB wind
  to form what we now observe as the intermediate shell.
  The expansion is approximately isotropic,
  but slightly elongated and lop-sided along axis~B. 
\item Soon thereafter,
  the fast stellar wind from the central star
  began to drive a shock into the now-ionized AGB wind
  to form the inner shell.
\item Around the same time, the orientation of the bipolar jets shifted to axis~A,
  which coincides with the major axis of the inner shell.
  The most recent ejections along this axis occurred roughly \SI{2000}{yr} ago.
\item Also around the same time,
  the unipolar bullet ejections flipped their direction by \ang{180}
  and began a sequence of outflow episodes along axis \AxM{E}
  to form what we now observe as the blueshifted knot complexes.
  The most recent ejection along this axis occurred roughly \SI{1000}{yr} ago.
\item Roughly \SI{2000}{yr} ago the intermediate shell became completely ionized,
  followed by the rapid ionization of the halo,
  which has led to the acceleration of the inner halo.
\end{enumerate}
Note that these timings rely on kinematic timescales,
which assume constant velocity outflows.
Any significant acceleration or deceleration could modify the numbers,
but is unlikely to change the relative ordering of events.

\subsection{Comparison with other planetary nebulae}
\label{sec:comp-with-other}

\begin{figure*}
  \centering
  \includegraphics[width=\linewidth]{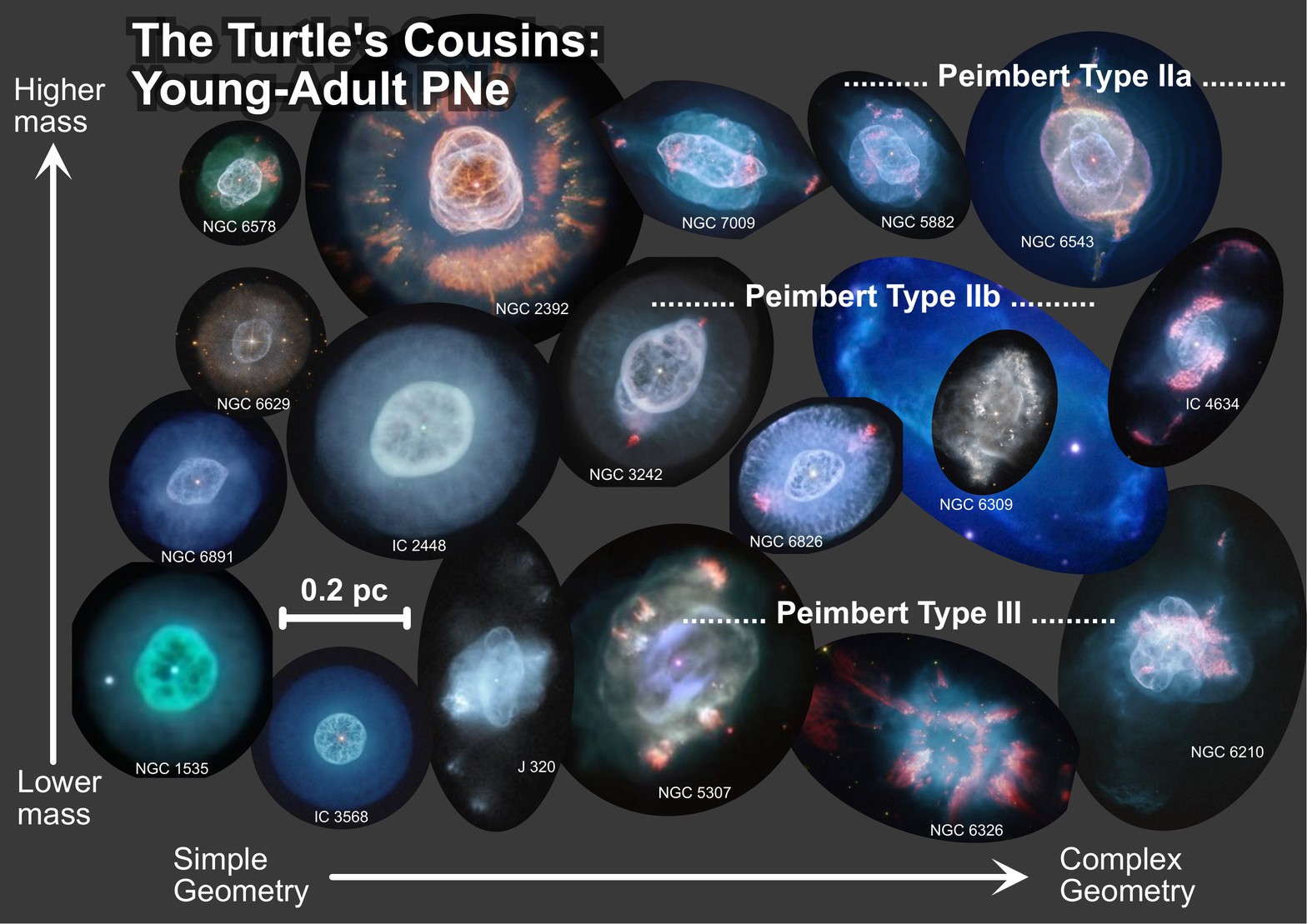}
  \caption[]{
    A selection of planetary nebulae that are similar to NGC~6210
    in terms of central star effective temperature and atmospheric abundances,
    as well as nebular radius, luminosity and ionizing optical depth.
    All nebulae are shown at the same physical scale (see scale bar of \SI{0.2}{pc}).
    Each row corresponds to a different Peimbert type, from bottom to top:
    Type~III, Type~IIb, Type~IIa,
    in decreasing order of stellar population age
    and increasing order of stellar mass.
    Within each row, the nebulae are arranged from left to right
    in order of increasing geometric complexity.
    Image credits: \textit{HST} and Judy Schmidt\footnotemark{}
    for all except
    NGC~1535 (Mount Lemmon SkyCenter, University of Arizona)
    and outer portions of NGC~6309 \citep{Rubio:2015a}.
  }
  \label{fig:cousins}
\end{figure*}

Planetary nebulae as a class are extremely diverse,
so we will concentrate our comparison on those nebulae that are similar to NGC~6210,
either in morphology,
or in other physical characteristics, such as size, brightness, ionization,
and chemical composition.

For morphological similarity, we use the sample of nebulae from \citet{Bear:2017a},
identified based on deviations from axisymmetry or point symmetry.
Of the 89 nebulae in this sample, 24 have central star effective temperatures and luminosities listed in the catalog of \citet{Weidmann:2020a},
and these are shown as open triangle symbols in Figure~\ref{fig:hr-diagram},
with those listed as ``Triple progenitor'' shown by larger symbols 
and those listed as ``Likely triple'' shown by smaller symbols.
It can be seen that the putative triple-star sources are widely distributed along the evolutionary tracks,
implying that the effects of triple interactions
can be seen for a wide variety of nebula ages and central star masses.
It is notable, however, that only one of the triple sources has a luminosity below \SI{300}{\lsun},
whereas such low-luminosity central stars represent a significant fraction 
in other samples of well-measured sources,
such as the ``Golden Astrometry PNe'' sample \citep{Gonzalez-Santamaria:2019a}.
Such sources are typically old, with very extended nebulae, so their relative absence may be due to the signatures of triple-interaction becoming increasingly ``washed out'' as the nebula expands.
On the other hand, it may also be that asymmetries in such nebulae tend to be ascribed to interaction with the interstellar medium instead of triple interactions,
and increasing incompleteness of the \citet{Weidmann:2020a} catalog for fainter stars is another likely contributing factor.

\newcommand\G[2]{\ensuremath{\mathrm{PN\,G#1{#2}}}}
For physical (non-morphological) similarity, we select a primary sample of Peimbert Type~III nebulae from the catalog of \citet{Quireza:2007a} that satisfy all the following criteria:
\begin{enumerate}[1.]
\item Moderate stellar effective temperature \(\SI{50}{kK} < \teff < \SI{100}{kK}\).
\item Matter-bounded, optically thin nebula: ratio of \ion{He}{1} to \ion{H}{1} Zanstra temperature \(\ge 1.3\) \citep{Phillips:2003c}.
\item Within a distance of \SI{4}{kpc} of the Earth
  (distances are taken from \citealp{Frew:2016a}). 
\item Similar nebular physical size to NGC~6210: \(\SI{0.07}{pc} < R < \SI{0.2}{pc}\),
  where \(R\) is measured on the minor axis of the intermediate shell.
\item Central star with H-rich spectral classification, principally O(H). \citep{Weidmann:2020a}
\end{enumerate}
This yields five sources in addition to the Turtle:
NGC~1535 (\G{206.4}{-40.5}),
IC~3568 (\G{123.6}{+34.5}),
Jonckheere~320 (\G{190.3}{-17.7}),
NGC~5307 (\G{312.3}{+10.5}),
and NGC~6326 (\G{338.1}{-08.3}).
The central stars are plotted on the HR diagram as orange symbols in Figure~\ref{fig:hr-diagram}
and the nebulae themselves are illustrated in the bottom row of Figure~\ref{fig:cousins},
all at the same physical scale.

\footnotetext{Originals of processed \textit{HST} images by Judy Schmidt are available at
  \url{https://www.flickr.com/photos/geckzilla/},
  licensed under Creative Commons CC BY 2.0
  \url{https://creativecommons.org/licenses/by/2.0/}.}
We also select a secondary sample of Peimbert Type~IIb,
which otherwise meet the 5 criteria given above,
yielding another seven sources:
NGC~6891 (\G{054.1}{-12.1}),
NGC~6629 (\G{009.4}{-05.0}),
IC~2448 (\G{285.7}{-14.9}), 
NGC~3242 (\G{261.0}{+32.0}), 
NGC~6826 (\G{083.5}{+12.7}),
and NGC~6309 (\G{009.6}{+14.8}),\footnote{
  We relax criterion 5 to admit NGC~6309, which has spectral classification ``O(He)?''.  
}
and IC~4634 (\G{000.3}{+12.2}).
Applying the same criteria to Type~IIa nebulae yields more than 10 sources,
from which we cherry pick five of the most famous:
NGC~6578 (\G{010.8}{-01.8}), 
NGC~2392 (\G{197.0}{+17.3}), 
NGC~7009 (\G{037.7}{-34.5}), 
NGC~5882 (\G{327.8}{+10.0}), 
and NGC~6543 (\G{096.4}{+29.9}).\footnote{
  Again, we relax criterion 5 to admit NGC~6543, which has spectral classification ``Of-WR(H)''.
  We also relax criterion 4 to admit NGC~2392, which is slightly larger than the other nebulae (\(R = \SI{0.22}{pc}\)).
} 
Type~II nebulae 
have low \(z\)-heights and peculiar velocities,
indicating membership of a younger, thin-disk stellar population \citep{Quireza:2007a}.
Type~IIb have similar abundances to Type~III,
while type~IIa are moderately enriched in N and He. 
On average, the progenitor mass should increase in sequence from Type~III,
through IIb, to IIa,
but there is a large overlap between the types \citep{Stanghellini:2018a}
and by filtering on both \teff{} and \(R\) we are probably restricting our samples to a narrow range of masses.%
\footnote{
  The highest mass progenitors (\(M > \SI{2}{\msun}\)) are associated with Peimbert Type~I nebulae,
  but no such nebulae from the \citet{Quireza:2007a} catalog satisfy our combined constraints.
  This may well be because the faster evolutionary timescale of higher mass stars on the post-AGB heating track
  (see \S~\ref{sec:comparison-with-post})
  means that \teff{} exceeds \SI{100}{kK} while the nebula is still compact,
  \(R < \SI{0.07}{pc}\).
}

We call the nebulae in these samples ``Young-adult PNe''
because they are an intermediate evolutionary stage between ``Young PNe''
(which are ionization bounded,
with \(\teff < \SI{50}{kK}\) and \(\oiii/\Ha < 1\), \citealp{Sahai:2011a})
and the fully mature nebulae whose central stars have reached their maximum \teff{} of \num{100} to \SI{200}{kK} and are about to descend the white dwarf cooling track.
As can be seen in Figure~\ref{fig:cousins},
all of these nebulae are built from the same basic architectural elements:
a bright inner shell (rim), and a fainter intermediate shell.%
\footnote{
  Although the inner shell is rather diffuse in two of the Type~III nebulae:
  NGC~5307 and NGC~6326.
}
Many also, like the Turtle, have larger-scale, faint ionized halos,
but these are not shown in the figure.
However, the details of the morphology can vary greatly:
two of the Type~III nebulae, NGC~1535 and IC~3568,
have a very regular, spherical appearance,
albeit with low amplitude irregularities in the inner shell.
Three of the Type~IIb nebulae also show a regular structure,
but these tend to be elliptical rather than spherical,
especially in the inner shell.
The remaining two-thirds of the sample nebulae all show
bipolar or multipolar structures.
In most cases, there is one principal axis of symmetry,
which corresponds both to the elongation of the inner shell and dominant low-ionization knots.
Others, such as J320 and NGC~5307 have multiple axes,
but even in these cases the structures are approximately point symmetric.
\citet{Chong:2012a} propose that a very wide range of morphologies can be reproduced using three pairs of lobes,
and it is indeed true that nearly all of our sample can be accommodated under their scheme,
with the exception of the Turtle,
which is a clear outlier in terms of the extremity of its asymmetries.

In some cases, the apparent differences in morphology are due to viewing angle effects.
An example is NGC~2392 where morpho-kinematic modeling \citep{Garcia-Diaz:2012a} shows that the nebula is highly elongated,
similar to NGC~7009 and NGC~6543,
but appears much more circular than those nebulae because its symmetry axis is close to the line of sight.
In most other cases, however, the differences in morphology are intrinsic.
Longslit spectroscopy reveals fast collimated outflows in many of the nebulae with lobes
\citep{Rechy-Garcia:2020a},
whereas most apparently circular or mildly elliptical nebulae show no evidence for such outflows
\citep{Corradi:2007a},
indicating that they truly are approximately spheroidal. 
It is particularly instructive to compare the Turtle with the almost-spherical NGC~1535,
since the central stars of these two nebulae are almost perfect twins \citep{Herald:2011a},
with closely similar effective temperature, surface gravity, and stellar wind mass-loss rate and velocity.
The basic parameters of the nebulae
(radii, densities, chemical composition, and velocities of the shells and halo)
are also similar.\footnote{
  Minor exceptions are that NGC~1535 is slightly higher excitation,
  as measured by \heii{}/\Hb{} \citep{Barker:1989a},
  and has a higher carbon abundance \citep{Kwitter:1996a},
  although both nebulae are oxygen-rich with \chem{C/O < 1}.
}
Despite these similarities,
NGC~1535 is totally lacking the fast lobes,
low-ionization knots, and asymmetries that are so prominent in the Turtle.
The implication is that none of these phenomena are intrinsic or necessary
for the formation and expansion of the main planetary nebula shell.


\section{Conclusions}
\label{sec:conclusions}

We have carried out detailed velocity mapping of the Turtle Nebula, NGC~6210,
via longslit spectroscopy at 34 slit positions in \Ha, \heii, \oiii, and \nii{} lines.
By combining the spectroscopy with proper motion measurements from HST imaging,
we derive the three-dimensional structure and kinematics
of the multiple components that comprise this complex, asymmetrical nebula.
Our preferred systemic heliocentric velocity for NGC~6210 is \SI{-39.3}{km.s^{-1}},
based on the centroid velocity of the slowly expanding halo. 

We find five separate ejection axes in the nebula,
which show an intricate interplay between symmetries and asymmetries.
The first two axes, A and B, which are approximately perpendicular to one another,
describe the major axes of
the inner shell (rim) and intermediate shell, respectively.
The intermediate shell is the most massive component of the nebula (\SI{0.135}{\msun})
and probably represents the bulk of the mass lost by the central star at the end of its AGB phase.
It has a kinematic age of roughly \SI{3500}{yr},
which should roughly correspond to the length of time since the central star begain to ionize the nebula.
Its shape is approximately a prolate spheroid,
with 2:1 axis ratio and inclined at \ang{40} to the plane of the sky,
but spatially offset from the star.
The high-density inner shell dominates the \Ha{} flux from the nebula,
even though it is only half the mass of the intermediate shell.
The inner shell represents material that has been shock-compressed by the expansion of the fast stellar wind from the central star.
Its projected shape is approximately bipolar
and with axis of elongation (axis~A) at \ang{15} to the plane of the sky,
but its true shape must be more complex
since it shows a pronounced velocity gradient parallel to the major axis of the intermediate shell (axis~B).
Although the inner shell is geometrically well-centered on the central star,
it shows complex brightness asymmetries in both position and velocity,
with an opposite sense in \heii{} to that in \Ha{} and \oiii{}. 
The expansion velocity of the intermediate shell is \SI{35}{km.s^{-1}},
while the inner shell expands at \SI{50}{km.s^{-1}},
which is atypically fast for the rim of a double-shell planetary nebula.

The next two axes, C and D, are ejection axes of the high-ionization outer lobes of the nebula.
The lobes are bipolar flows with a high degree of point symmetry,
inclined at \ang{10} to \ang{20} to the plane of the sky.
They show outflow velocities of \num{40} to \SI{60}{km.s^{-1}} close to the axis
and slightly lower velocities in a wider angle flow, with bow-shock like morphologies at their tips.
We propose that they are the result of collimated bipolar fast winds or jets from the central star system,
which shock and entrain material from the intermediate shell and ionized halo of the nebula. 
The kinematic timescales imply that axis~C and D were active around \num{3000} to \SI{5000}{yr} ago, around the same time as the ``birth'' of the nebula.
An additional lobe outflow is aligned with axis~A
(the axis of elongation of the inner shell),
with a higher speed of \SI{90}{km.s^{-1}}.
Axes~A and C are almost coincident on the plane of the sky,
but can be clearly separated kinematically in the spectra.
The axis~A outflow is younger than the other lobes,
with a kinematic age of \num{2000} to \num{3000} years.
Although its high-ionization emission shows the same point symmetry as the other lobes,
it is unique in having a low-ionization compact component (N knot),
which is present on one side only.

The final axis, E, is the ejection axis of the low-ionization knot complexes,
which dominate the appearance of the nebula in the \nii{} line.
Axis~E is highly inclined (\(\approx\ang{70}\)) to the plane of the sky
and the low-ionization outflow is less well collimated than the lobes.
These combine to make the axis hard to identify in images,
although it is clearly revealed in the longslit spectra.
Furthermore, while A, B, C, and D are symmetry axes,
axis~E would be better described as an \emph{asymmetry} axis,
since the kinematic timescales are utterly different between the redshifted and blueshifted sides.
The red semi-axis (predominantly N and W of the central star)
was active with outflow velocities of \SI{30}{km.s^{-1}}
before or around \SI{3500}{yr} ago.
Uncertainties in the proper motion measurements mean that we cannot tell
if the ejection began before the planetary nebula phase or not.
In contrast, the blue semi-axis (predominantly S and E)
was active with outflow velocities of \SI{50}{km.s^{-1}}
between \num{1000} 1nd \SI{3000}{yr} ago,
making it the kinematically youngest component of the nebula.
The density of the low-ionization knots is too high for them to be explained by shocked ionized halo material,
so we propose that they represent dense bullets ejected from the central star system.
Their total mass is \SI{0.06}{\msun}, or about 30\% of the mass of the nebular shells. 

NGC~6210 is the product of a relatively low-mass (\(\approx \SI{1.25}{\msun}\)) progenitor
and its post-AGB central star is halfway along its high-luminosity heating track.
In some ways it is very similar to much simpler nebulae such as NGC~1535 and IC~3568,
but its rich variety of bipolar and monopolar outflows,
combined with marked asymmetries in the nebular shells,
set it apart from most other nebulae.
An interacting triple star system is a natural explanation for the Turtle's many peculiarities,
but the exact mechanism is unknown
and direct evidence for multiplicity of the central star is still lacking.
We hope that our detailed elucidation of the Turtle's complicated history of anisotropic mass loss will provide inspiration for future modeling efforts.

\section*{Acknowledgements}
Based upon observations carried out at the Observatorio Astronómico Nacional on the Sierra San Pedro Mártir (OAN-SPM), Baja California, México. 
We thank the daytime and night support staff at the OAN-SPM for facilitating and helping obtain our observations.
In particular, we would like to thank the telescope operators
Gabriel García (deceased), Felipe Montalvo, Salvador Monroy, and Gustavo Melgoza
for their skilful support throughout the numerous observing runs that contributed to the results presented here.
We are grateful for financial support provided by
\foreignlanguage{spanish}{
  Dirección General de Asuntos del Personal Académico,
  Universidad Nacional Autónoma de México},
through grants
\foreignlanguage{spanish}{
  Programa de Apoyo a Proyectos de Investigación
  e Inovación Tecnológica}
IN107019 and IN103519,
and to the Mexican
\foreignlanguage{spanish}{
  Consejo Nacional de Ciencia y Tecnología}
through grant A1-S-15140.
Scientific software and databases used in this work include
SAOImage~DS9\footnote{\url{https://sites.google.com/cfa.harvard.edu/saoimageds9}} \citep{Joye:2003a},
SIMBAD, Vizier, and Aladin Lite from Strasbourg Astronomical Data Center (CDS)\footnote{\url{https://cds.u-strasbg.fr}},
the Hong Kong/AAO/Strasbourg H-alpha planetary nebula database\footnote{\url{http://planetarynebulae.net/EN/hash.php}},
and the following Python packages:
numpy, pandas, astropy, matplotlib, seaborn, pyneb, pyflct, astrodrizzle.

\section*{Data availability}
\label{sec:data-availability}

The data underlying this article are available from the
\foreignlanguage{spanish}{San Pedro Mártir} Kinematic Catalogue of Galactic Planetary Nebulae
at \url{http://kincatpn.astrosen.unam.mx}
and from the Mikulski Archive for Space Telescopes
at \url{https://dx.doi.org/10.17909/t9-9g3w-nd31}.

\bibliography{turtle-paper-refs}
\appendix

\section{Density calibration}
\label{sec:density-calibration}

\begin{figure}
  \includegraphics[width=0.8\linewidth]
  {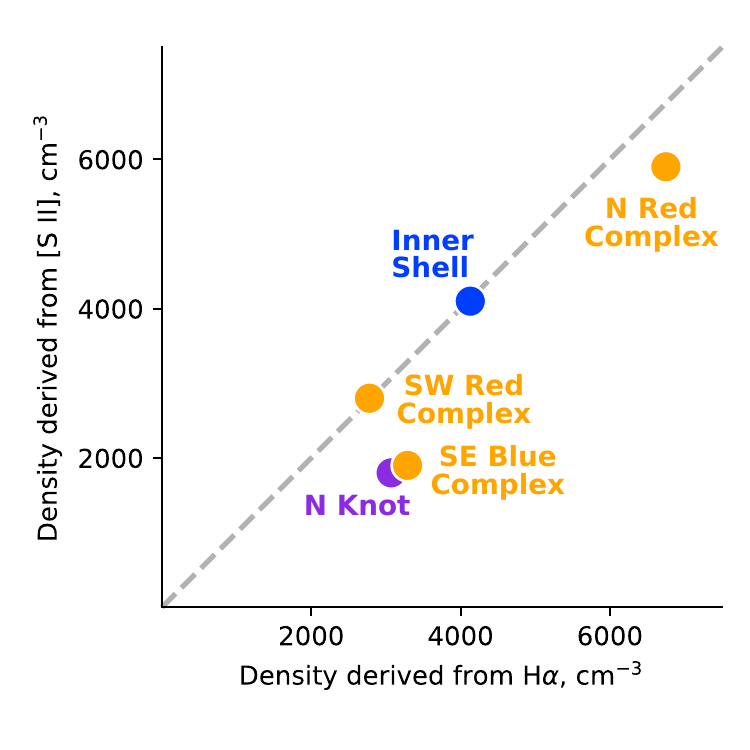}
  \caption{
    Comparison between densities derived from \Ha{} surface brightness and densities derived from the \sii{} 6716/6731 line ratio for the small number of features that are detected in our single MES-SPM \sii{} slit spectrum.
    The dashed line indicates equality of the two estimates.
  }
  \label{fig:density-calibration}
\end{figure}

As a check on our \Ha{}-derived densities, we also calculate densities from MES-SPM observations of the red \sii{} doublet for a single slit position (slit~s of Figure~\ref{fig:slit-positions}).
This passes through five distinct spatio-kinematic features in the inner nebula,
allowing us to measure the line ratio \(I(6716)/I(6731)\) for each
(unfortunately, \sii{} emission from the outer lobes and halo is undetectably weak).
Because the relative spectrograph efficiency between the doublet wavelengths is not well-determined,
we have calibrated the line ratios by forcing the average value over the entire slit to be \(I(6716)/I(6731) = 0.60\), in agreement with the spectrophotometry of \citet{Liu:2004a}.
Results are shown in Figure~\ref{fig:density-calibration},
calculated with PyNeb \citep{Luridiana:2015a} using atomic data from \citet{Podobedova:2009a} and \citet{Tayal:2010a}.
It can be seen that there is a reasonable agreement between the two density diagnostics,
which justifies our decision not to use any filling factor correction in equation~\eqref{eq:sha-density}.
The need for a filling factor is obviated by carefully calculating the line-of-sight depth of each emission component separately.

\section{Ionization structure of the nebula}
\label{sec:ioniz-struct-nebula}

There is a marked difference in appearance of the nebula between
high-ionization lines such as \oiii{} (Fig.~\ref{fig:proper-motions-oiii})
and low-ionization lines such as \sii{}, \oii{} and \nii{} (Fig.~\ref{fig:proper-motions-nii}),
implying that there are strong variations in the degree of ionization.
The appearance in \Ha{} is broadly similar to \oiii{},
which suggests that the doubly ionized ions, such as \chem{O^{2+}} and \chem{N^{2+}},
are the dominant stages,
whereas the singly ionized ions, such as \chem{O^{+}} and \chem{N^{+}},
have low abundance in most of the nebula.
This in turn implies that the nebula must be optically thin in the \chem{H^0} and \chem{He^0} ionizing continua,
which is also consistent with the fact that the \nii{} emission is \emph{not} concentrated in an outer ring,
as would be expected for an optically thick nebula
\citetext{for example, the Ring Nebula, \citealp{ODell:2013b}}.

\begin{figure}
  \includegraphics[width=\linewidth]
  {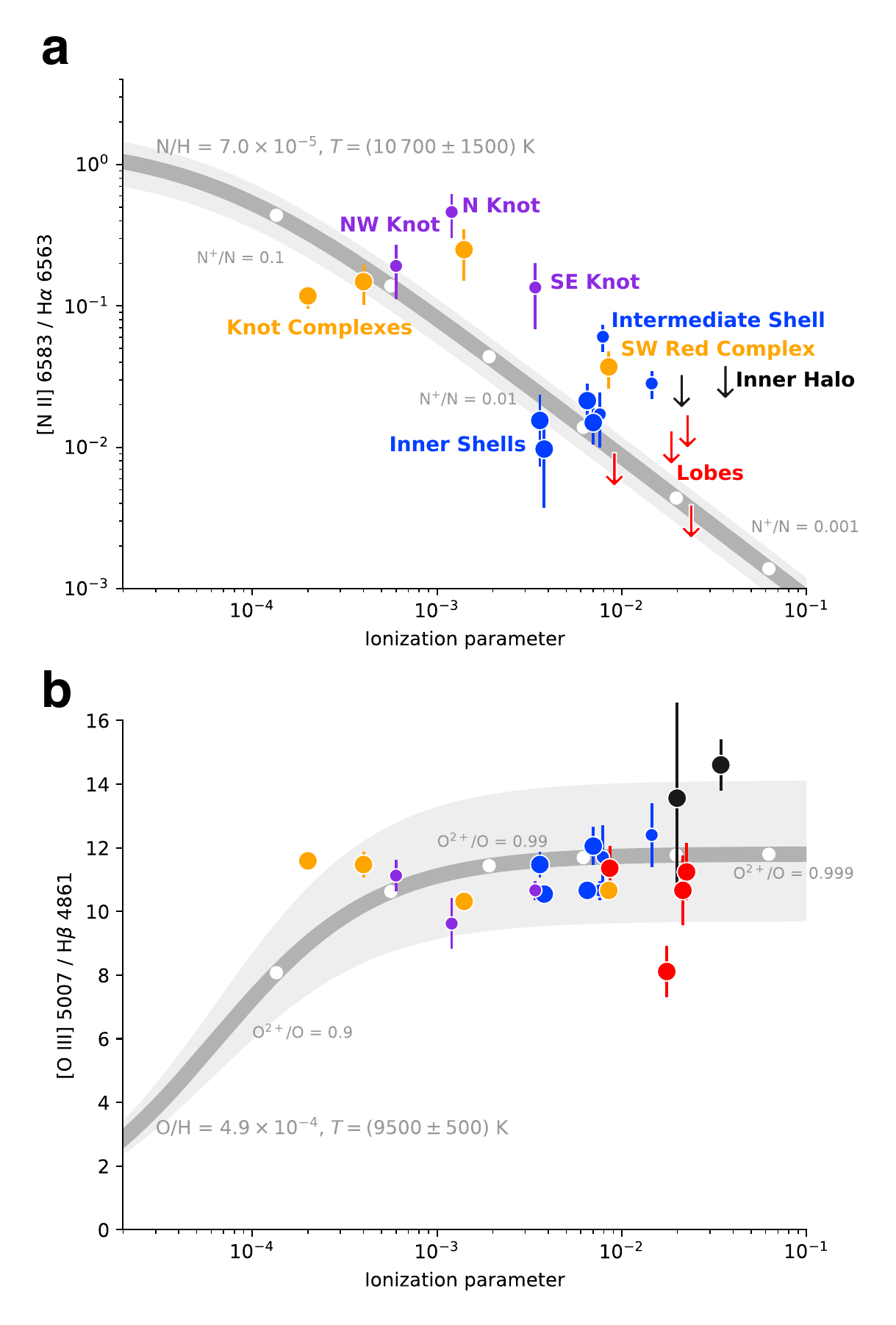}
  \caption{
    Diagnostic line ratios plotted against ionization parameter for different nebular features:
    (a)~\nii/\Ha, (b)~\oiii/\Hb.
    Yellow symbols show knot complexes and purple symbols show individual knots, as labeled.
    Large blue symbols show the inner shell, and small blue symbols show the intermediate shell.
    Red symbols show the outer lobes and black symbols show the halo.
    Since the \nii{} line is not detected from the lobes and halo,
    only upper limits are available for these components in panel~a.
    The ionization parameter is calculated from the \Ha{} surface brightness and the three-dimensional reconstruction, as explained in the text.
    The thick gray line in each panel shows a toy model of fixed elemental abundance and temperature as labeled,
    with varying ion fraction as determined by optically thin photoionization equilibrium as a function of ionization parameter.
    The gray shaded area shows the effect of varying the model temperature by \SI{\pm 1500}{K} (panel a) and \SI{\pm 500}{K} (panel b).
  }
  \label{fig:line-ratios}
\end{figure}

In an optically thin nebula, the flux of ionizing photons at each point is \(\Fion = Q / 4\pi R^{2}\),
where \(Q\) is the ionizing photon luminosity.
The local ionization parameter is defined as \(\ionpar = \Fion / n c\), where \(c\) is the speed of light. 
We have calculated \(\ionpar\) for each of the features listed in Table~\ref{tab:3d}
and list in column~5 of Table~\ref{tab:summary} the range of values found for each component.
In this calculation, we use the density of each feature as calculated from the \Ha{} surface brightness (\S~\ref{sec:density-structure})
and assume \(Q = \SI{4.9e47}{s^{-1}}\), which we derive from the CSPN stellar atmosphere models of \citet{Krticka:2020a} using the stellar parameters of \citet{Herald:2011a}.
In Figure~\ref{fig:line-ratios} we plot the line ratios \nii{} \Wav{6583}/\Ha{} and \oiii{} \Wav{5007}/\Hb{} against the ionization parameter for each feature.
The \nii{} and \oiii{} surface brightness were determined from HST images, as outlined in \S~\ref{sec:density-structure}.

It is apparent from the table and figure that there is a wide range of ionization parameters in the nebula.
The steep decline in density with radius (faster than \(R^{-2}\)) means that the ionization parameter tends to increase with radius, although the lowest values are associated with the knot complexes, owing to their relatively high densities as compared with other features at similar radii.
Figure~\ref{fig:line-ratios} shows that the \nii/\Ha{} ratio varies by two orders of magnitude and is clearly anti-correlated with \ionpar.
The \oiii/\Hb{} ratio, on the other hand, shows no correlation with \ionpar{} and indeed varies little within the nebula.
This is what is expected in an optically thin nebula, where the doubly ionized metals predominate.

As an example, gray solid lines in Figure~\ref{fig:line-ratios} show a toy photoionization model,
in which we assume
\(
\chem{O^{2+}} / \chem{O^+} = \chem{N^{2+}} / \chem{N^+} = A \, \ionpar ,
\)
where \(A \approx \num{1.6e4}\) is a dimensionless constant,
whose value we fix by requiring the model to approximately fit the \(\nii/\Ha\) observations.\footnote{
  Setting the \chem{N} and \chem{O} ion fractions to be equal in this equation is tantamount to assuming an \chem{N^+/O^+} ionization correction factor (ICF) of unity \citep{Kingsburgh:1994a}.
  Fits to a large suite of photoionization models
  \citetext{equations~[14--16] of \citealp{Delgado-Inglada:2014b}}
  imply \(\chem{ICF(N^+/O^+)} = 1.4 \pm 1.0\) for the nebular parameters of NGC~6210,
  whereas spectrophotometry of the whole nebula \citep[Table~7]{Pottasch:2009a} yields  \(\chem{ICF(N^+/O^+)} = 0.78\).
  Both of these estimates are sufficiently close to unity for our purposes.
}
In the model, we calculate the line emissivities using \textsc{Pyneb},
assuming electron temperatures and abundances as marked on the figures,
which are based on the values derived by \citet{Pottasch:2009a}.
Selected ionic fractions are indicated by white dots, which show that,
even in the \nii{}-brightest knots,
the \chem{N^+/N} ratio never exceeds \num{0.1}.

One caveat here is that the measured line ratios correspond to the integrated emission along the line of sight, whereas, 
in the case of the knot complexes,
there is spectroscopic evidence for ionization stratification
(see final paragraph of \S~\ref{sec:knot-complexes}),
which means that the \nii{} and \oiii{} emission may come from distinct zones.%
\footnote{
  The \Ha{} and \Hb{} emission would come from both zones but the \oiii{} zone would predominate due to its larger emission measure.
}
It is therefore possible that \emph{locally} \chem{N^+/N} may reach unity,
but that the effect is masked by integration along the line of sight.

Although we have demonstrated that the nebula as a whole is optically thin in the \chem{H^0}- and \chem{He^0}-ionizing continua,
this is clearly not the case for the \chem{He^+}-ionizing continuum (\(h\nu > \SI{54.4}{eV}\)),
as evidenced by the compact distribution of \heii{} emission at the inner edge of the high-ionization shell (see Figs.~\ref{fig:heii-shell-annotated}, \ref{fig:heii-shell-components}, and \ref{fig:heii-shell-velocity-axes}).
This is consistent with the presence of an ionization-bounded \chem{He^{2+}} Strömgren sphere,
with radius \(\approx \SI{0.04}{pc}\),
which traps all the \chem{He^+}-ionizing radiation.
An interesting consequence is that,
even though the ionization parameter increases with distance from the central star,
there is likely to be no \chem{O^{3+}} in the outer nebula since the photoionization of \chem{O^{2+}} requires \(h\nu > \SI{54.9}{eV}\).
Indeed, our measurements (Fig.~\ref{fig:line-ratios}b) show that \oiii{}/\Hb{} remains high in the lobes and halo,
consistent with \chem{O^{2+}/O \approx 1}.

\bsp	
\label{lastpage}

\end{document}